\documentclass[aps,prd,twocolumn,superscriptaddress,groupedaddress, nofootinbib,preprintnumbers]{revtex4}
\usepackage{bm}
\usepackage[dvipdfmx]{graphicx}
\usepackage{latexsym,amsmath,amssymb,graphicx,booktabs}
\usepackage{hyperref}
\usepackage{placeins}
\usepackage{subfigure}
\usepackage{mathtools}
\usepackage[normalem]{ulem}

\usepackage{aas_macros}
\usepackage{amssymb}
\usepackage{amsmath}
\usepackage{amsfonts}
\usepackage{upgreek}
\usepackage{latexsym}
\usepackage{appendix}
\usepackage{dsfont}

\usepackage[export]{adjustbox}

\newcommand\spart{\;\raise1.0pt\hbox{/}\hskip-6pt\partial}
\newcommand\spartb{\;\overline{\raise1.0pt\hbox{/}\hskip-6pt\partial}}

\newcommand{\bfB}{\boldsymbol{B}}

\newcommand{\bfE}{\boldsymbol{E}}

\newcommand{\bfJ}{\boldsymbol{J}}

\newcommand{\gag}{g_{\rm a\gamma}}
\newcommand{\mass}{m_{\rm a}}

\newcommand{\beq}{\begin{equation}}
\newcommand{\eeq}{\end{equation}}
\newcommand{\bea}{\begin{eqnarray}}
\newcommand{\eea}{\end{eqnarray}}
\newcommand{\unit}[1]{\,\rm{#1}}
\newcommand{\diff}{{\mathrm d}}

\definecolor{mygreen}{RGB}{0,130,0}
\definecolor{myorange}{RGB}{255,165,0}

\begin{document}

\title{Axion dark matter search from terrestrial magnetic fields at extremely low frequencies}

\author{Atsushi Nishizawa}
\email{atnishi@hiroshima-u.ac.jp}
\affiliation{Physics Program, Graduate School of Advanced Science and Engineering, Hiroshima University, Higashi-Hiroshima, Hiroshima 739-8526, Japan}
\affiliation{Astrophysical Science Center, Hiroshima University, Higashi-Hiroshima, Hiroshima 739-8526, Japan}
\affiliation{Research Center for the Early Universe (RESCEU), Graduate School of Science, The University of Tokyo, Tokyo 113-0033, Japan}
\author{Atsushi Taruya}
\affiliation{Center for Gravitational Physics and Quantum Information, Yukawa Institute for Theoretical Physics, Kyoto University, Kyoto 606-8502, Japan}
\affiliation{Kavli Institute for the Physics and Mathematics of the Universe, Todai Institutes for Advanced Study, The University of Tokyo, (Kavli IPMU, WPI), Kashiwa, Chiba 277-8583, Japan}
\author{Yoshiaki Himemoto}
\affiliation{Department of Liberal Arts and Basic Sciences, College of Industrial Technology, Nihon University, Narashino, Chiba 275-8576, Japan}
\vspace{1 em}
\date{\today}

\begin{abstract}
The natural environment of the Earth can act as a sensitive detector for dark matter in ultralight axions. When axions with masses between $1\times10^{-15}\unit{eV}$ and $1\times10^{-13}\unit{eV}$ pass through the Earth, they interact with the global geomagnetic field, generating electromagnetic (EM) waves in the extremely low-frequency range ($0.3$--$30\unit{Hz}$) through axion-photon coupling. This paper is one of a series of companion papers for~\cite{Taruya:2025zql}, focusing on the data analysis method and search results for an axion signal. Utilizing the theoretical predictions of axion-induced EM spectra from a companion study, we analyzed long-term observational data of terrestrial magnetic fields in this frequency band to search for axion-induced signals. Our analysis identified 65 persistent signal candidates with a signal-to-noise ratio (SNR) greater than 3. Aside from these candidates, we placed a new upper bound on the axion-photon coupling parameter, significantly refining the previous constraint from CAST by at most two orders of magnitude down to $g_{a\gamma} \lesssim 4\times10^{-13} \unit{GeV}^{-1}$ for the axion mass around $3 \times 10^{-14}\unit{eV}$.
\end{abstract}
\preprint{YITP-25-55}

\maketitle

\section{Introduction}

Numerous independent astronomical observations suggest that the majority of matter in the Universe consists of an invisible component known as dark matter (DM). Since DM does not fit within the framework of the standard model of particle physics, its fundamental nature remains one of the most profound enigmas in both cosmology and fundamental physics. Among various candidates for dark matter, axions were originally proposed to preserve the CP symmetry in quantum chromodynamics (QCD) without requiring fine-tuning of a CP-violating coupling parameter, addressing the so-called strong-CP problem~\cite{peccei_quinn,Weinberg:1977ma,Wilczek:1977pj}. While there exist other axion-like particles~\cite{Arvanitaki:2009fg}, which are predicted by string theory and can span a broad mass range. These axions and axion-like particles (which we collectively refer to as axions) have long been central to research as prominent ultralight dark matter candidates~\cite{Preskill:1982cy,Abbott:1982af,Dine:1982ah}, with possible masses ranging from $10^{-23}\unit{eV}$ to $1\unit{eV}$ (see, e.g., \cite{Marsh_review2016} for a review).

The axion DM can couple with electromagnetic (EM) fields, described by the interaction Lagrangian 
\beq
\mathcal{L}_{\rm int}=\frac{\gag}{4}\,a\,F_{\mu\nu}\tilde{F}^{\mu\nu}\;,
\eeq 
where $a$ is the axion field, $F_{\mu\nu}$ is the EM field strength tensor with its dual given by $\tilde{F}^{\mu\nu}=\epsilon^{\mu\nu\alpha\beta}F_{\alpha\beta}/2$, and $\gag$ represents the coupling coefficient, referred to as the axion-photon coupling. Despite its weak coupling, the interaction offers a direct way to search for axions in laboratory experiments~\cite{Sikivie_2021review,Adams:2022pbo}. In addition, astronomical observations provide a variety of opportunities to search for the presence of axions through modulation of observed photons, altering the lifetimes of astronomical objects, and introducing new phenomena~\cite{DiLuzio:2020wdo,Galanti:2022ijh}. The absence of these signatures, therefore, allows us to constrain the coupling parameter $\gag$.

A novel and independent probe for axion dark matter is terrestrial EM fields in the extremely low-frequency (ELF) band. Coupled with the static geomagnetic field, coherently oscillating axions permeating the entire Earth can generate monochromatic EM waves at a frequency corresponding to the axion mass $\mass$. The use of such an environment has been investigated in Ref.~\cite{Arza:2021ekq}, obtaining a constraint on $\gag$ in the mass range of $2\times10^{-18}\,{\rm eV}\lesssim \mass\lesssim 7\times10^{-17}\,{\rm eV}$ from the data of the magnetometer network on Earth. More recently, the heavier mass range, $\mass\lesssim 4\times10^{-15}\,{\rm eV}$, has been constrained with the high-sampling measurement data of terrestrial magnetic fields~\cite{Friel:2024shg}. There are other efforts constraining the coupling with direct measurements in the mass range, $10^{-15}\,{\rm eV}$--$10^{-12}\,{\rm eV}$~\cite{CAST:2017uph,CAST:2024eil,Sulai:2023zqw,Oshima:2023csb,Bloch:2023wfz}. 

In this paper, we focus on the EM fields at the frequencies, $0.3-30$\,Hz, corresponding to the axion mass of $1\times10^{-15}\,{\rm eV}\leq \mass\leq 1\times10^{-13}\,{\rm eV}$. For the frequency band, the wavelength can be comparable to the Earth circumference. In contrast to the frequency range of $f\lesssim 1$\,Hz studied in Refs.~\cite{Arza:2021ekq,Friel:2024shg}, EM waves generated near the Earth's surface at certain frequencies resonate between the Earth's surface and the ionosphere and show spectral peaks in the frequency band of $1\unit{Hz}$ $\lesssim f \lesssim 100\unit{Hz}$. Therefore, a proper theoretical modeling of the atmospheric cavity resonator that accounts for altitude-dependent conductivity is required to predict a finite axion-induced EM amplitude. 

This paper is one of a series of companion papers, focusing on the data analysis method and search results for an axion signal, placing also the constraint on the coupling parameter $\gag$ using long-term monitoring data of terrestrial magnetic fields. In addition to this work, a separate paper presents the theoretical formulation and the prediction of the induced EM waves \cite{TNH_Theory}, while a letter paper highlights the key findings from both theoretical and data analysis studies \cite{Taruya:2025zql}.
We show that the most stringent bound on the axion-photon coupling among direct observational means is obtained across the mass range of $1\times10^{-15}{\rm eV}$--$1\times10^{-13}{\rm eV}$. 

The organization of the paper is as follows. In Sec.~\ref{sec:theoretical-modeling}, we summarize the theoretical prediction of the axion-induced EM spectrum, which has been derived in the companion paper~\cite{TNH_Theory}. Our data analysis method is described in Sec.~\ref{sec:DA-method}, and the data used and the results obtained (an axion signal estimator, SNR, and axion signal candidates) are presented in Sec.~\ref{sec:DA-results}. From the amplitude distribution of an axion signal estimator, we obtain an observational constraint on the axion-photon coupling strength in Sec.~\ref{sec:observational-constraint}. Finally, Sec.~\ref{sec:conclusions} is devoted to Conclusions. In this paper, we adopt the units of $c=\hbar=1$.

\section{Axion-induced signal in terrestrial $B$-fields} 
\label{sec:theoretical-modeling}

In the presence of axions, the Maxwell equations are modified, and there appear effective charge and current through the axion-EM coupling~ \cite{Sikivie_2021review,TNH_Theory}:
\begin{align}
    &\nabla\cdot\bigl(n^2\,\bfE\bigr)=g_{\rm a\gamma} \nabla\cdot\bigl(\,a\,\bfB\bigr),
    \label{eq:Maxwell1}
    \\
    &\nabla\cdot\bfB=0,
    \label{eq:Maxwell3}
    \\
    &\nabla\times\bfE + \partial_t\bfB=0,
    \label{eq:Maxwell4}
    \\
    &\nabla\times \bfB-\partial_t\bigl(n^2\,\bfE\bigr)=-g_{\rm a\gamma} \left[ \nabla\times\bigl(a\,\bfE\bigr)+\partial_t\bigl(a\,\bfB\bigr) \right],
    \label{eq:Maxwell2}    
\end{align}
where $n$ is the complex refractive index.

If axions constitute DM, they possess a non-relativistic velocity of approximately $v_{\rm DM} \sim 10^{-3}$. In this case, terms involving the spatial gradient, $|\nabla a| \sim \mass  v_{\rm DM} a$, are generally negligible compared to those involving the time derivative, $|\partial_t a| \sim \mass a$. As a result, the primary modification in Eq.~\eqref{eq:Maxwell2} arises solely through the effective current, which is expressed as \( \bfJ_{\rm eff} = -\gag \, \partial_t (a \bfB) \).

The axion field equation
\begin{align}
\partial_t^2\,a -\nabla^2a +\mass^2\,a=-g_{\rm a\gamma}\,\bfE\cdot\bfB \;,
\label{eq:axion_field_EoM}
\end{align}
is in general solved with the Maxwell equations above. However, since the backreaction to axion amplitude is sufficiently small in the current situation, the source term in Eq.~\eqref{eq:axion_field_EoM} is negligible. Using the same approximation that the spatial gradient is much smaller than the time derivative, we have a simple solution oscillating coherently with $\mass$
\begin{align}
    a=a_0\,e^{-i\mass t}.
    \label{eq:coherent_axion}
\end{align}
Therefore, the axions are predominantly coherent and produce an alternating current with a frequency
\beq
f_{\rm a}=\frac{\mass}{2\pi} \simeq2.4\,\left(\frac{\mass}{10^{-14}\,{\rm eV}}\right)\unit{Hz}\;,
 \label{eq:frequency_axion}
\eeq
when coupled with a static magnetic field. 
The amplitude $a_0$ is related to the local dark matter density $\rho_{\rm DM}$ by
\beq
a_0 = \frac{\sqrt{2 \rho_{\rm DM}}}{\mass} \;.
\eeq
Since the de Broglie wavelength of the axion is estimated to be 
\beq
\lambda_{\rm DM}=\frac{1}{\mass v_{\rm DM}} \simeq 1\,{\rm AU}\,\left( \frac{10^{-14}\,{\rm eV}}{\mass} \right) \left( \frac{10^{-2}}{v_{\rm DM}} \right) \;,
\eeq
we can take the field value $a_0$ to be constant over the Earth's geometry for the axion mass of our interest. 

In the presence of alternating current, the modified Amp\'ere-Maxwell law suggests that monochromatic EM waves are produced with its frequency determined by Eq.~\eqref{eq:frequency_axion}. The bandwidth of the produced EM waves is estimated to be $\Delta f/f_{\rm a}\sim v_{\rm DM}^2\sim10^{-6}$. Thus, the EM waves induced by the coherent axion manifest as a prominent and sharp spike in frequency, and the signal would steadily exist over the coherence time given by $\lambda_{\rm DM}/v_{\rm DM}$.

As a representative static magnetic field, we consider the geomagnetic field, $\bfB_{\rm geo}$, characterized predominantly by a dipole configuration, with its strength typically of $|\bfB_{\rm geo}|\sim 25-65\,\mu$T \cite{IGRF13_2021}, and give a quantitative prediction of the axion-induced EM waves. 
Although our primary interest lies in the magnetic fields rather than the electric fields, the electric field equation is easier to handle when imposing boundary conditions. Moreover, once we obtain a solution for the electric fields, a solution for the magnetic fields can easily be derived from Eq.~\eqref{eq:Maxwell4}.

The electric field equation is obtained by substituting Eq.~\eqref{eq:Maxwell2} into the rotation of Eq.~\eqref{eq:Maxwell4}. Replacing the magnetic field $\bfB$ in the electric field equation with the geomagnetic field $\bfB_{\rm geo}$, we obtain  
\begin{align}
    & \nabla^2\bfE - n^2\,\partial^2_t\bfE-\nabla(\nabla\cdot\bfE)= -\gag\,(\partial_t^2\,a)\,\bfB_{\rm geo}.  
\label{eq:wave_eq}
\end{align}
For the geomagnetic field $\bfB_{\rm geo}$, we adopt the IGRF-13 model of the Earth's magnetic field \cite{IGRF13_2021}, which provides the harmonic coefficients up to $\ell=13$. 
Note that allowing the refractive index to spatially vary, the electric field is not divergence-free even in the absence of charge [see Eq.~\eqref{eq:Maxwell1}], and thus the third term at the left-hand side in Eq.~\eqref{eq:wave_eq} becomes non-vanishing. In this paper, we are particularly interested in the axion-induced EM waves in the presence of geomagnetic field. We shall thus consider a monochromatic EM wave with the frequency, $f_{\rm a}=\mass/(2\pi)$ in Eq.~\eqref{eq:frequency_axion}. In this case, the refractive index is related to the conductivity $\sigma$ through
\begin{align}
    n^2(r)=1+i\,\frac{\sigma(r)}{2\pi f_{\rm a}} \;.
    \label{eq:n(r)}
\end{align}
Here we consider the conductivity as a function of the radius $r$ from the center of the Earth. It is known that the radial profile of atmospheric conductivity significantly affects the amplitude and spectral features of the extremely low-frequency EM waves.

Since the Earth interior is highly conductive, we can treat the Earth surface as a perfect conducting boundary, and impose the condition that the non-radial direction of the electric fields should vanish, i.e., $\bfE_\parallel=0$ at the Earth radius, $r=R_{\rm E}$. On the other hand, for the atmospheric conductivity, while it is small ($\sim 10^{-12}$\,S/m) near the Earth's surface, it increases significantly to $\sim 10^{-2}$\,S/m at the altitude of the ionosphere, $r\sim R_{\rm E}+100$\,km, e.g., \cite{Cole_Pierce1965,Kudintseva_etal2016,Nickolaenko_etal2016}. Although this is still lower than the typical conductivity inside the Earth ($\sim10$\,S/m), the EM fields exhibit a diffusion-like behavior and one expects them to decay rapidly above the ionosphere. Following the treatment in the literature (e.g., Refs.~\cite{Greifinger_Greifinger1978,Sentman1990}), we thus impose the boundary condition that allows only upgoing waves at the altitude of the upper atmosphere, $r\gg R_{\rm E}$.

With the boundary conditions specified above, we solve the electric field equation in Eq.~\eqref{eq:wave_eq} by expanding it in the vector spherical harmonics, which separates the angular and radial dependencies. The radial component is governed by an ordinary differential equation that admits an analytical solution when the refractive index $n(r)$ or the conductivity profile $\sigma(r)$ is divided into spatially constant segments. The solution for each segment consists of a homogeneous part and a term sourced from the axion-induced alternating current, with the coefficients of the homogeneous solution left undetermined. These are later determined by enforcing the boundary conditions and ensuring continuity between segments. Consequently, a global solution valid at all segments is constructed, and as a result, the amplitude of the electric field becomes linearly proportional to the axion-photon coupling, $\gag$. More details on the construction of solution as well as the properties of EM fields can be found in Ref.~\cite{TNH_Theory}.

Once the solution of the axion-induced electric field is obtained, the magnetic field is computed from Eq.~\eqref{eq:Maxwell4} (assuming a monochromatic EM wave):
\beq
\bfB =-\frac{i}{\mass}\nabla\times\bfE \;.
\eeq
Then, the resultant magnetic field expression is also linearly proportional to the coupling $\gag$, with its non-zero component appearing in the non-radial direction, that is, in the direction tangent to the Earth's surface. In the next section,
We used the measurement data of the terrestrial magnetic fields to search for the axion-induced signal. This is because the magnetic fields are easier to measure and long-lasting data from observatory around the world are publicly available.

\begin{figure}[t]
\includegraphics[width=8.2cm,angle=0]{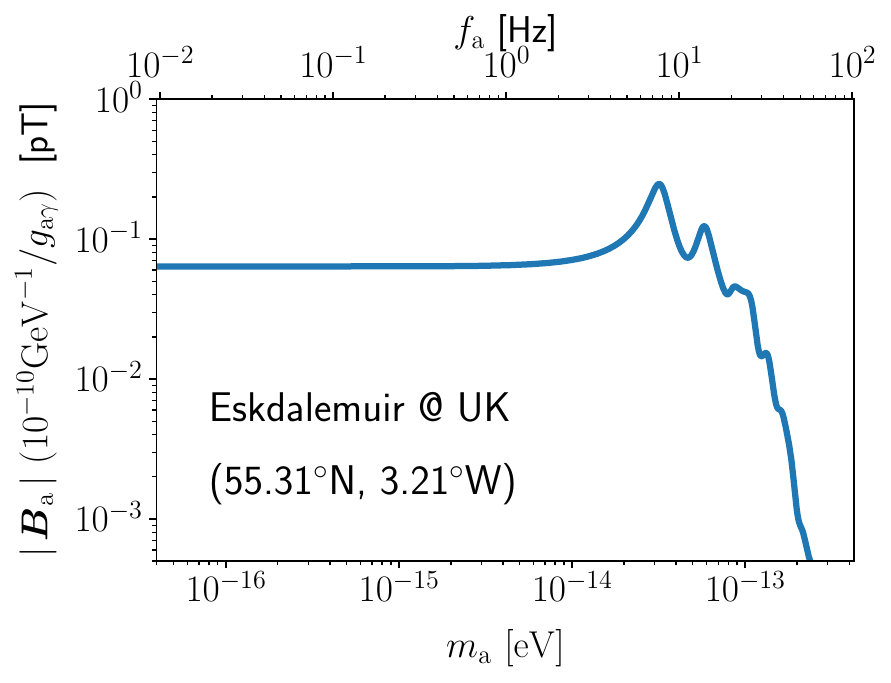}
\vspace*{-0.3cm}
\caption{Expected magnetic field amplitude induced by the coherently oscillating axion, $|{\boldsymbol B}_{\rm a}|$, at Eskdalemuir observatory. The amplitude is normalized by the axion-photon coupling strength of $10^{-10}\,$GeV$^{-1}$. The result is shown as a function of the axion mass $\mass$
(lower) and the frequency $f_{\rm a}$ (upper) of the induced EM waves.
\label{fig:Response_axion_B-field_Eskdalemuir}
}
\end{figure}

Figure~\ref{fig:Response_axion_B-field_Eskdalemuir} illustrates the predicted magnetic field amplitude in the non-radial mode as a function of $\mass$ or the frequency $f_{\rm a}$ of the induced EM signal, at the Eskdalemuir observatory, UK, $(55.31^\circ\,{\rm N},\,3.21^\circ\,{\rm W})$, where the long-term monitoring data at high frequencies up to $100\,$Hz is available (see Sec.~\ref{sec:DA-method}). 
Here, in solving the EM field equation, we used the model of atmospheric conductivity profile tabulated in Ref.~\cite{Kudintseva_etal2016}. In Fig.~\ref{fig:Response_axion_B-field_Eskdalemuir},
we assume that the axions constitute dark matter having the local density of $\rho_{\rm DM}=0.3\,{\rm GeV}\,{\rm cm}^{-3}$, 
with the coupling parameter of $g_{a\gamma}=10^{-10}$\,GeV$^{-1}$. There are prominent peaks, known as the Schumann resonance~\cite{Schumann1952a, Schumann1952b} (see also  Refs.~\cite{Jackson_1998,Nickolaenko_Hayakawa2002,Simoes_etal2012,Nickolaenko_Hayakawa2013}), which is caused by EM waves trapped between the Earth's surface and the ionosphere, forming a resonant cavity. The first peak appears
at the frequency $f_{\rm a}\sim7.8$\,Hz, corresponding to $\mass\sim3\times10^{-14}$\,eV. 

The typical amplitude is estimated from the modified Amp\'ere-Maxwell law, Eq.~\eqref{eq:Maxwell2}. Equating the terms in Eq.~\eqref{eq:Maxwell2} and assuming the characteristic scale of the produced EM waves of $R$ yield $|\bfB_{\rm a}|\sim \gag \mass a_0\,R|\bfB_{\rm geo}|$, that is,
\begin{align}
    |\bfB_{\rm a}|& \sim 0.3\,{\rm pT} \left(\frac{\gag}{10^{-10}\,{\rm GeV}^{-1}}\right)
    \left(\frac{\rho_{\rm DM}}{0.3\,{\rm GeV}\,{\rm cm}^{-3}}\right)^{1/2}
    \nonumber
    \\
    &\times\left(\frac{R}{R_{\rm E}}\right)\left(\frac{|\bfB_{\rm geo}|}{50\,\mu{\rm T}}\right),
\label{eq:rough_estimation_axion_B_field}
\end{align}
where the Earth radius is $R_{\rm E} = 6371\unit{km}$. This estimation is consistent with Ref.~\cite{Arza:2021ekq}. On the other hand, at $\mass\gtrsim10^{-13}$\,eV, its amplitude is sharply suppressed, reflecting the dipole nature of the geomagnetic fields, where the higher multipoles are sufficiently small. Although the amplitude in Eq.~\eqref{eq:rough_estimation_axion_B_field} is more than seven orders of magnitude smaller than that of the geomagnetic fields, it is still accessible with a high-precision magnetometer even for commercial use. Further, at the frequencies $f=0.3-30$\,Hz, the major background noise is a random superposition of transient EM waves, which can be discriminated from a persistent axion signal having a sharp spectral feature by using a long-term monitoring data of terrestrial magnetic fields. 

\section{Data analysis method}
\label{sec:DA-method}

Magnetic fields measured by a magnetometer are composed of background fields $\bfB_{\rm b}$ such as Schumann resonance and magnetometer noise and an axion-induced signal $\bfB_{\rm a}$:
\beq
\bfB(t) = \bfB_{\rm b}(t) + \bfB_{\rm a}(t) \;.
\eeq   
The Fourier transform of the magnetic field is defined by
\beq
\bfB(t) \equiv \int \diff f\, \tilde{\bfB}(f)\, e^{-2\pi i f t} \;.
\eeq
Since the axion signal is proportional to the coupling constant $g_{{\rm a}\gamma}$ and is assumed to be coherent and monochromatic during the observation time, we can separate them and define the normalized amplitude of the axion-induced magnetic fields, $\boldsymbol{R}$, as
\beq 
\bfB_{\rm a}(t) \equiv \boldsymbol{R} (f_{\rm a}) g_{{\rm a}\gamma} e^{-2\pi i f_{\rm a} t} \;,
\eeq 
where $f_{\rm a}$ is the frequency corresponding to axion mass, $m_{\rm a}$. We assume that $\boldsymbol{R} (f_{\rm a})$ is time-independent and the axion-induced signal is stationary. Then the Fourier component is given by
\bea
\tilde{\bfB}(f) &\equiv& \int \diff t \, \bfB(t)\, e^{2\pi i f t} \nonumber \\
&=&  \tilde{\bfB}_{\rm b}(f) + \boldsymbol{R} (f_{\rm a}) g_{{\rm a}\gamma} \delta(f-f_{\rm a}) \;. \nonumber \\
\eea
Assuming that the magnetic fields are stationary in a data segment, which typically lasts for several hours, and defining the power spectra
\bea
\langle \tilde{\bfB}^\ast(f) \tilde{\bfB}(f^{\prime}) \rangle &\equiv& P (f) \delta(f-f^{\prime}) \;, \\
\langle \tilde{\bfB}_{\rm b}^\ast(f) \tilde{\bfB}_{\rm b}(f^{\prime}) \rangle &\equiv& P_{\rm b} (f) \delta(f-f^{\prime}) \;, 
\eea
where $\langle \cdot \rangle$ is the ensemble average,
we have
\bea
P (f) \delta(f-f^{\prime}) &=& P_{\rm b} (f) \delta(f-f^{\prime}) \nonumber \\
&+& | \boldsymbol{R} (f_{\rm a})|^2 g_{{\rm a}\gamma}^2 \delta(f-f_{\rm a}) \delta(f^{\prime}-f_{\rm a}) \;. \nonumber \\
\eea
Here we ignored a cross talk between $\tilde{\bfB}_{\rm b}(f)$ and $\boldsymbol{R} (f_{\rm a})$ at $f=f_{\rm a}$ because there is no correlation between sources for $\tilde{\bfB}_{\rm b}(f)$ and $\boldsymbol{R} (f_{\rm a})$. We define the single-sided power spectrum of magnetic fields
\bea
S(f) &\equiv& 2 P (f) \nonumber \\
&\approx& S_{\rm b} (f) + \frac{2}{T_{\rm obs}} | \boldsymbol{R} (f_{\rm a})|^2 g_{{\rm a}\gamma}^2 \left\{ \delta (f-f_{\rm a}) \right\}^2 \;, \nonumber \\
\label{eq:sf}
\eea
where the factor of two at the first line arises from the contributions from positive and negative frequencies. At the second line, since the observation time $T_{\rm obs}$ is finite, we used $\delta(0) \approx T_{\rm obs}$ (See e.g.~Ref.~\cite{Allen:1997ad} for the definition). Also we replaced the double-sided power spectrum $P_{\rm b} (f)$ with the single-sided power spectrum $S_{\rm b} (f)$.  


We divide the time series data of the magnetic fields into segments of smaller period $T_{\rm seg}$, typically several hours for stationarity of the data, and try to subtract the Schumann magnetic fields from the data at the frequency of the axion signal, $f=f_{\rm a}$.

\begin{figure}[t]
\includegraphics[width=7cm,angle=0]{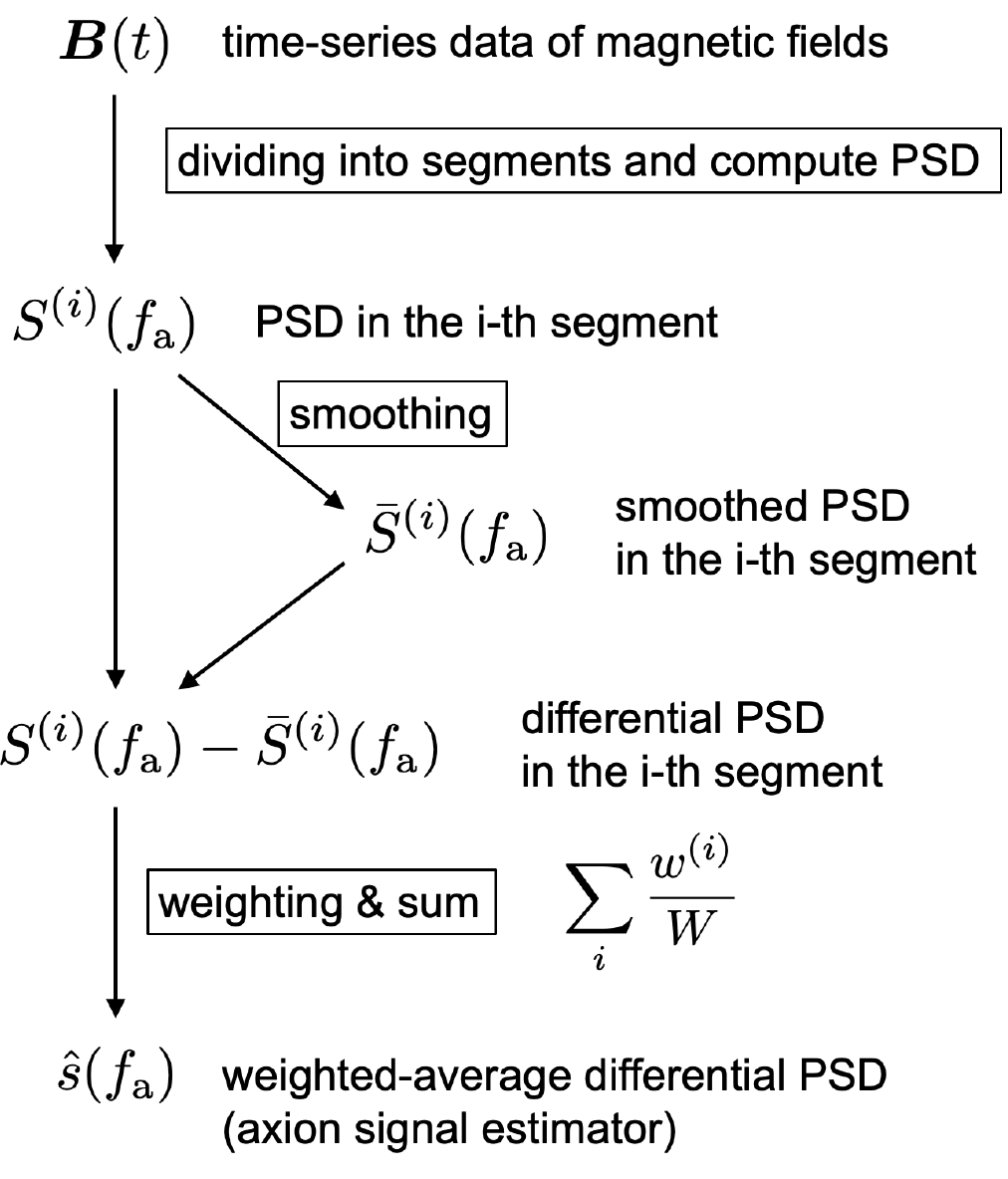}
\caption{Flow chart of the analysis computing the axion signal estimator.
\label{fig:data-analysis-flow-chart}
}
\end{figure}

We consider an axion signal at $f=f_{\rm a}$ in the $i$-th segment:
\beq
S^{(i)}(f_{\rm a}) = S_{\rm b}^{(i)} (f_{\rm a}) + 2 T_{\rm seg} |\boldsymbol{R} (f_{\rm a})|^2 g_{{\rm a}\gamma}^2 \;.
\eeq
The power spectrum of the Schumann magnetic fields at $f=f_{\rm a}$ can be estimated from the power spectrum around $f=f_{\rm a}$ by smoothly interpolating it. Denoting the smooth spectrum of each segment by $\bar{S}^{(i)} (f_{\rm a})$ and subtracting it from $S_{\rm b}^{(i)} (f_{\rm a})$, 
which we call the $i$-th segment differential data, we define an estimator of the axion line signal at $f=f_{\rm a}$ by summing up all differential data segments multiplied by the weights, $w^{(i)}$, based on the square inverse of mean $S_{\rm b} (f_{\rm a})$
\beq
\hat{s}(f_{\rm a}) \equiv s_{\rm b} (f_{\rm a}) + 2 T_{\rm seg} |\boldsymbol{R} (f_{\rm a})|^2 g_{{\rm a}\gamma}^2 \;,
\label{eq:signal-weighted-ave}
\eeq
where
\bea
s_{\rm b} (f_{\rm a}) &\equiv& \sum_i \frac{w^{(i)}}{W} \left\{ S_{\rm b}^{(i)} (f_{\rm a}) - \bar{S}^{(i)} (f_{\rm a}) \right\}  \;, \label{eq:sb} \\
W &\equiv& \sum_i w^{(i)} \;.
\eea
The definition of the weight $w^{(i)}$ is introduced later. Since the weighted average and the subtraction of smooth spectrum are commutable, the spectral subtraction can be performed once after the weighted average. The above analysis procedure is summarized in Fig.~\ref{fig:data-analysis-flow-chart}.

From the definition of the estimator, if the axion signal is sharp enough, the differential spectrum, $s_{\rm b} (f_{\rm a})$, is statistically random about zero. The axion signal is always positive only if the signal amplitude is larger than the smoothed spectrum. 

Writing the variance of noise as $[ \Delta \hat{s} (f_{\rm a}) ]^2 \equiv {\rm Var} \left[ s_{\rm b} (f_{\rm a}) \right]$,
from Eq.~\eqref{eq:signal-weighted-ave}, we define SNR:
\bea
{\rm SNR} &\equiv& \sqrt{\frac{\langle\hat{s}(f_{\rm a})\rangle}{\Delta \hat{s}(f_{\rm a})}}
\label{eq:SNR-definition}\\
&=&\sqrt{\frac{2 T_{\rm seg}}{\Delta \hat{s} (f_{\rm a})}}\, |\boldsymbol{R} (f_{\rm a})| g_{a\gamma} \;.
\label{eq:SNR}
\eea

When segment data are equally weighted, that is, $w^{(i)}=1$, the noise variance is reduced to 
\beq
\left[ \Delta \hat{s} (f_{\rm a}) \right]^2 = \frac{1}{N_{\rm seg}^2} \left\langle \sum_i^{N_{\rm seg}} \left\{ S_{\rm b}^{(i)} - \bar{S}^{(i)} \right\}^2 \right\rangle \;.
\eeq
Combining with Eq.~\eqref{eq:SNR}, we find that the SNR is improved proportional to $N_{\rm seg}^{1/4}$ as the observational time increases, recovering the standard scaling~\cite{Nakatsuka:2022gaf}.

\section{Data and analysis results}
\label{sec:DA-results}

\subsection{Data}

We use the publicly-available magnetic field data deposited at the British Geological Survey~\cite{BGS-data}. The data had been taken by two induction coils (CM11E1) at the Eskdalemuir observatory, UK~\cite{Beggan-Musur-2018} from September 1, 2012 to November 4, 2022. There are two channels, North-South (CH1) and East-West (CH2), with the sampling frequency, $100\unit{Hz}$. The raw data are given in digitizer units and need to be converted to picotesla (pT) by the conversion factors, $3.491\times 10^{-6}\unit{V}/{\rm count}$ for CH1 and $3.475\times 10^{-6}\unit{V}/{\rm count}$ for CH2 and the frequency-dependent calibration factors between $10^{-3}\unit{Hz}$--$10^2\unit{Hz}$ in Table~\ref{tab:calibration-factor} in Appendix~\ref{app:calibration}. The response of the magnetometer decreases at low frequencies, by 5\% at $10^{-2}\unit{Hz}$ and more than 90\% at $10^{-3}\unit{Hz}$. For the reason, we set the lowest frequency for the use of the data to $10^{-3}\unit{Hz}$. We have also checked the absolute calibration by comparing the data with those measured with an independent instrument, and found that the data we use are well calibrated throughout the period between 2012 and 2022. The detail of the comparison is provided in Appendix~\ref{app:calibration}.

\subsection{Estimator for axion signals}

For the purpose of searching for axion signals, the orientation of the measured magnetic fields is helpful to distinguish the signal from noise \cite{TNH_Theory}. However, the systematics and stability are not well-understood, and we decided to use only the signal power spectral density (PSD), computed from the two channels, CH1 and CH2. Furthermore, to search for a sharp line signal, stably long-lasting data are necessary. Therefore, we use only the data lasting more than one month. Then each sequential data are divided into 8-hour segments. The length of a data segment is chosen so that the frequency width of an axion signal at $35\unit{Hz}$, that is, $\Delta f_{\rm a} \sim (\Delta v/c)^2 f_{\rm a} \sim 3.5\times 10^{-5}\unit{Hz}$, is included in a single frequency size of $1/(8\unit{h}) \approx 3.5 \times 10^{-5}\unit{Hz}$. As the geomagnetic fields are rapidly damped above $30\unit{Hz}$, the choice of frequency bin size hardly impacts the sensitivity above $30\unit{Hz}$. Axion signals at the frequencies below $30\unit{Hz}$ are well within a single frequency bin. This is not an optimal choice, considering the coherent time of axion dark matter, particularly at lower frequencies. However, there are 30 second-long transient noises with huge amplitude (saturated at $\sim 10^8\unit{pT}$). They prevent us from taking longer data segments. To avoid the contamination from the transient noise, we fix the data segment size to 8 hrs. The number of 8-hour data segments that we used for our analysis is shown in Table~\ref{tab:amount-of-data}. The total number of the data segments is $N_{\rm seg}=9909$.

\begin{table}[tb]
\begin{center}
\begin{tabular}{|c| c| c|}
\hline
Year & Number of data segments & $w^{(Y)}/W$ \\
\hline
2012 & 176 & 0.0247 \\
2013 & 1028 & 0.1137 \\
2014 & 581 & 0.0455 \\
2015 & 1007 & 0.0684 \\
2016 & 1097 & 0.0962 \\
2017 & 976 & 0.0766 \\
2018 & 984 & 0.1483 \\
2019 & 968 & 0.1126 \\
2020 & 1095 & 0.1022 \\
2021 & 1094 & 0.1532 \\
2022 & 903 & 0.0595 \\
all years & 9909 & 1 \\
\hline
\end{tabular}
\end{center}
\caption{The amount and quality of data used for the analysis. The number of data segments is counted in the unit of 8 hours. $w^{(Y)}$ is the sum of $w^{(i)}$ in each year and $W$ is the sum of $w^{(Y)}$.
\label{tab:amount-of-data}
}
\end{table}

Given the PSD of the magnetic fields for each data segment, we compute the estimator of an axion signal, $\hat{s}(f_{\rm a})$ in Eq.~\eqref{eq:signal-weighted-ave}. We stack the PSDs multiplied by the weight $w^{(i)}$, which is defined by the inverse square of mean $S_{\rm b} (f_{\rm a})$ computed from the spectrum between $1.1\unit{Hz}$ and $9.9\unit{Hz}$\footnote{The frequency range is chosen to average a flat spectrum in the most sensitive frequency band  up to $10\unit{Hz}$ (see Fig.~\ref{fig:Response_axion_B-field_Eskdalemuir}).}, and then subtract the smooth component obtained by applying a quadratic Butterworth filter with a sampling frequency of $0.05 \unit{Hz}$. Dividing by the sum of the weights for data segments, $W$, we obtain the weighted average differential PSD, $\hat{s}(f_{\rm a})$. With these procedure, the sharp line of an axion signal is constructively stacked and the spectral features of the Schumann resonance are subtracted (but not perfectly). 

As written in the document on the data server~\cite{BGS-data}, at the integer-valued frequencies, there are artificial noise lines that appear irregularly and whose cause is unknown.  We veto those artificial line noises by removing the data in the range of $\pm 3.47\times 10^{-2}\unit{Hz}$ ($\pm 1000$ frequency bins) below and above the integer-valued frequencies. 

\begin{figure*}[t]
\begin{center}
\includegraphics[width=12.5cm]{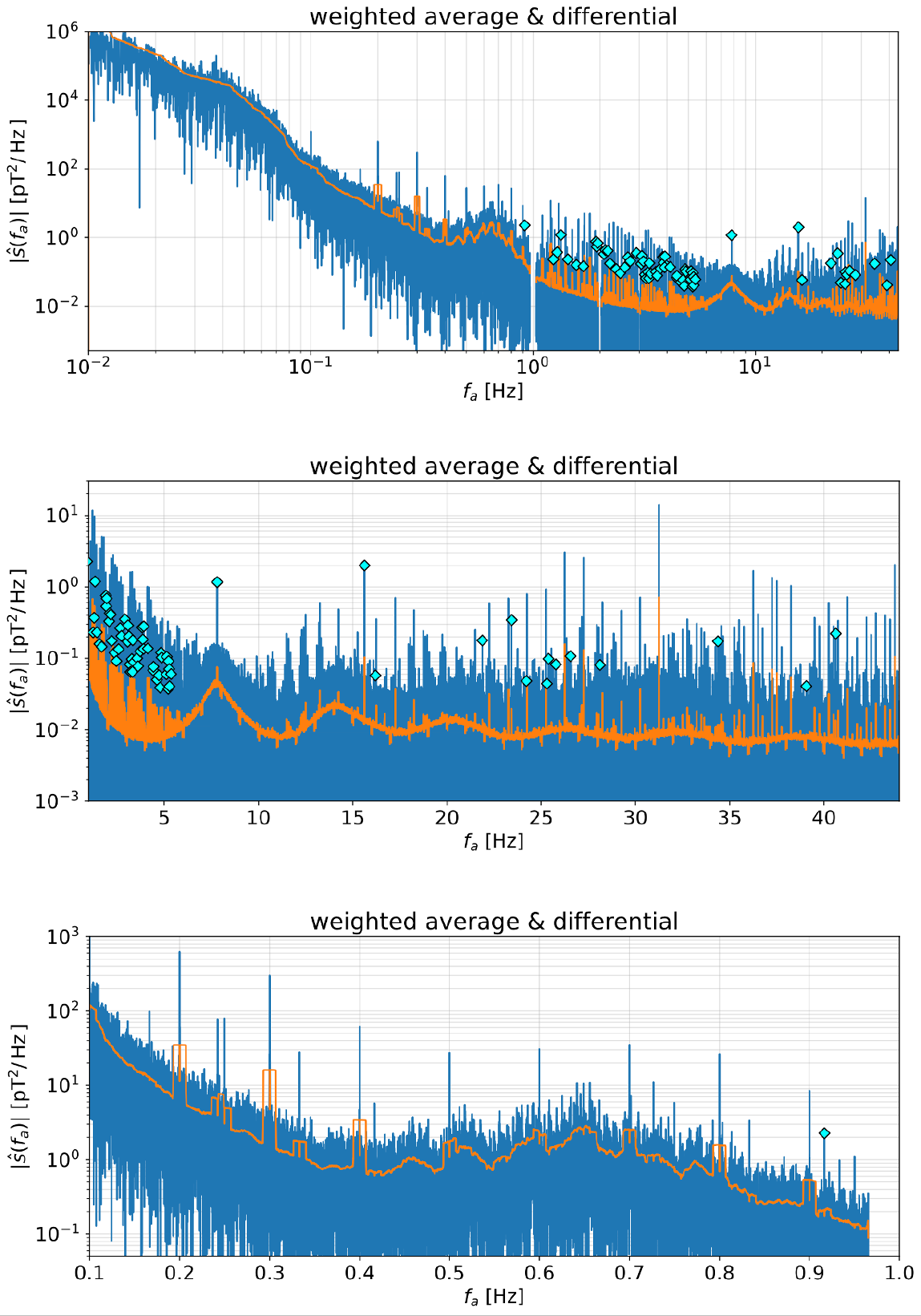}
\caption{Weighted-average differential PSD of the magnetic fields, stacked over all eight-hour data segments with noise-based weights and subtracted the smooth spectrum, $\hat{s}(f_{\rm a})$ in blue. The standard deviation is depicted in orange. The axion signal candidates are marked by diamonds in cyan. The second and third panels from the top are the same as the first one but in different frequency ranges.}
\label{fig:differential-averaged-PSD-8h}
\end{center}
\end{figure*}

The weighted-average differential PSD, $\hat{s}(f_{\rm a})$, is plotted in Fig.~\ref{fig:differential-averaged-PSD-8h}. The spectral amplitude (blue) increases toward low frequencies below $1\unit{Hz}$ at the timescales of storms, and is almost flat between $1\unit{Hz}$ and $\sim 100\unit{Hz}$ with Schumann resonance peaks~\cite{Constable:2023PEPI}. The second and third panels from the top are the same as the first one but in different frequency ranges. Spectral lines are concentrated above $1\unit{Hz}$. Below $1\unit{Hz}$, prominent lines appear at the frequencies of the multiples of $0.1\unit{Hz}$.

\subsection{SNR}

SNR is computed from Eq.~\eqref{eq:SNR-definition}. Note that the definition of the SNR is sensitive to the sign of the weighted average differential PSD, which plays a crucial role when we check the persistency of the signals. The standard deviation of noise, $\Delta \hat{s}(f_{\rm a})$, is evaluated from the data of $\pm 200$ frequency bins ($\pm 3.47\times 10^{-3}\unit{Hz}$) around $f=f_{\rm a}$ except for 3 frequency bins centered at $f=f_{\rm a}$, which is shown in orange in Fig.~\ref{fig:differential-averaged-PSD-8h}. With this treatment, the noise variance estimation does not include an axion signal candidate itself and automatically excludes line signals broader than a single frequency bin. For the procedure, we set the lower limit of $f_{\rm a}$ to $0.01\unit{Hz}$ to have the sufficient number of frequency bins for noise variance evaluation, while the upper limit of $f_{\rm a}$ is set to $44\unit{Hz}$ because the spectrum around and above $50\unit{Hz}$ is unreliable due to the Nyquist frequency.

\begin{figure*}[t]
\begin{center}
\includegraphics[width=12.5cm]{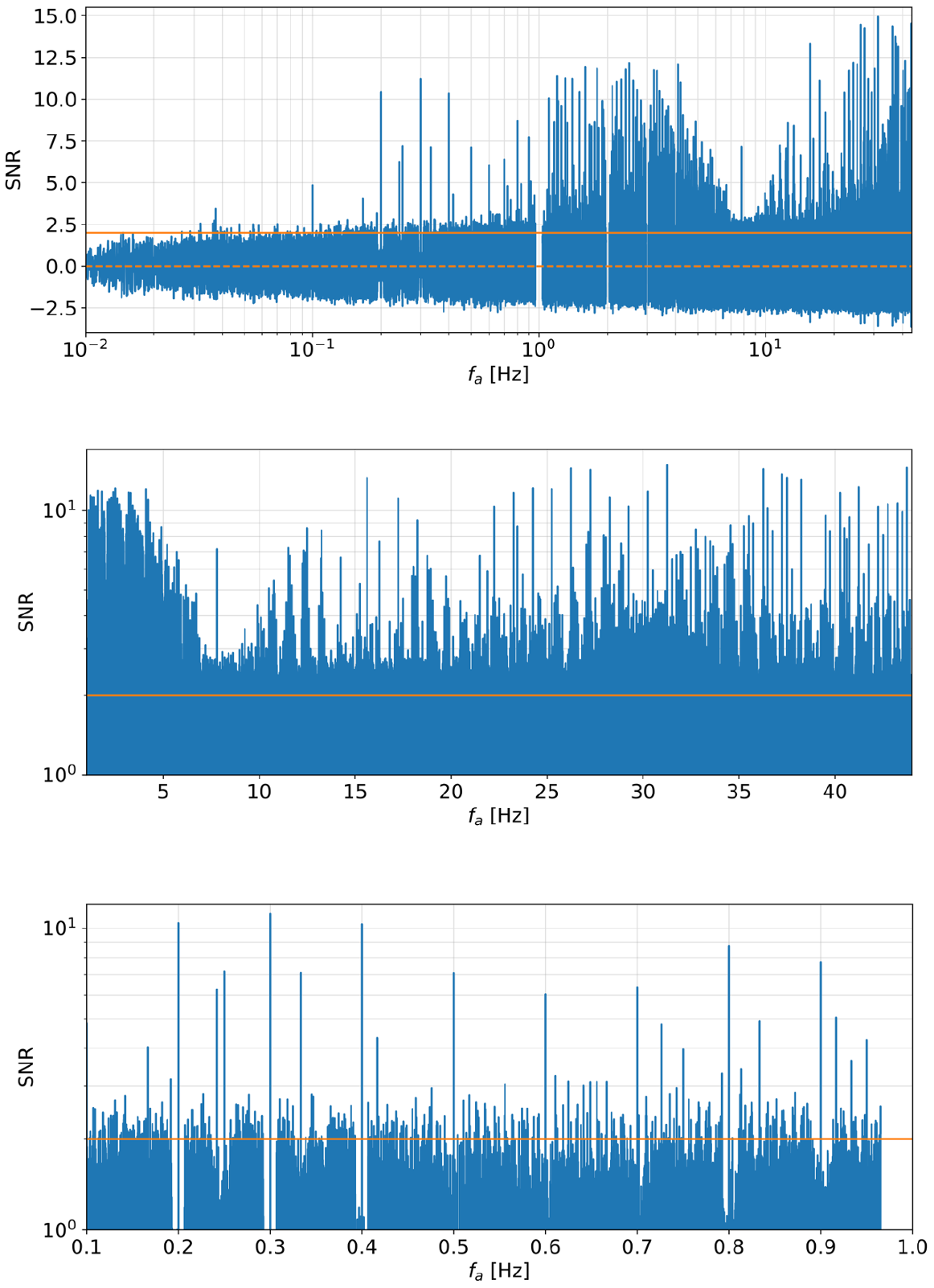}
\caption{SNR with 8-hour data segments from all-year data. The same plot but in the different ranges of the frequencies corresponding to axion masses. The horizontal solid and dashed lines are SNR$=2$ and SNR$=0$. The second and third panels from the top are the same as the first one but in different frequency ranges.}
\label{fig:SNR-8h-all}
\end{center}
\end{figure*}


\begin{table}[t]
\begin{center}
\begin{tabular}{|c| c| c| c|}
\hline
&& \& & \& excluding  \\
SNR threshold & all years & persistency & neat frequencies  \\
\hline
SNR $>14$ & 5 & 5 & 0  \\
SNR $>13$ & 9 & 9 & 1  \\
SNR $>12$ & 14 & 14 & 1 \\
SNR $>11$ & 32 & 31 & 1 \\
SNR $>10$ & 46 & 41 & 1 \\
SNR $>9$ & 60 & 49 & 2 \\
SNR $>8$ & 96 & 74 & 9 \\
SNR $>7$ & 132 & 99 & 13 \\
SNR $>6$ & 183 & 131 & 22 \\
SNR $>5$ & 257 & 172 & 31 \\
SNR $>4$ & 473 & 270 & 50  \\
SNR $>3$ & 821 & 375 & 65 \\
SNR $>2$ & 26912 & 660 & 342 \\ 
\hline
\end{tabular}
\end{center}
\caption{Number of signal candidates filtered by the conditions of SNR threshold, persistency, and not-neat frequency.}
\label{tab:number-of-signal-candidates}
\end{table}

In Fig.~\ref{fig:SNR-8h-all}, the SNR as a function of the frequency corresponding to axion mass is plotted. The second and third panels from the top are the same as the first one but in different frequency ranges. The SNRs are likely to be larger at higher frequencies because noise variance is smaller there. The signals above the SNR threshold are concentrated above $1\unit{Hz}$, particularly in the range of $1\unit{Hz}$--$6\unit{Hz}$.

Depending on the SNR threshold, we found a number of signal candidates listed at the second row in Table~\ref{tab:number-of-signal-candidates}. Filtering SNR$>2$ (SNR$>3$) for all-year data leaves 26912 (821) signal candidates. The candidates are mostly in a single frequency bin and are consistent with the sharpness of an axion signal. If a signal candidate is true, it must be persistent through all the data. We further set the condition for the candidates to be credible, requiring that the SNR for each-year data is above the weighted SNR threshold, ${\rm SNR} \times w^{(Y)}/W$, where $w^{(Y)}$ is the sum of $w^{(i)}$ in each year and is listed in Table~\ref{tab:amount-of-data}. The numbers of signal candidates after imposing the condition of persistency are also listed at the third row in Table~\ref{tab:number-of-signal-candidates}. Persistent filtering works efficiently for the candidates with low SNR and reduces the number of candidates drastically because noises may change its sign every year due to the subtraction of the smooth component in Eq.~\eqref{eq:sb} but an axion signal does not change. The second condition reduces the number of signal candidates from 26912 to 660 for SNR$>2$ and from 821 to 375 for SNR$>3$. 

To reduce the number of candidates further, we rule out the candidates at the frequencies of the multiples of $0.05\unit{Hz}$ (within $0.001\unit{Hz}$), called ``neat frequencies''. These signals are likely to be produced by the analogue filters or the digitizer~\cite{Ciaran-Beggan}. After excluding the candidates at the neat frequencies, the number of candidates is significantly reduced to 342 for SNR$>2$ and to 65 for SNR$>3$. The latter is listed in Table~\ref{tab:list-of-signal-candidates} and is shown with the filled diamonds in Fig.~\ref{fig:differential-averaged-PSD-8h}. They are concentrated above $1\unit{Hz}$, particularly in the range of $1\unit{Hz}$--$6\unit{Hz}$. The highest SNR signal has SNR$=13.342$ and is the only signal above SNR threshold of $10$.  

\begin{table*}[t]
\begin{center}
\begin{tabular}{|c|c|c|c|c|c|c|c|c|c|c|}
\hline
ID & frequency [Hz] & \; SNR \qquad & & ID & frequency [Hz] & \; SNR \qquad & & ID & frequency  [Hz] & \; SNR \qquad \\
\hline
1 & 0.9167 & 5.071 &&  23 & 3.1333 & 6.766 &&  45 & 5.1167 & 3.941 \\
2 & 1.2333 & 3.584 &&  24 & 3.1833 & 4.347 &&  46 & 5.1333 & 5.109 \\
3 & 1.2833 & 4.696 &&  25 & 3.2167 & 3.996 &&  47 & 5.2167 & 3.002 \\
4 & 1.3333 & 8.580 &&  26 & 3.2833 & 4.813 &&  48 & 5.2333 & 4.642 \\
5 & 1.4333 & 4.184 &&  27 & 3.3167 & 4.164 &&  49 & 5.2667 & 4.110 \\
6 & 1.5667 & 3.767 &&  28 & 3.3333 & 6.571 &&  50 & 5.2833 & 3.076 \\
7 & 1.6833 & 3.736 &&  29 & 3.3833 & 4.637 &&  51 & 5.3667 & 3.825 \\
8 & 1.9167 & 9.374 &&  30 & 3.5667 & 5.240 &&  52 & 7.8125 & 7.168 \\
9 & 1.9333 & 8.288 &&  31 & 3.6167 & 4.581 &&  53 & 15.6250 & 13.342 \\
10 & 1.9531 & 6.336 &&  32 & 3.7333 & 5.837 &&  54 & 16.2052 & 3.105 \\
11 & 2.0833 & 6.867 &&  33 & 3.8333 & 8.211 &&  55 & 21.8750 & 5.896 \\
12 & 2.1167 & 7.939 &&  34 & 3.8667 & 6.892 &&  56 & 23.4375 & 8.704 \\
13 & 2.1667 & 7.787 &&  35 & 3.9062 & 4.317 &&  57 & 24.2188 & 3.459 \\
14 & 2.2167 & 5.234 &&  36 & 3.9167 & 8.036 &&  58 & 25.3073 & 3.052 \\
15 & 2.3667 & 4.536 &&  37 & 4.1333 & 6.148 &&  59 & 25.3906 & 4.468 \\
16 & 2.4667 & 4.154 &&  38 & 4.4167 & 4.306 &&  60 & 25.7812 & 4.018 \\
17 & 2.5833 & 5.059 &&  39 & 4.4333 & 4.589 &&  61 & 26.5625 & 4.487 \\
18 & 2.6667 & 3.375 &&  40 & 4.6167 & 3.617 &&  62 & 28.1250 & 4.336 \\
19 & 2.7333 & 6.445 &&  41 & 4.7167 & 4.008 &&  63 & 34.3750 & 6.699 \\
20 & 2.9167 & 8.538 &&  42 & 4.7833 & 3.245 &&  64 & 39.0625 & 3.169 \\
21 & 3.0667 & 6.273 &&  43 & 4.8333 & 5.596 &&  65 & 40.6250 & 7.852 \\
22 & 3.0833 & 8.200 &&  44 & 4.8667 & 5.200 && & & \\
\hline
\end{tabular}
\end{center}
\caption{Frequencies and SNRs of persistent significant signal candidates above SNR$=3$ for all-year data and above the SNR threshold weighted by noise level for each-year data, removing the candidates of neat frequencies.}
\label{tab:list-of-signal-candidates}
\end{table*}

\subsection{Axion signal candidates}

\begin{figure*}[t]
\begin{center}
\includegraphics[width=8cm]{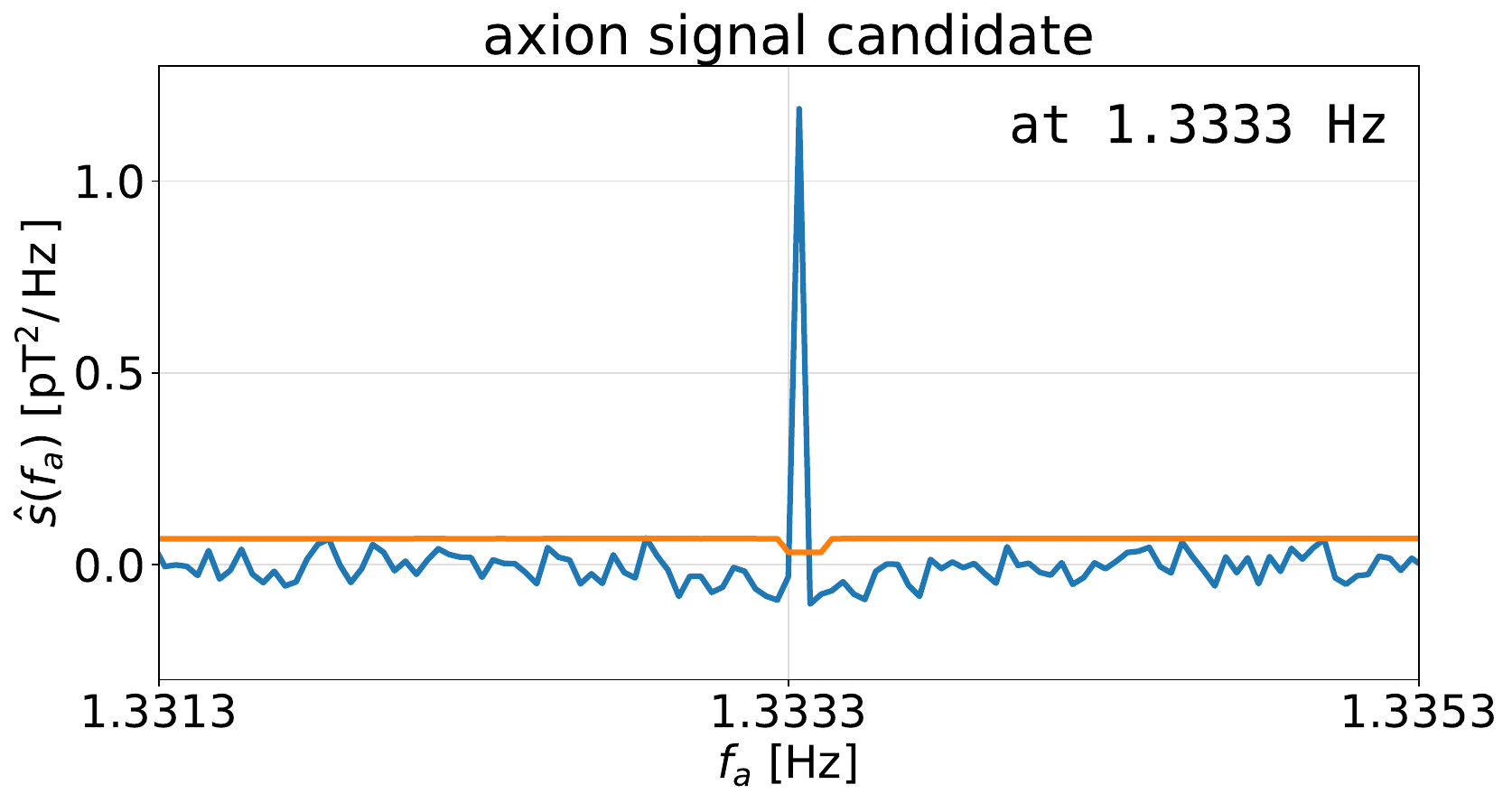}
\includegraphics[width=8cm]{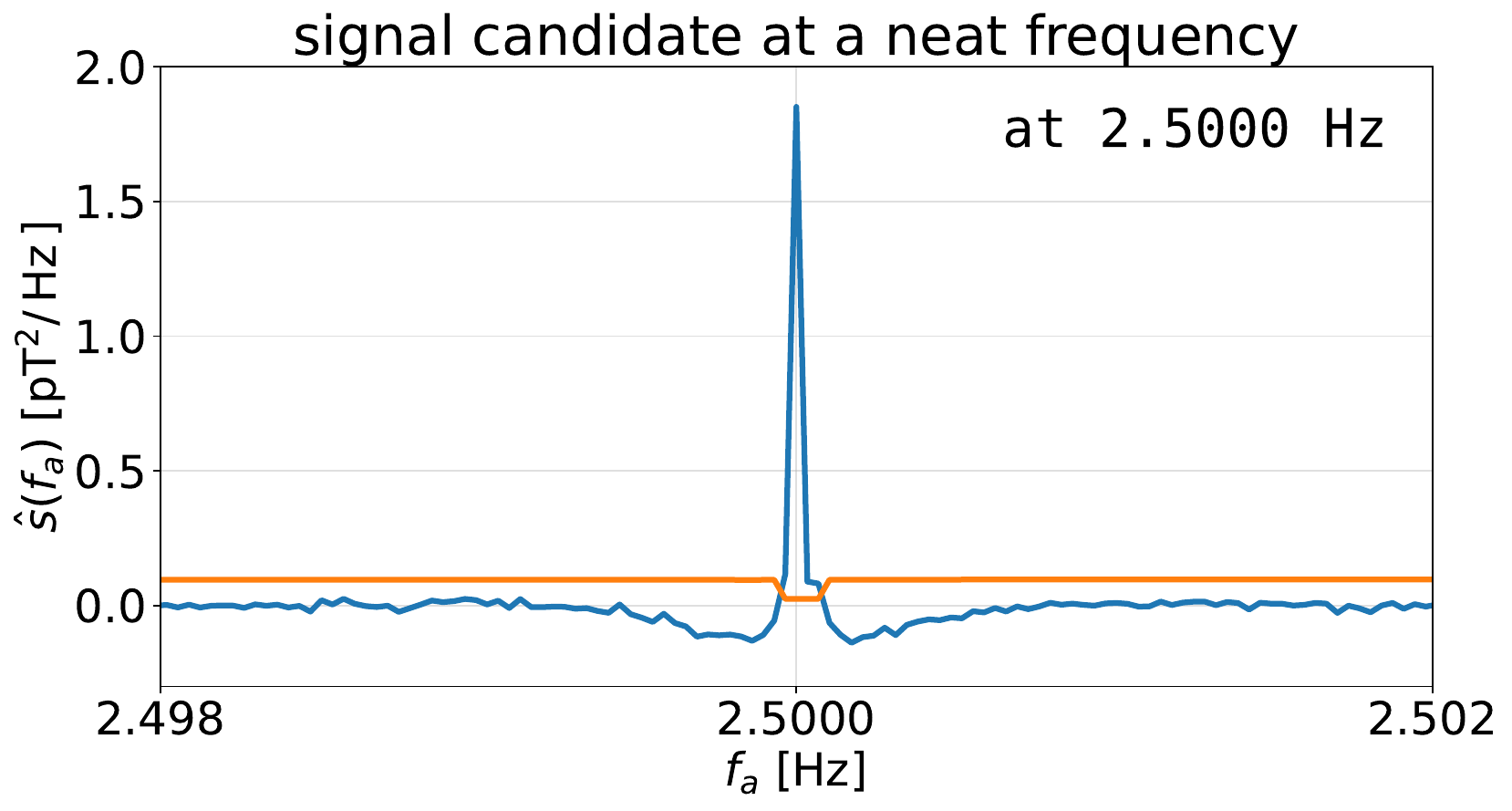}
\includegraphics[width=8cm]{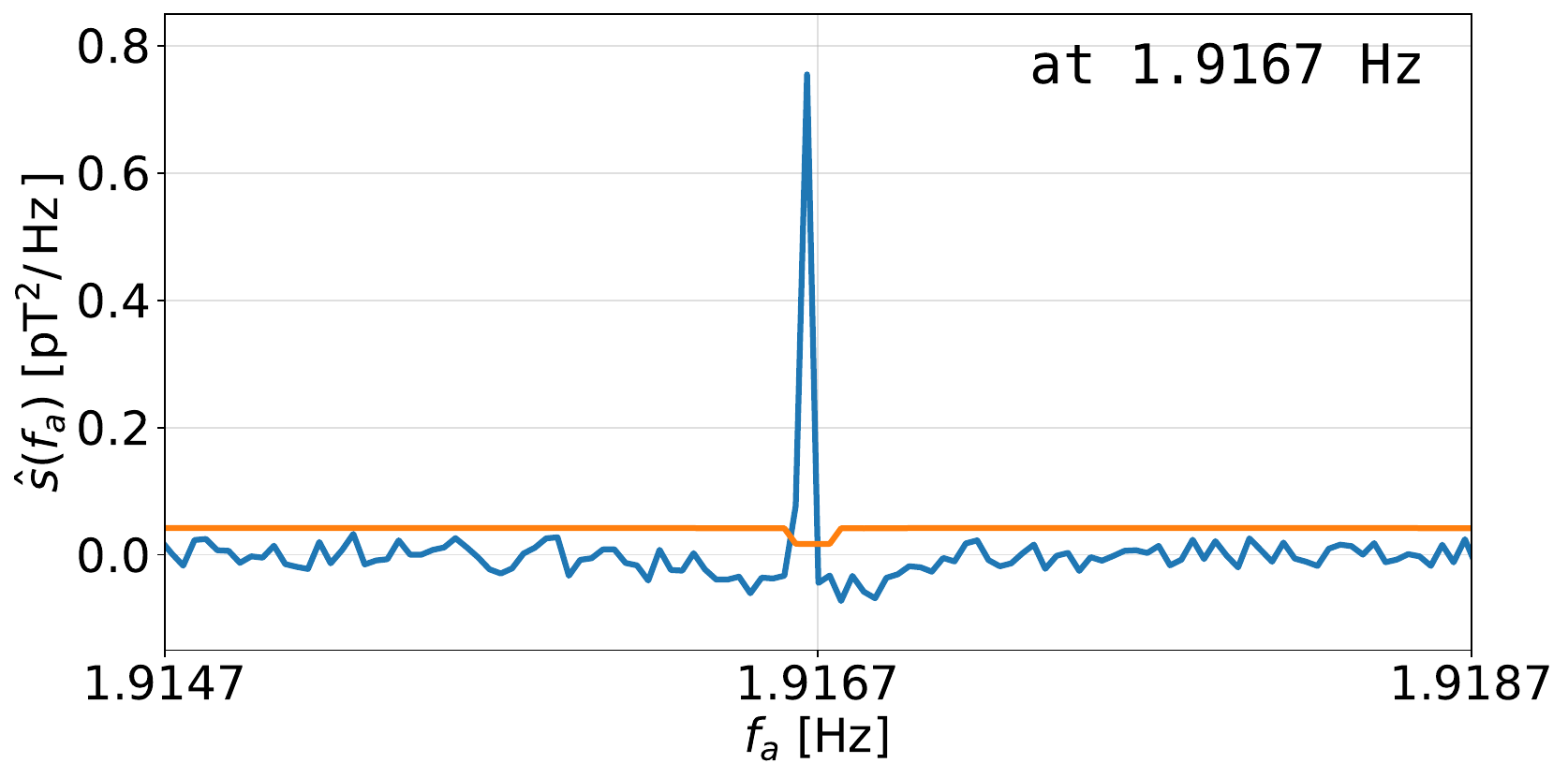}
\includegraphics[width=8cm]{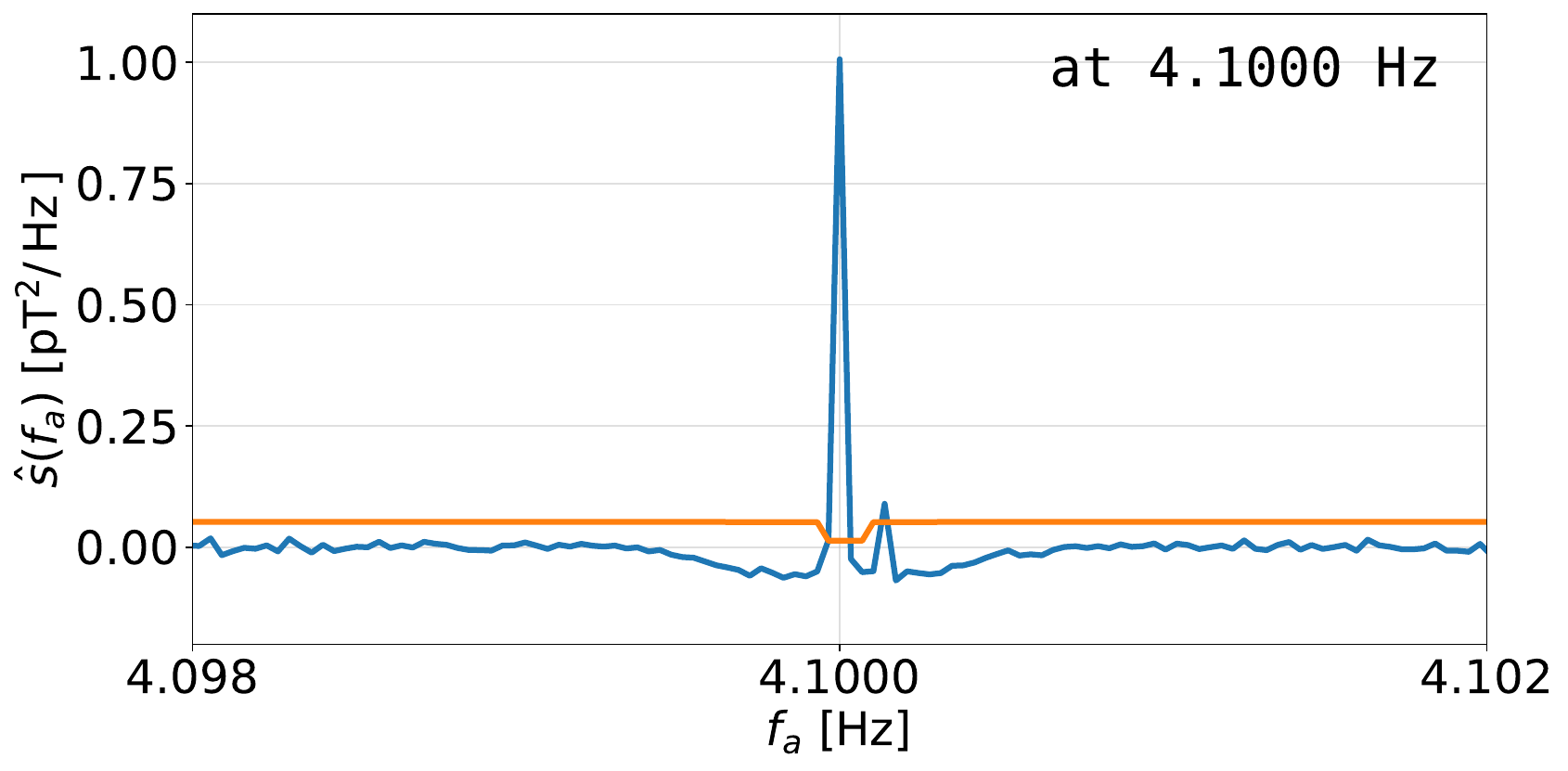}
\includegraphics[width=8cm]{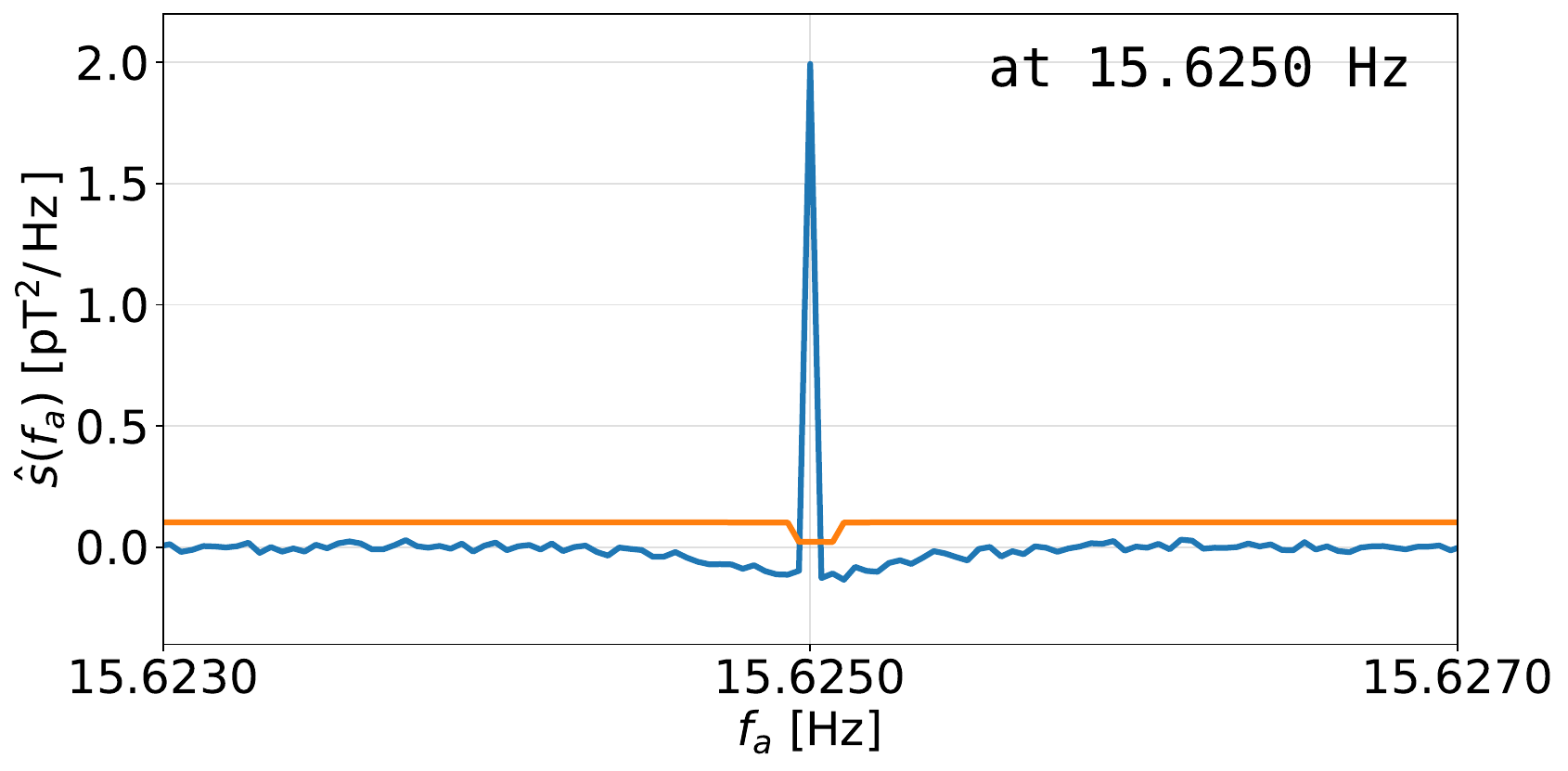}
\includegraphics[width=8cm]{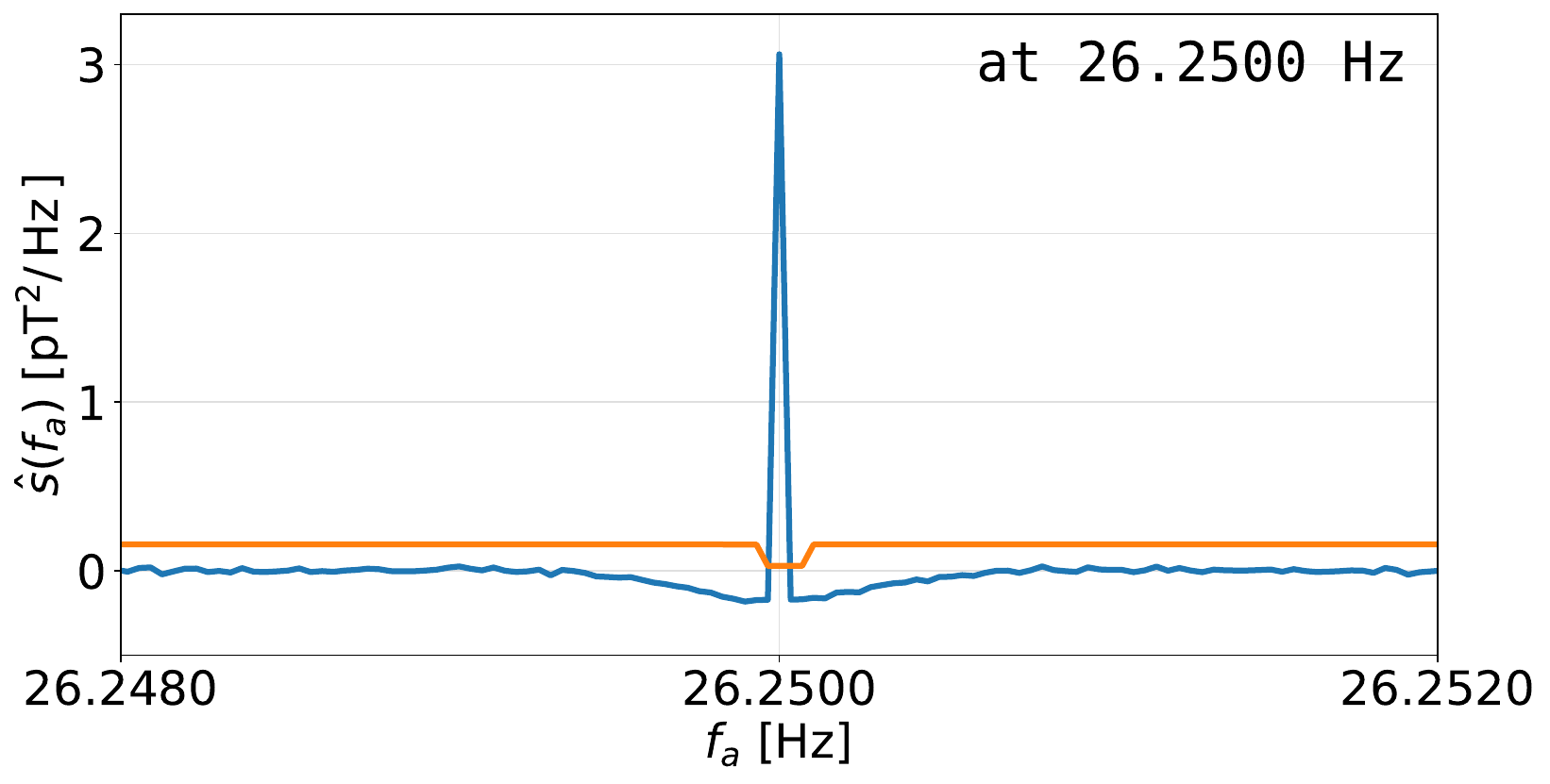}
\includegraphics[width=8cm]{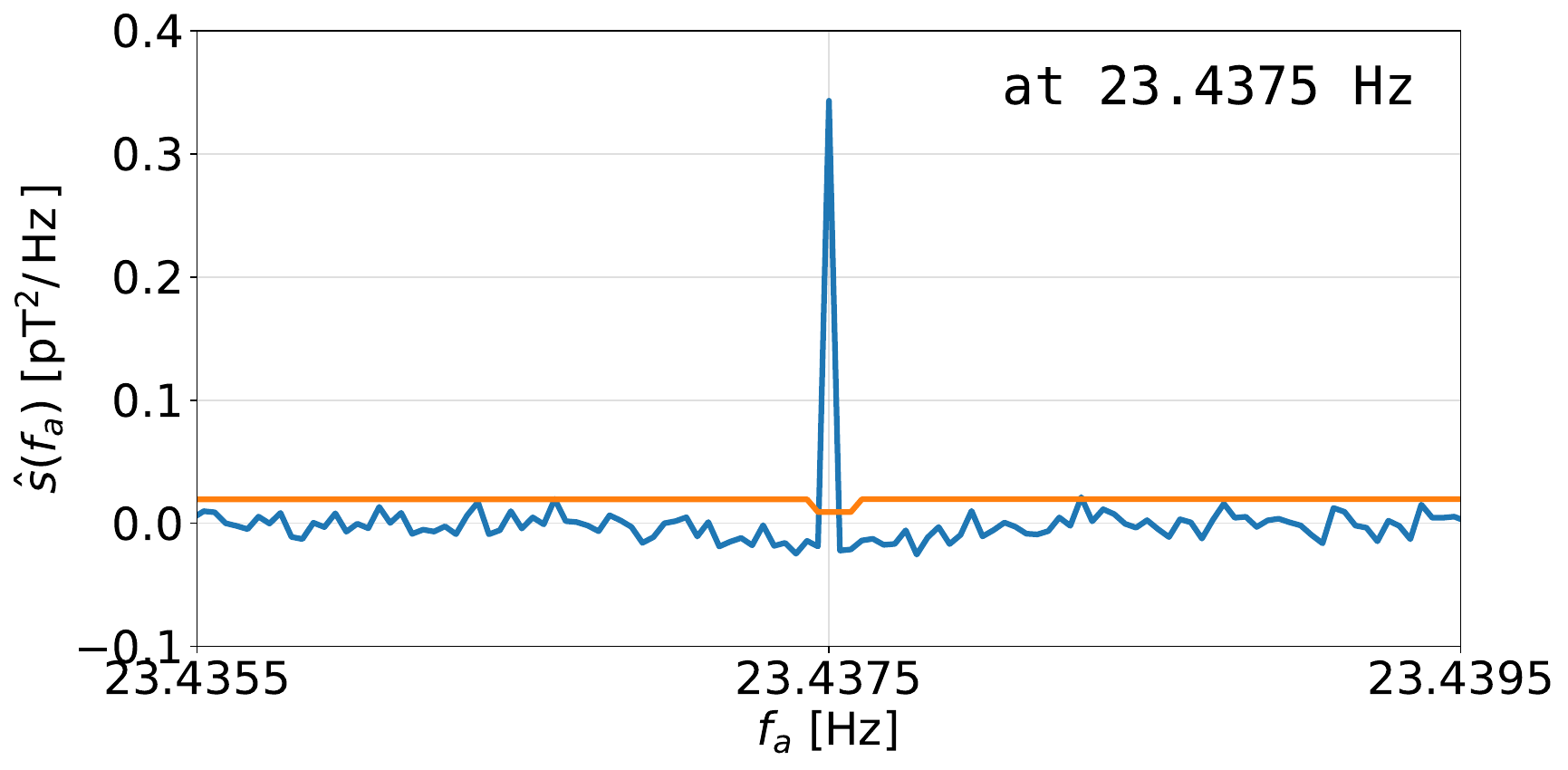}
\includegraphics[width=8cm]{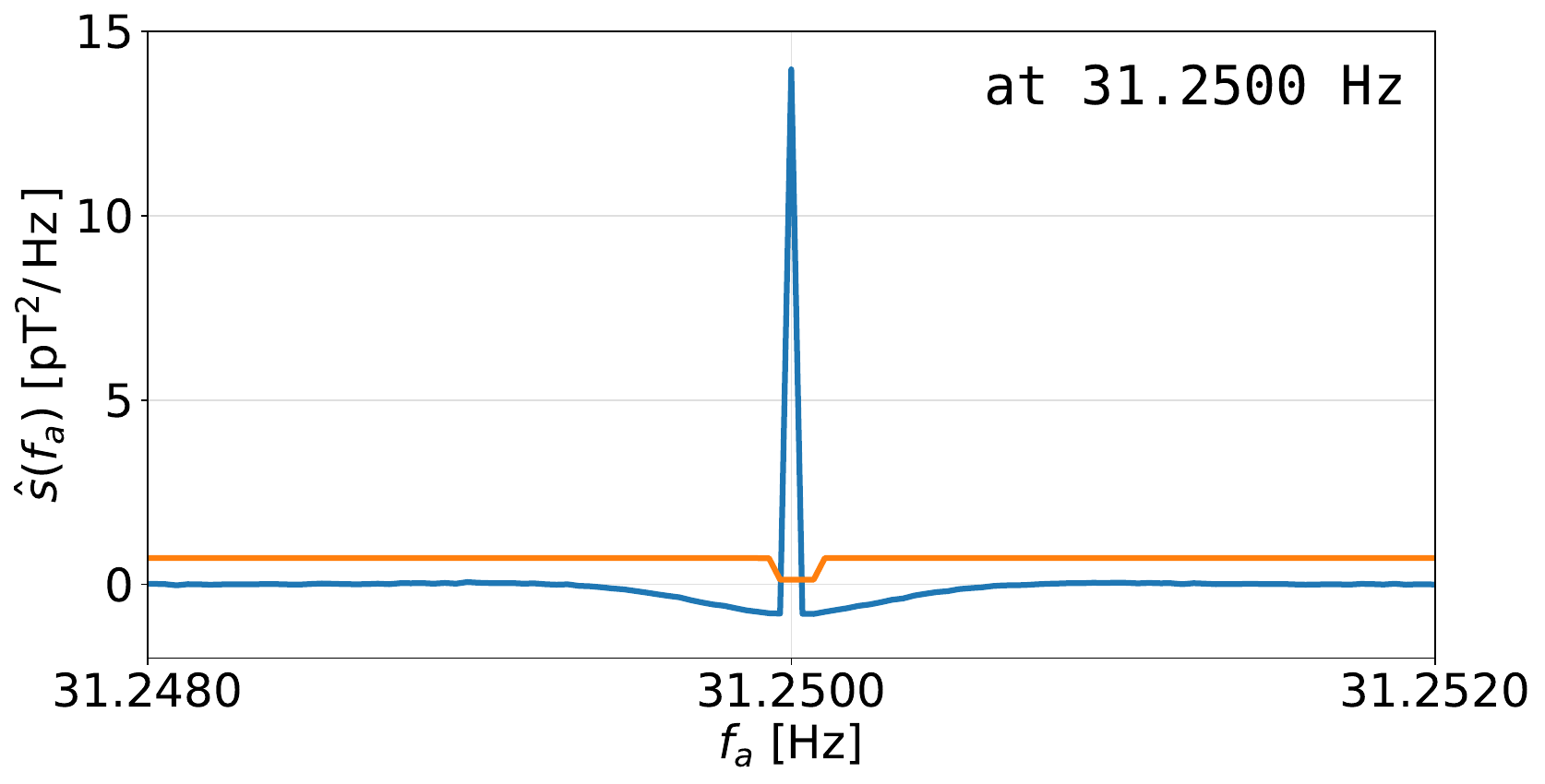}
\caption{Axion signal candidates at $1.3333\unit{Hz}$ (SNR$=8.580$), $1.9167\unit{Hz}$ (SNR$=9.374$), 
$15.6250\unit{Hz}$ (SNR$=13.342$), and $23.4375\unit{Hz}$ (SNR$=8.704$),
and the signal candidates at neat frequencies,
$2.5000\unit{Hz}$ (SNR$=12.152$), $4.1000\unit{Hz}$ (SNR$=12.073$), $26.2500\unit{Hz}$ (SNR$=14.460$), and
$31.2500\unit{Hz}$ (SNR$=14.932$). The left panels are axion signal candidates with top-4 SNRs and the right panels are the significant signal candidates rejected due to neat frequencies. The signal, $\hat{s}(f_{\rm a})$, is in blue and its standard deviation as noise is in orange.}
\label{fig:signal-candidates}
\end{center}
\end{figure*}

For further investigation of the axion signal candidates, we plot in Fig.~\ref{fig:signal-candidates} the top-four-SNR axion signal candidates and four large-SNR noise-like candidates at neat frequencies. They are in one or two frequency bins, consistent with the frequency width of the axion signal, and have no remarkable difference in spectral features.
There are shallow dips around the sharp lines. This is because a smoothed spectrum is computed by applying a quadratic Butterworth filter with a sampling frequency of $0.05 \unit{Hz}$ and is subtracted from the original spectrum in Eq.~\eqref{eq:sb}. Note that noise variance does not differ significantly for the signal candidates but the ranges of the vertical axis are different, depending on their SNRs.

As mentioned in the previous section, the candidates at neat frequencies are likely to be produced by the analogue filters or the digitizer~\cite{Ciaran-Beggan}. Since we have no additional information to veto the axion signal candidates further, all the signal candidates that pass our three signal criteria should be kept as possible axion signals for future searches. On the other hand, one should investigate the signal candidates deeply by studying their noise properties or should check their consistency with the magnetic field data at other observatories.

In Table~\ref{tab:list-of-signal-candidates-low-freq}, we list the signal candidates with SNR$>2$ at low frequencies (below $1\unit{Hz}$). In the previous study using SuperMAG 1-second sampling data~\cite{Friel:2024shg}, there were three signal candidates with modest significance. However, none of the candidates in Table~\ref{tab:list-of-signal-candidates-low-freq} is identified with those found in the previous search. Although our sensitivity is slightly worse than that in the previous study at $0.2630\unit{Hz}$, which is one of the signal candidate frequencies in the previous search, we did not find any signal candidate at the frequency and constrain the existence of axions with coupling strength larger than $g_{{\rm a} \gamma} \lesssim 1.3\times10^{-11}\,{\rm GeV}^{-1}$ at 95\% credible level (CL).    

\begin{table}[h]
\begin{center}
\begin{tabular}{|c|c|c|}
\hline
ID & frequency [Hz] & \; SNR \qquad  \\
\hline
A & 0.4339 & 2.030 \\
B & 0.4764 & 2.957 \\
C & 0.5275 & 2.364 \\
D & 0.8851 & 2.626 \\
E & 0.9167 & 5.071 \\
\hline
\end{tabular}
\end{center}
\caption{Frequencies and SNRs of persistent significant signal candidates above SNR$=2$ (below $1\unit{Hz}$) for all-year data and above the SNR threshold weighted by noise level for each-year data, removing the candidates of neat frequencies.}
\label{tab:list-of-signal-candidates-low-freq}
\end{table}


\section{Observational constraint}
\label{sec:observational-constraint}

Except for 342 frequencies that the axion signal candidates (SNR$>2$) were found, the data are consistent with noise and we obtain the upper limit on the axion coupling strength in the range of $0.43-43.9$\,Hz. The 95\% CL upper limit on the axion-photon coupling strength is obtained from
\beq
\int_{-\infty}^{\hat{s}_{\rm obs}(m_{\rm a})} \diff \hat{s}\, p[\hat{s}(m_{\rm a})|g_{\rm a\gamma}] = 0.05 \;,
\eeq
where $\hat{s}_{\rm obs}$ is the observed value of the weighted-average differential PSD, which can be converted to the axion coupling strength by Eq.~\eqref{eq:signal-weighted-ave} and the theoretical curve from Fig.~\ref{fig:Response_axion_B-field_Eskdalemuir}. $p[\hat{s}(m_{\rm a})|g_{\rm a\gamma}]$ is the probability density distribution of the weighted-average differential PSD in the presence of axions, which is obtained from the data $\hat{s}$ of the surrounding $\pm200$ bins, excluding the three bins around $f=f_{\rm a}$ and adding an axion signal in Eq.~\eqref{eq:signal-weighted-ave} at $f=f_{\rm a}$. Since the frequency bin of the axion signal and the frequency bins of noise sampling do not overlap, they do not affect each other. As shown in Appendix~\ref{app:noise-dist}, the distribution of the noise amplitude, $\hat{s}$, is well approximated by the Gaussian distribution with zero mean, except for the frequency range of $1.1$--$3.9\unit{Hz}$, indicating that the 95\% CL upper limit obtained in this analysis roughly corresponds to twice the standard deviation of the Gaussian distributions.

\begin{figure*}[t]
\includegraphics[width=12cm,angle=0]{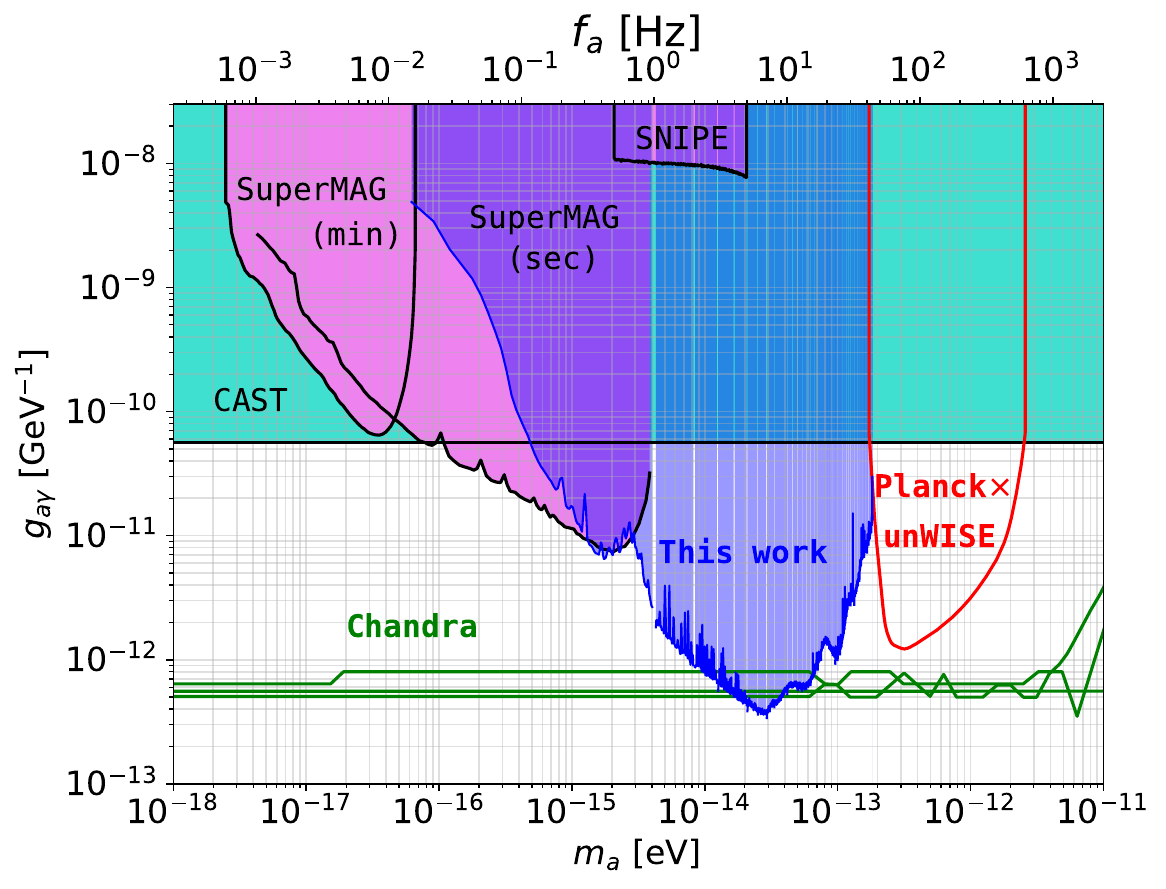}
\caption{Constraint (95\% CL) on the axion-photon coupling $\gag$ from the long-term monitoring data of magnetic fields at Eskdalemuir observatory (this work, blue). Other excluded regions are from SuperMAG~\cite{Arza:2021ekq,Friel:2024shg}, SNIPE~\cite{Sulai:2023zqw}, Planck and unWISE blue galaxy sample~\cite{Goldstein:2024mfp}, CAST~\cite{CAST:2017uph,CAST:2024eil}, and Chandra assuming $g_{\rm a\gamma}=g_{\rm ae}$~\cite{Reynes:2021bpe,Reynolds:2019uqt,Dessert:2021bkv,Ning:2024ozs}. The data are taken from \cite{AxionLimits}. The constraints by direct means are filled and that by astrophysical model-dependent means are not filled.}
\label{fig:constraint_axion_photon_coupling}
\end{figure*}

The upper limit on the axion-photon coupling is shown in Fig.~\ref{fig:constraint_axion_photon_coupling}. Around the axion mass of $3\times 10^{-14}\unit{eV}$, the upper limit is the tightest and is improved from that by CAST~\cite{CAST:2017uph,CAST:2024eil} by about two orders of magnitude, and exceeds that obtained from astrophysical X-ray observations by Chandra~\cite{Reynes:2021bpe,Reynolds:2019uqt,Dessert:2021bkv,Ning:2024ozs}. In the lower mass range, our constraint is weakened because of fixing the data segment size to 8 hrs and using broader frequency resolution, irrespective of the coherent time of axions. The constraint can be stronger by optimizing the data segment size or the frequency bin resolution for each axion mass and could be comparable to the upper limit from SuperMAG~\cite{Arza:2021ekq,Friel:2024shg}. In the higher mass range above $10^{-14}\unit{eV}$, the constraint is also weakened due to the suppression of the magnetic field response to axions. Interestingly, our new constraint closes the window between the excluded regions from SuperMAG and the Planck and unWISE blue galaxy sample~\cite{Goldstein:2024mfp}, though the astrophysical model-dependent constraints from Chandra~\cite{Reynes:2021bpe,Reynolds:2019uqt,Dessert:2021bkv,Ning:2024ozs} already exist.

Finally, we comment on the so-called stochastic effect~\cite{Centers_etal2019,Lisanti_etal2021}, which is caused by random realizations of axion field amplitude during the measurement time and could loosen the upper limit on the axion coupling. This effect is relevant when the measurement time is shorter than the coherence time of axions. In our analysis, we computed the axion field amplitude from the average density of dark matter and assumed the amplitude constant in time. However, since the measurement time is more than 10 times longer than the coherent time in the sensitive range of axion mass, the stochastic effect is negligible~\cite{Nakatsuka:2022gaf}. More quantitatively, we use 9909 8-hr data segments and the fraction of the good quality data is $\sim 10$\%, resulting in the effective measurement time of $\sim 300$ days. In the mass range heavier than $2\times 10^{-15}\unit{eV}$ or frequency higher than $0.4\unit{Hz}$, the condition above is satisfied and the stochastic effect is negligible. Therefore, our constraint improved from SuperMAG and CAST in the mass range above $2\times 10^{-15}\unit{eV}$ is not affected by the stochastic effect.  

\section{Conclusions}
\label{sec:conclusions}

In this paper, we utilize the terrestrial EM waves in extremely low-frequency bands as a powerful probe for ultralight axion dark matter. We search for the characteristic spectral feature induced by coherently oscillating axions interacting with the Earth's magnetic fields in the long-term monitoring data of terrestrial magnetic fields. For the SNR threshold of 2 (3), we found 342 (65) axion signal candidates in the range of $0.43-43.9$\,Hz ($0.91-40.7$\,Hz). These candidates cannot be vetoed further due to the lack of additional information to identify their causes. At other frequencies at which signal candidates were not found, using the theoretical prediction of the induced EM waves that properly takes the finite conductivity of the atmosphere into account, we place the most stringent upper bound on the axion-photon coupling strength at $10^{-15}\,{\rm eV}\lesssim \mass\lesssim 10^{-13}\,{\rm eV}$ among the direct observational means. 

The follow-up measurements by future ground-based experiments such as DANCE \cite{Obata:2018vvr} or X-ray astronomical observations by Athena \cite{Sisk-Reynes:2022sqd} would play a crucial role to confirm or disprove the signal candidates. In this paper, we focused on axions as a representative ultralight dark matter, but the methodology for data analysis can be extended to other types of ultralight dark matter coupled with EM fields. An example worth considering for future study would be dark photon dark matter \cite{Nelson_Scholtz2011,Fabbrichesi_etal2020}, and its kinetic mixing can be constrained with the same data set we used~\cite{Fedderke_etal2021,Fedderke_etal2021_superMAG}.

\begin{acknowledgments}
We are grateful to C. Beggan for  kindly providing the data at the Eskdalemuir observatory and for his valuable comments. 
This work was supported in part by JSPS KAKENHI Grant Numbers JP20H05861 and JP21H01081 (AT), JP23K03408, JP23H00110, and JP23H04893 (AN), JP21K03580 (YH).
\end{acknowledgments}

\appendix 

\section{Calibration of the data}
\label{app:calibration}

\begin{figure*}[h]
\begin{center}
\includegraphics[width=8.5cm]{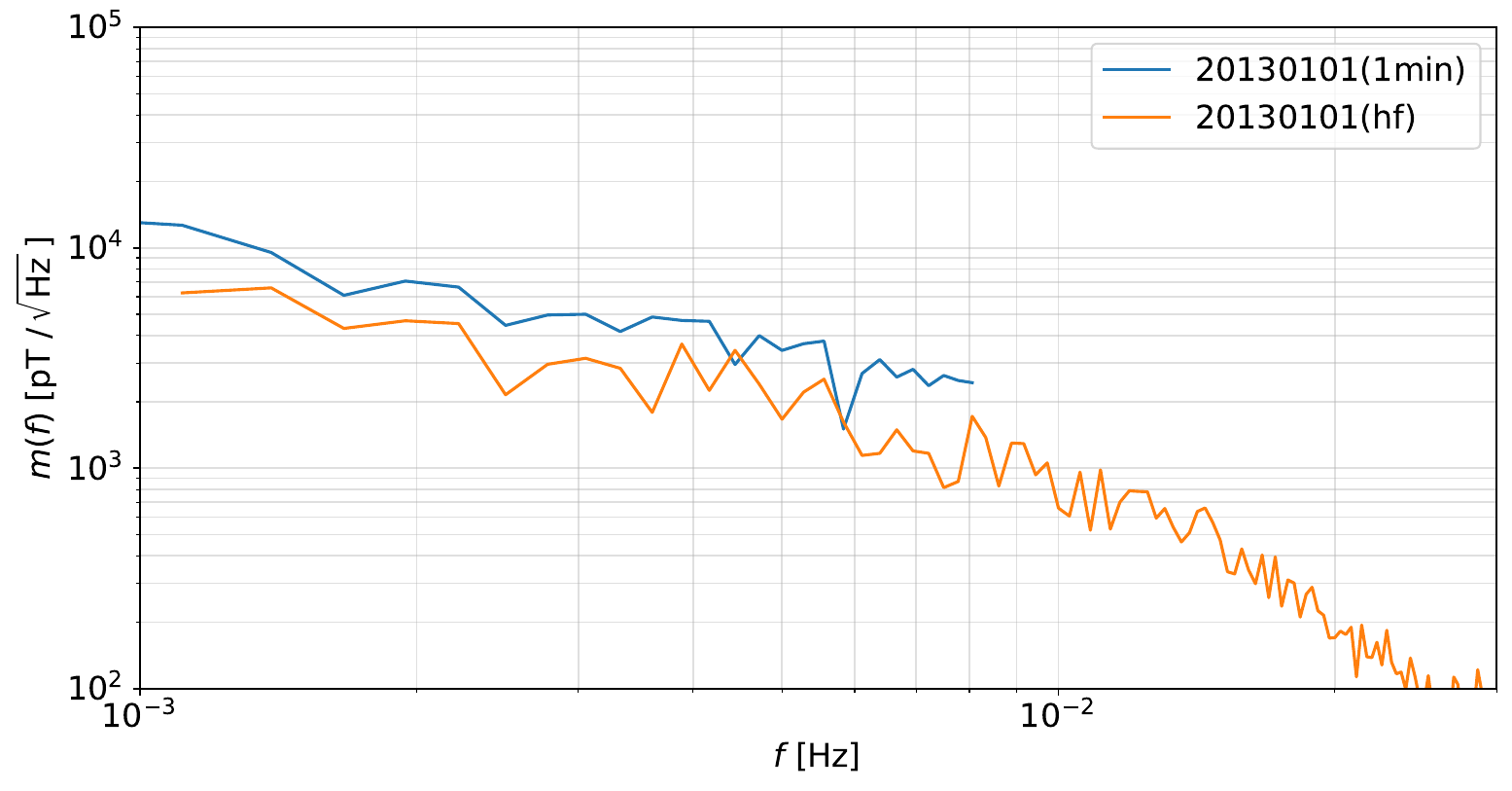}
\includegraphics[width=8.5cm]{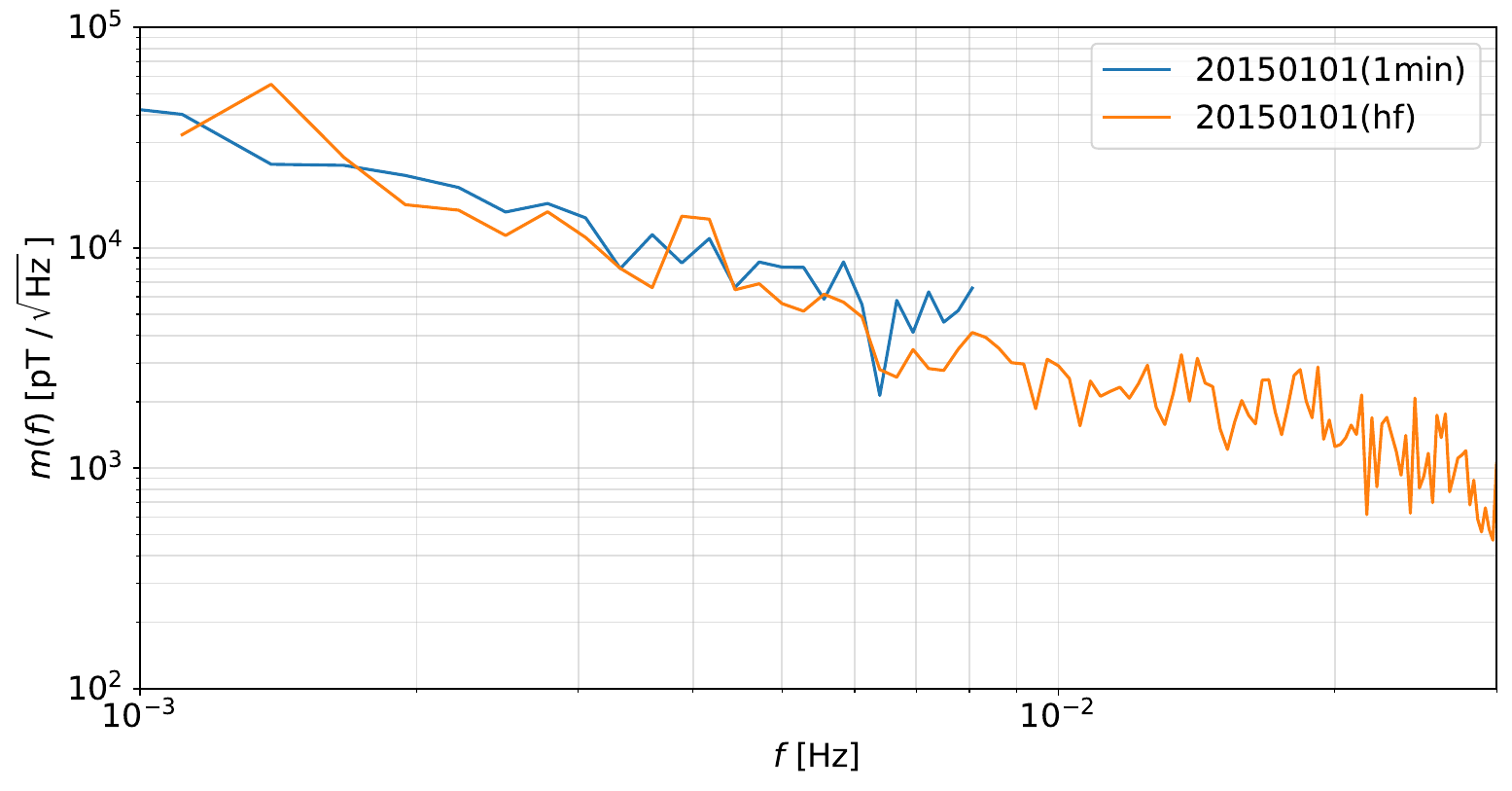}
\includegraphics[width=8.5cm]{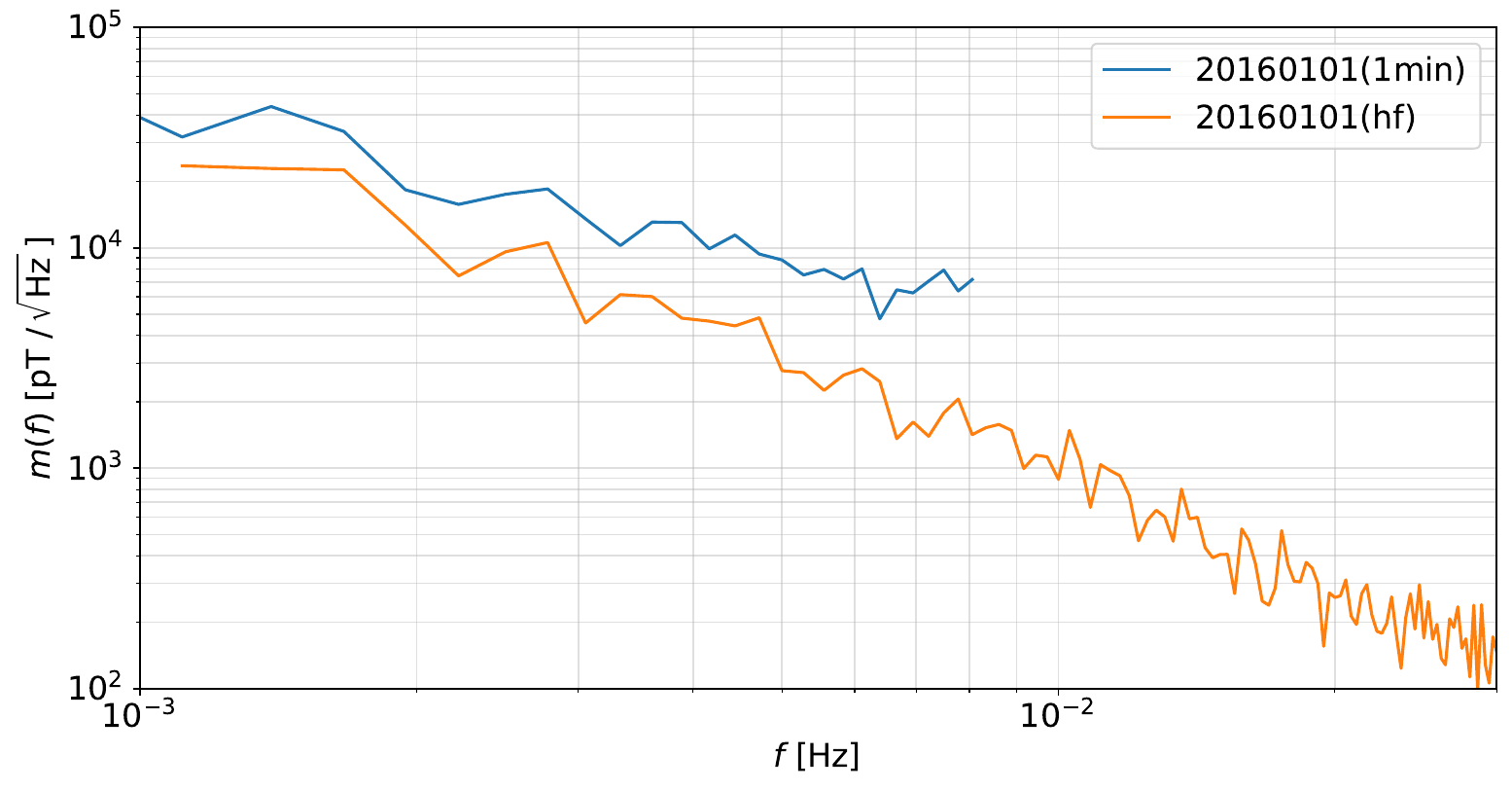}
\includegraphics[width=8.5cm]{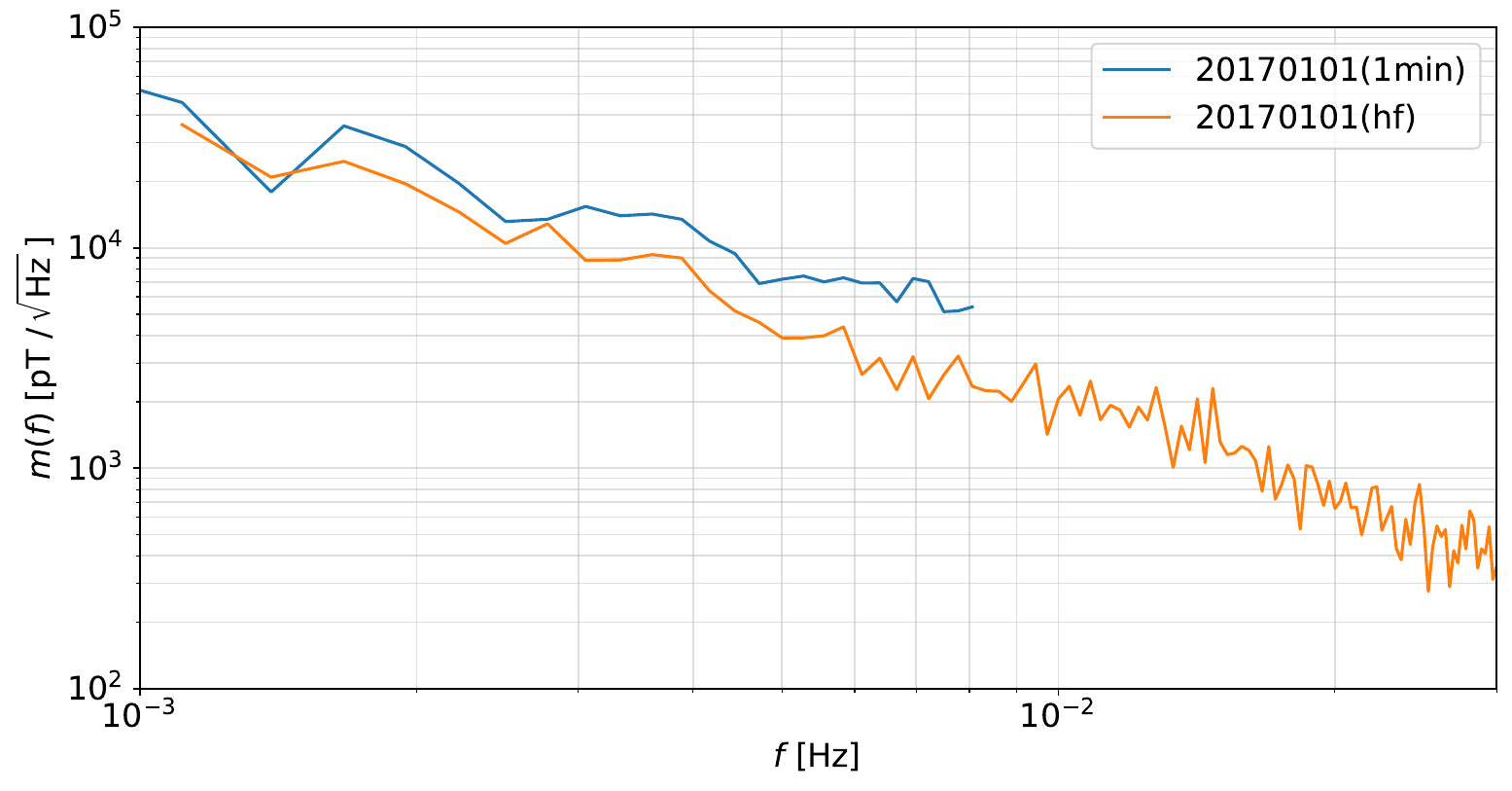}
\includegraphics[width=8.5cm]{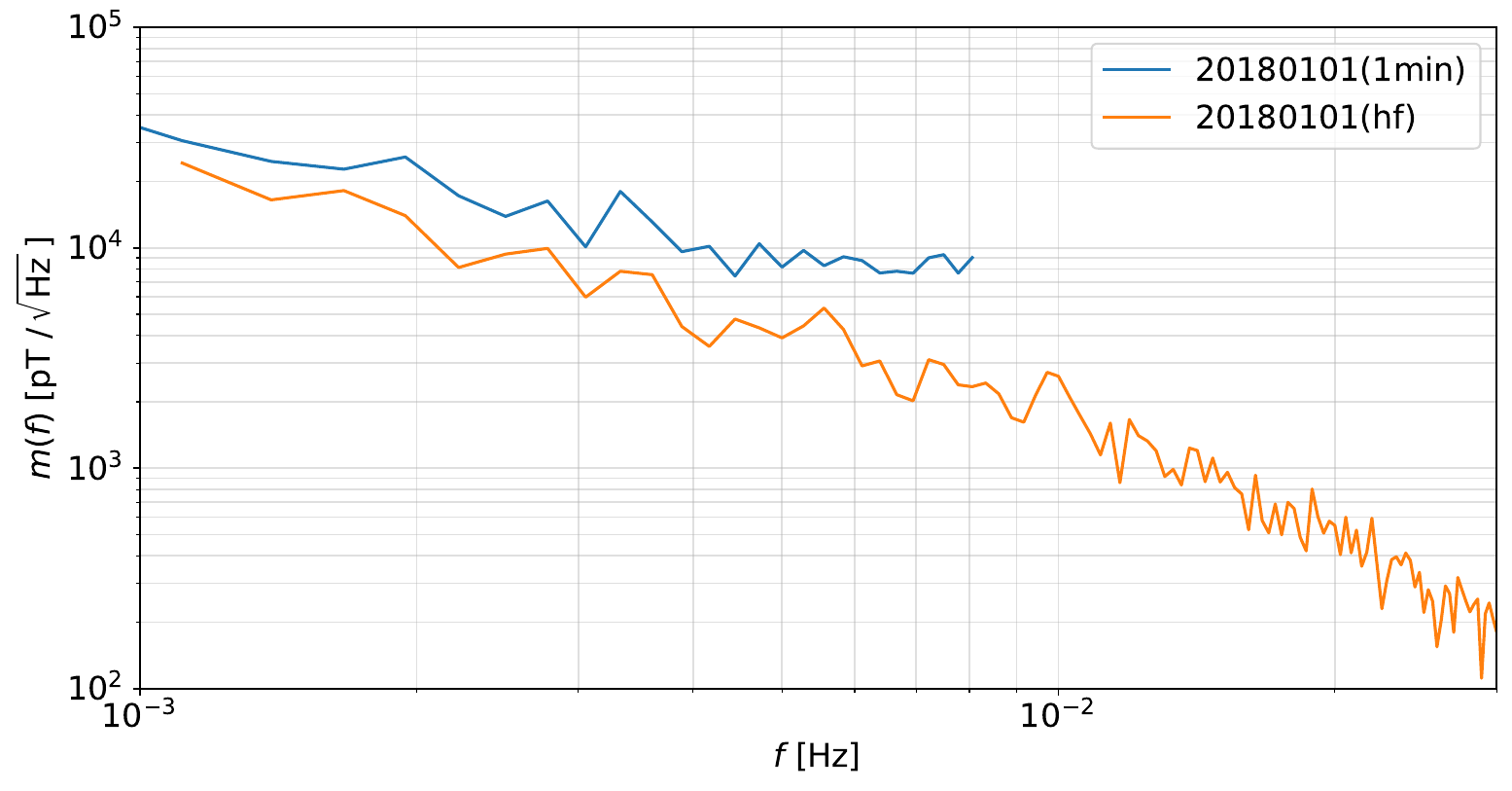}
\includegraphics[width=8.5cm]{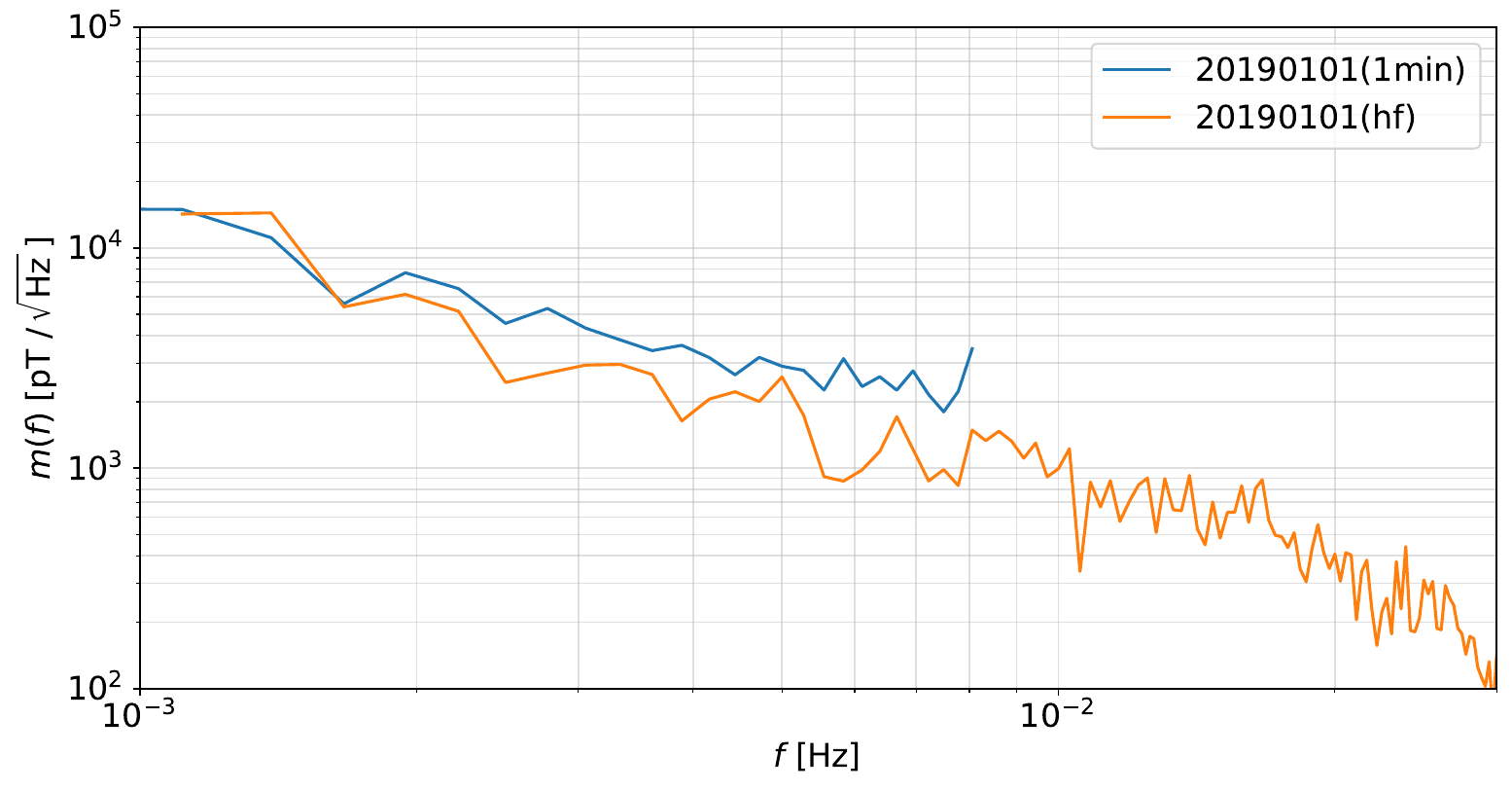}
\includegraphics[width=8.5cm]{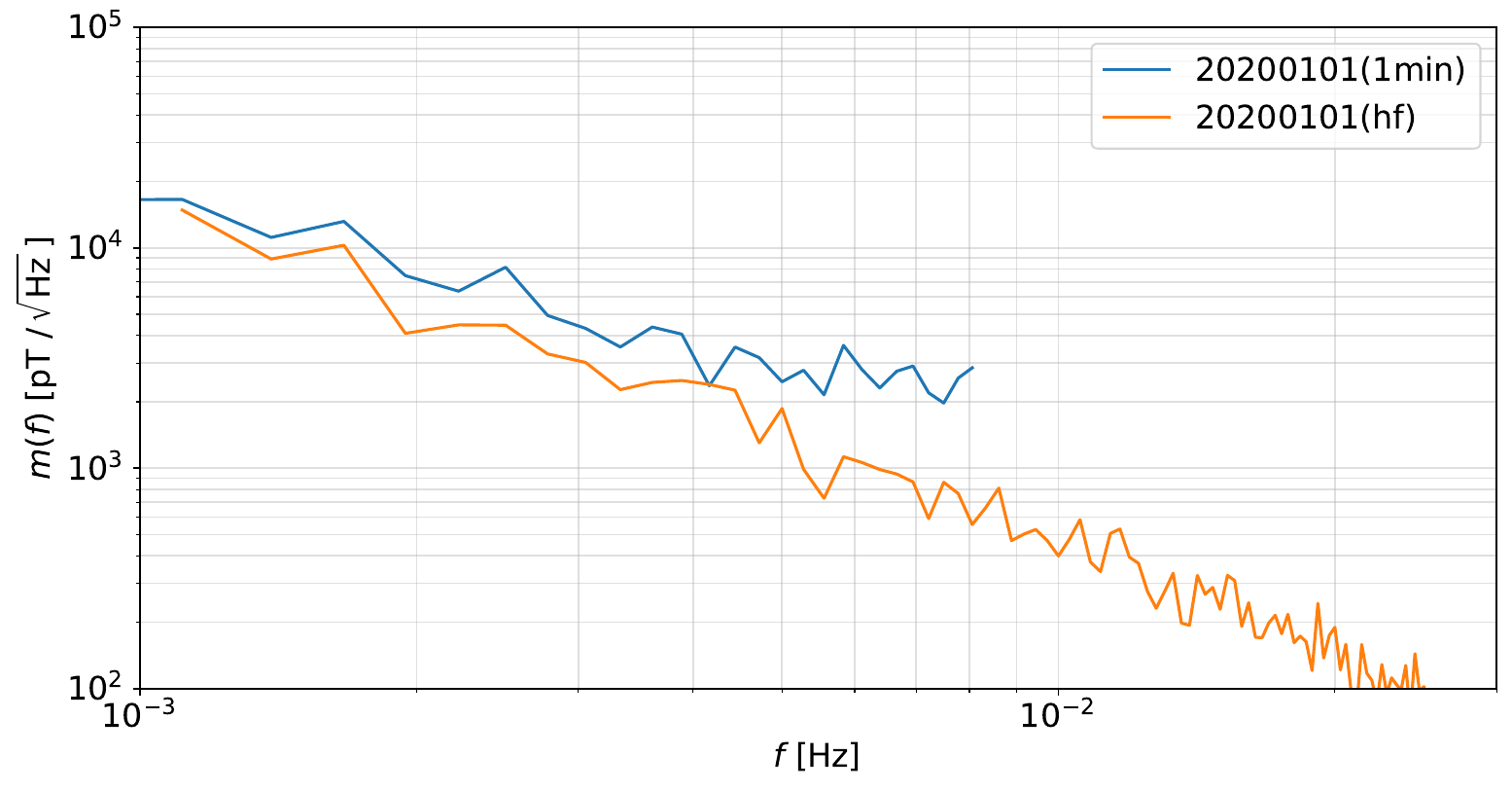}
\includegraphics[width=8.5cm]{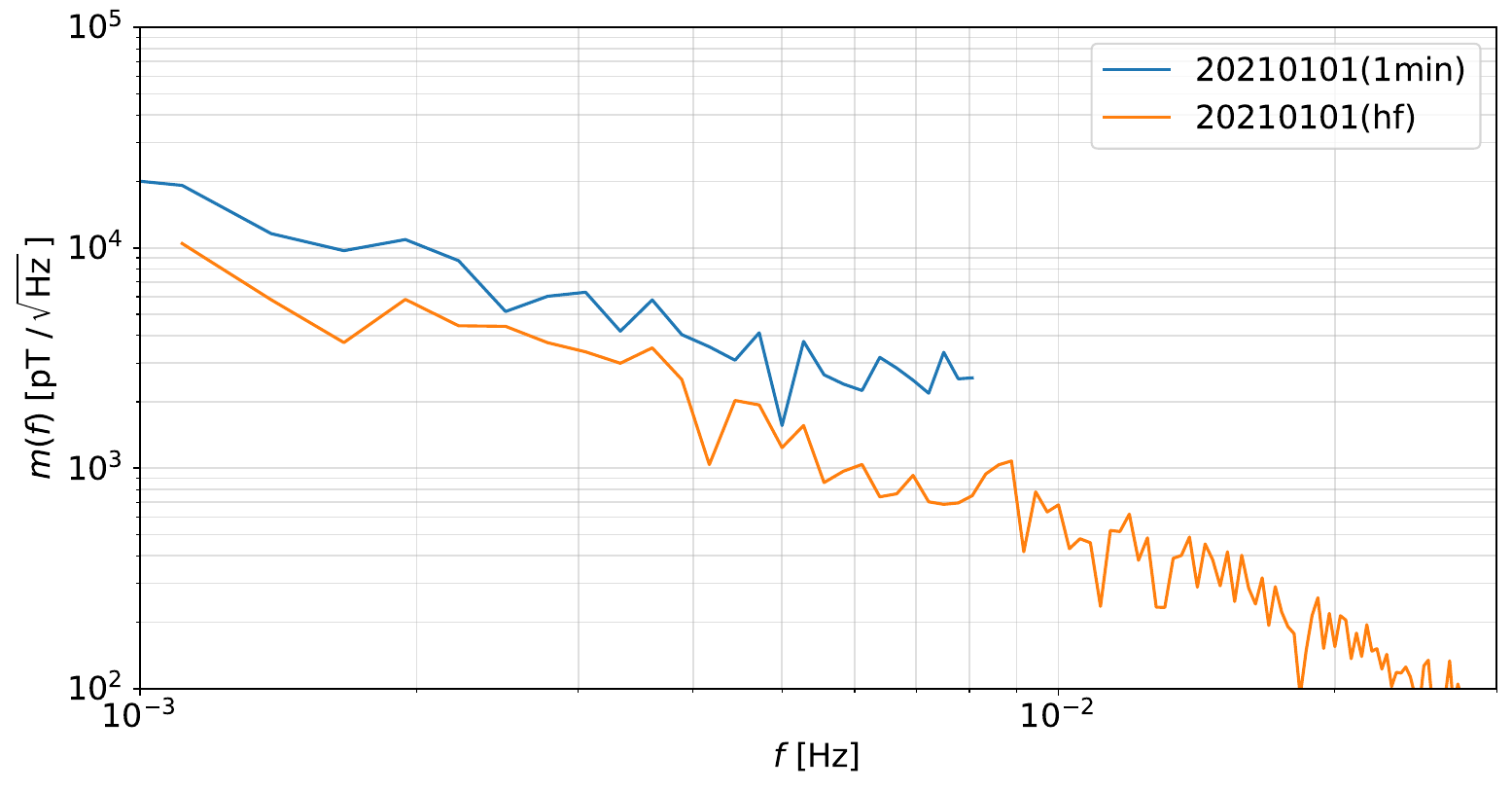}
\caption{Comparison of INTERMAGNET 1-minute sampling data (blue) and 100\,Hz-sampling data from the British Geological Survey (orange) at the Eskdalemuir observatory. Each data point is the 3/24-quantile amplitude from the minimum among all 1-hour data on January 1 of each year (to avoid the transient loud noises).}
\label{fig:calibration-comparison}
\end{center}
\end{figure*}

\begin{figure*}[h]
\begin{center}
\includegraphics[width=8cm]{calibration-20200101.pdf}
\includegraphics[width=8cm]{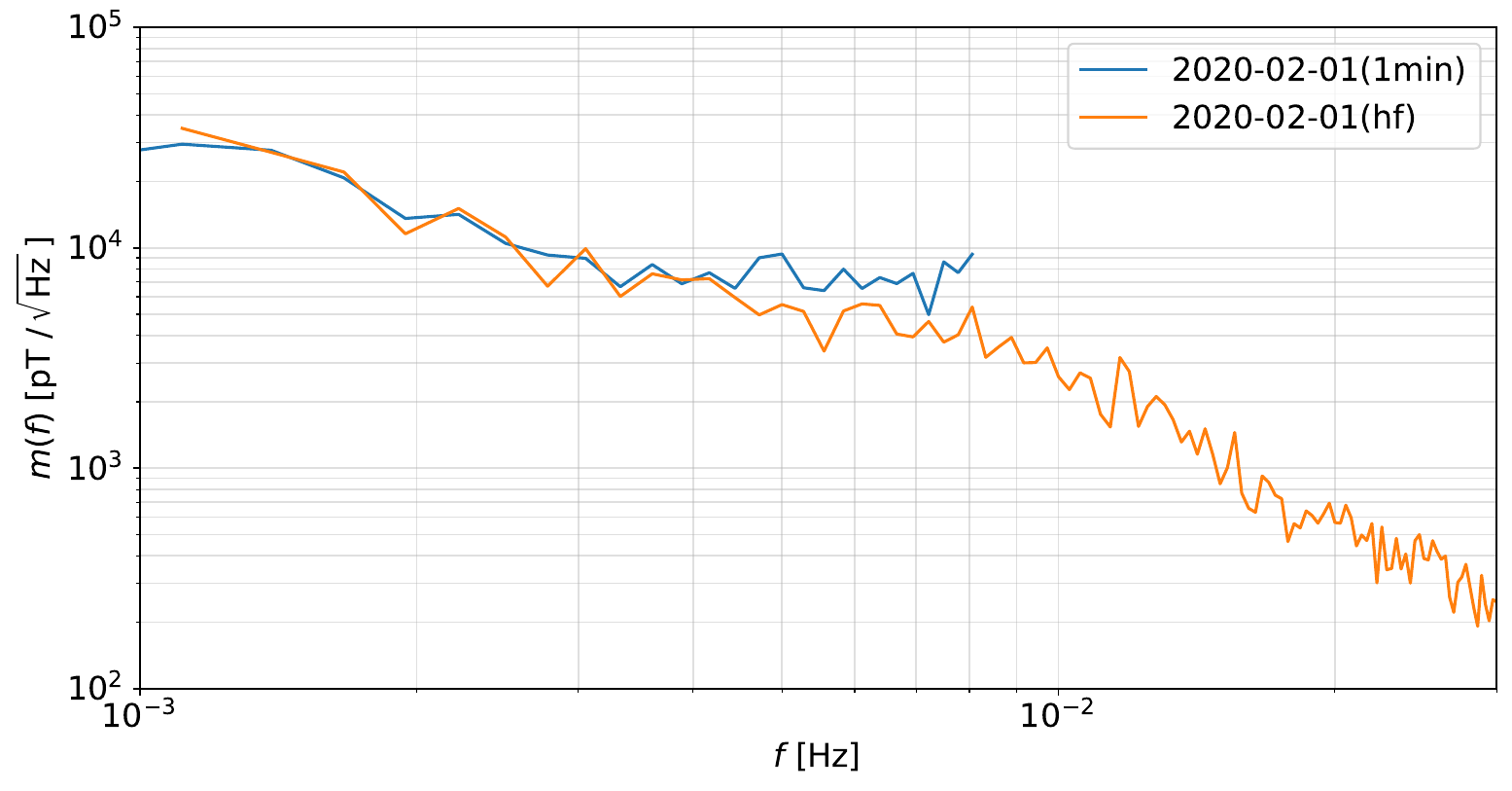}
\includegraphics[width=8cm]{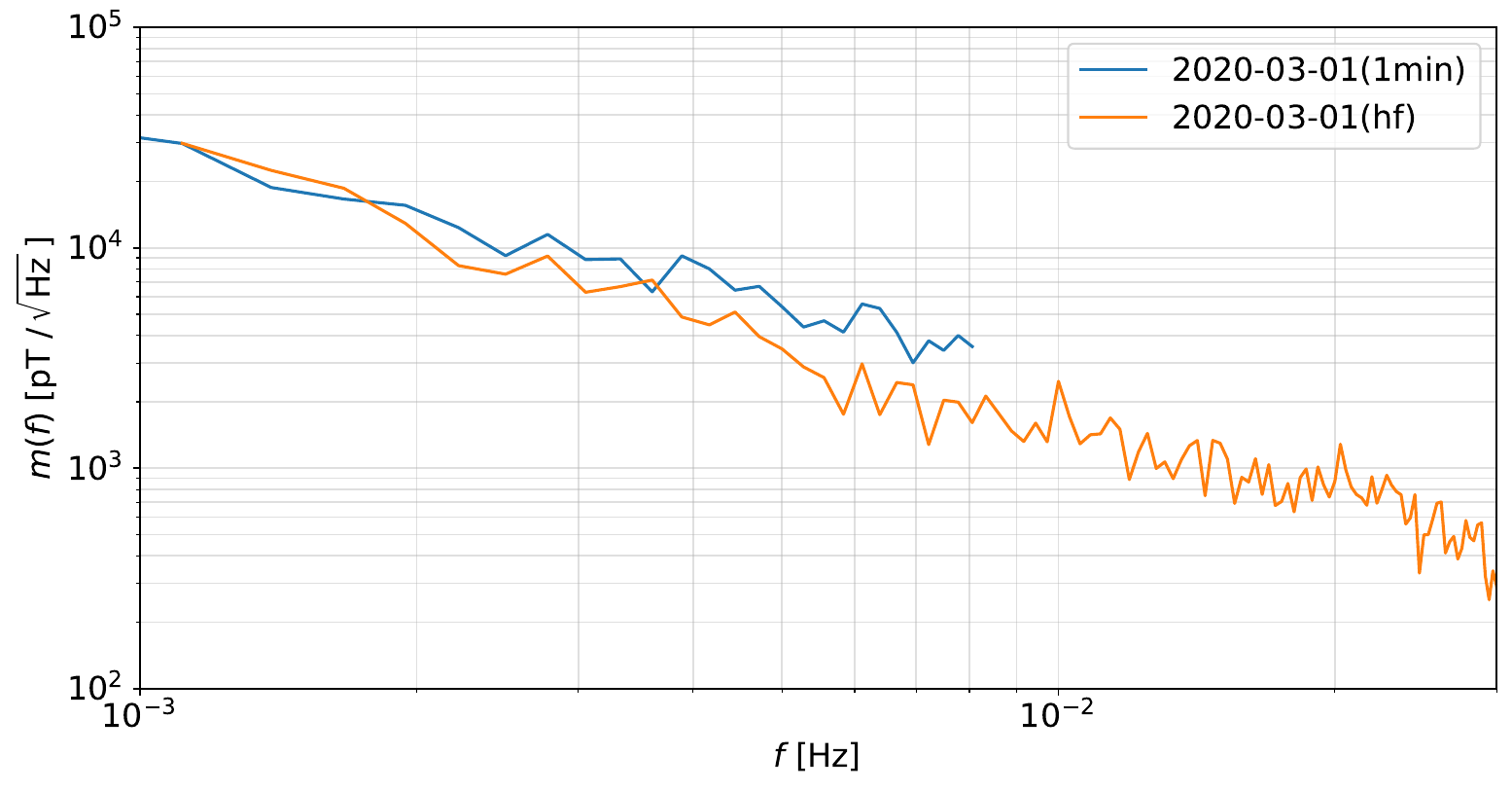}
\includegraphics[width=8cm]{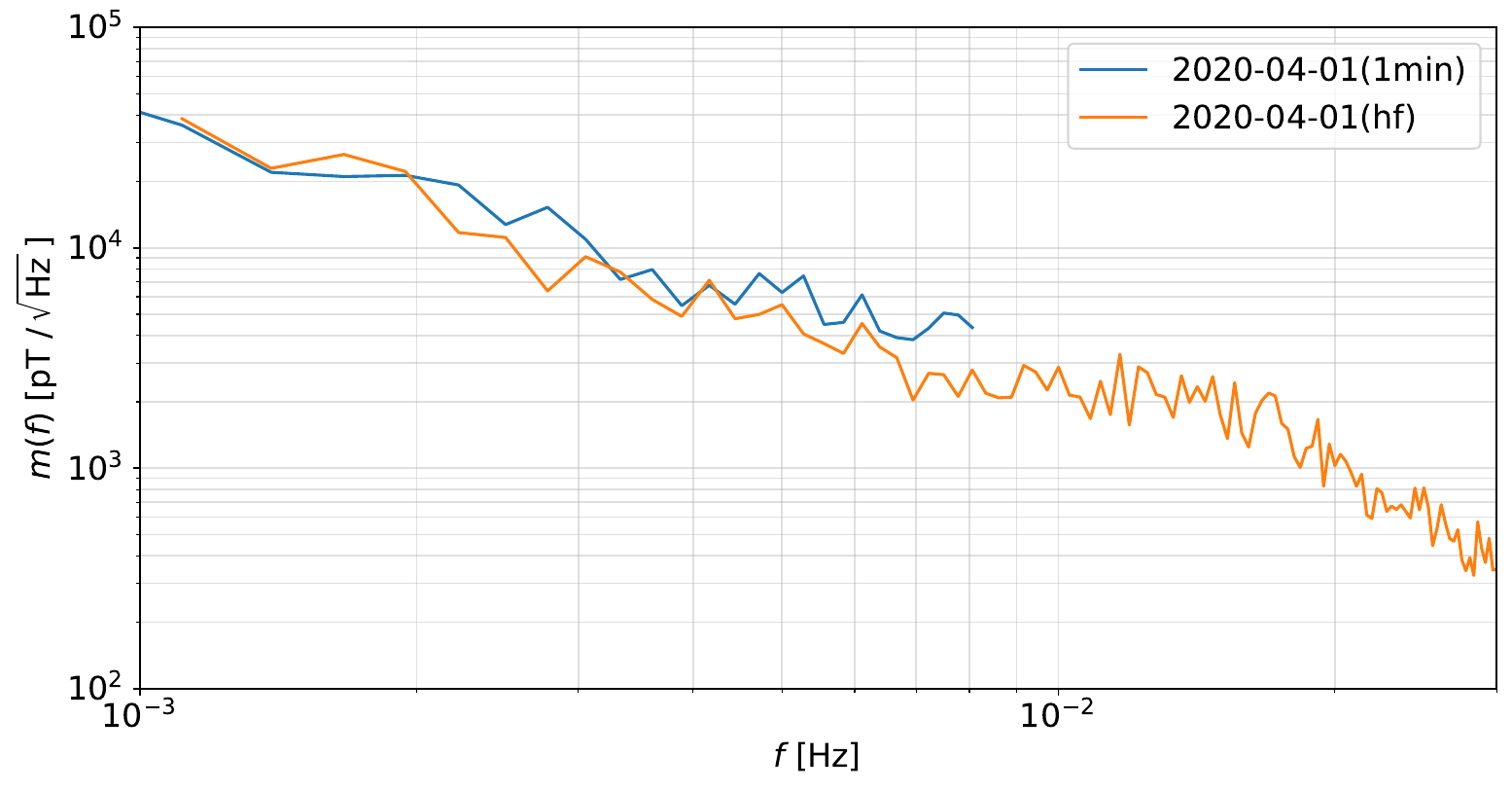}
\includegraphics[width=8cm]{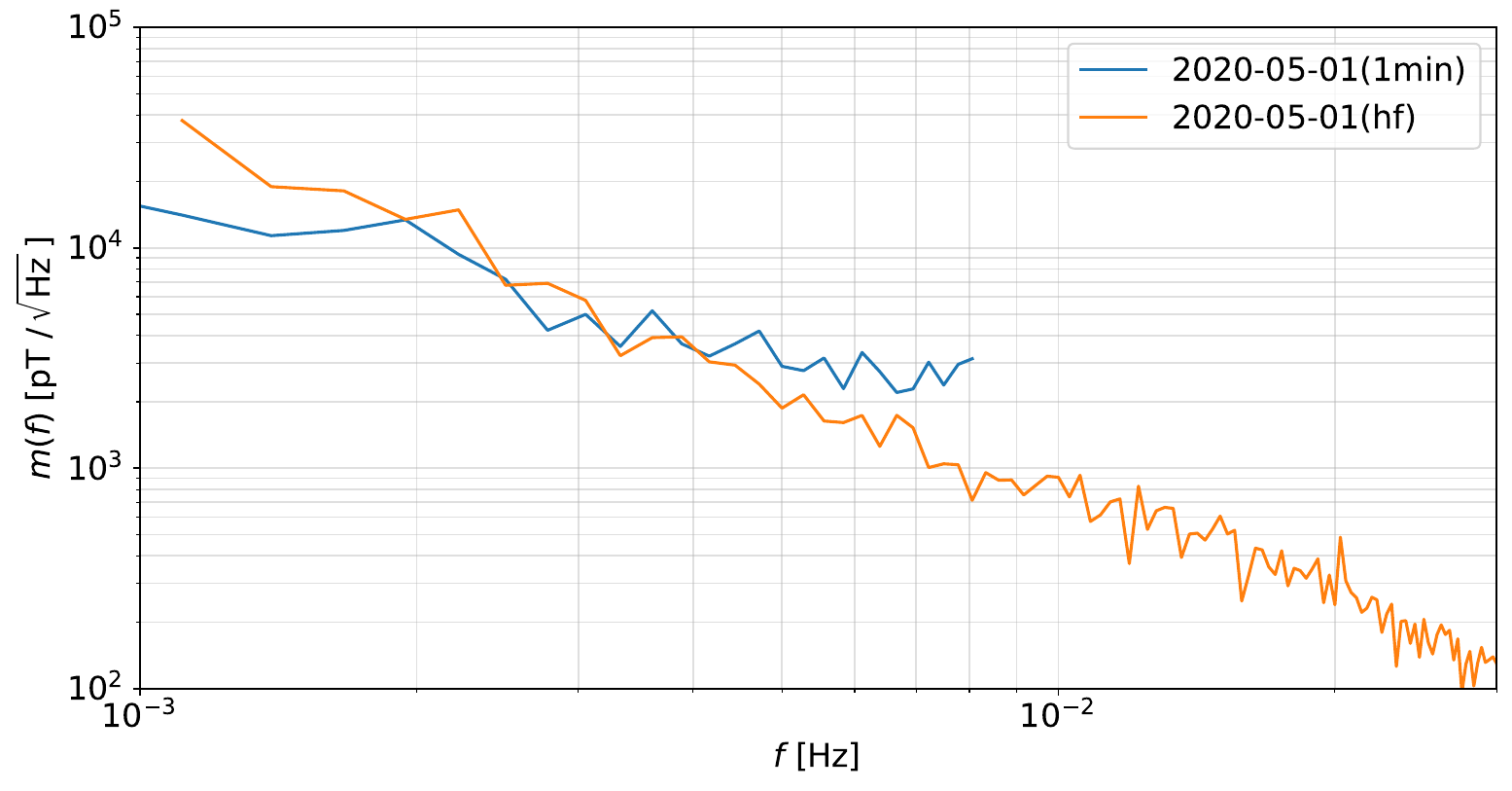}
\includegraphics[width=8cm]{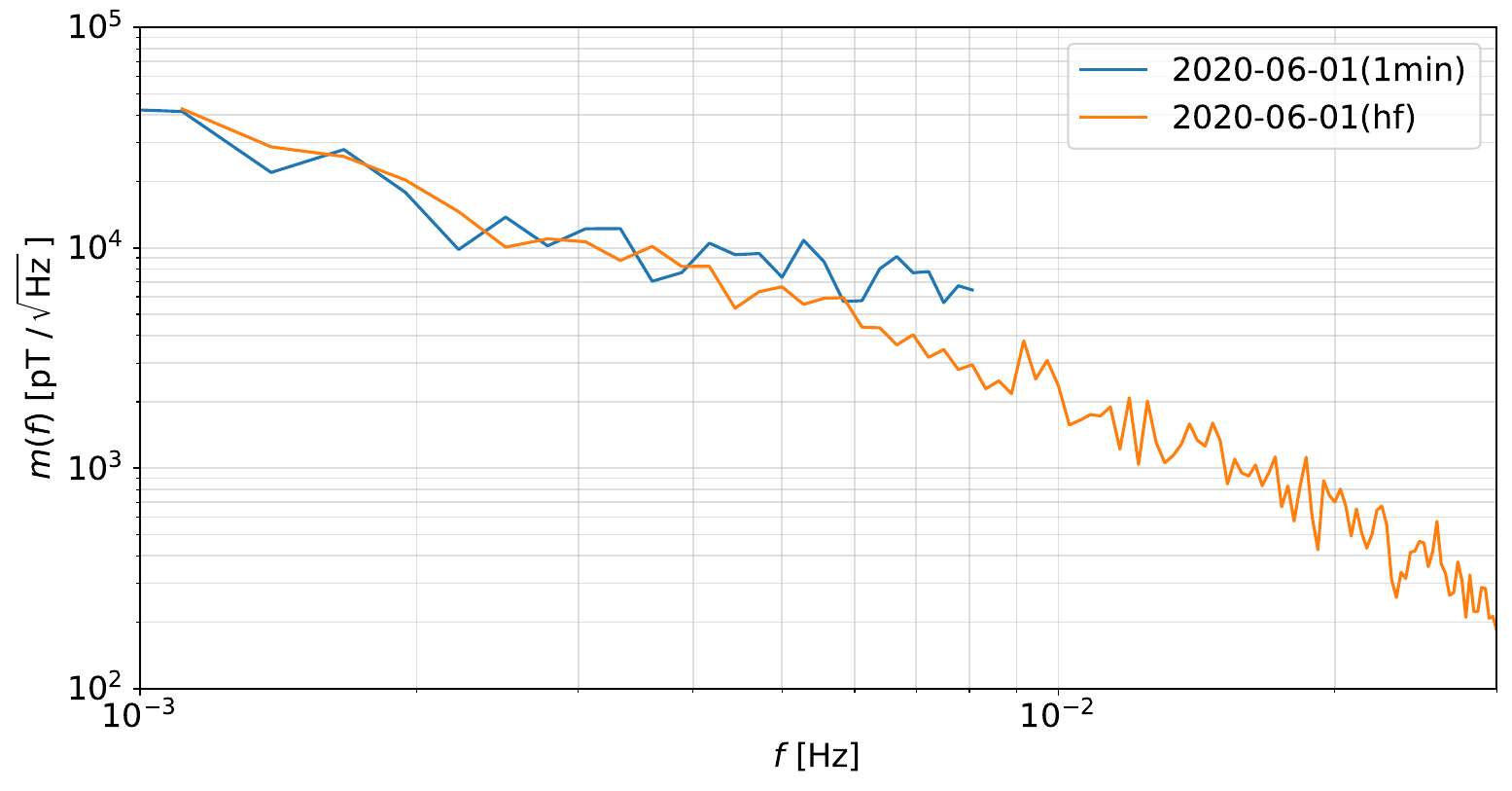}
\includegraphics[width=8cm]{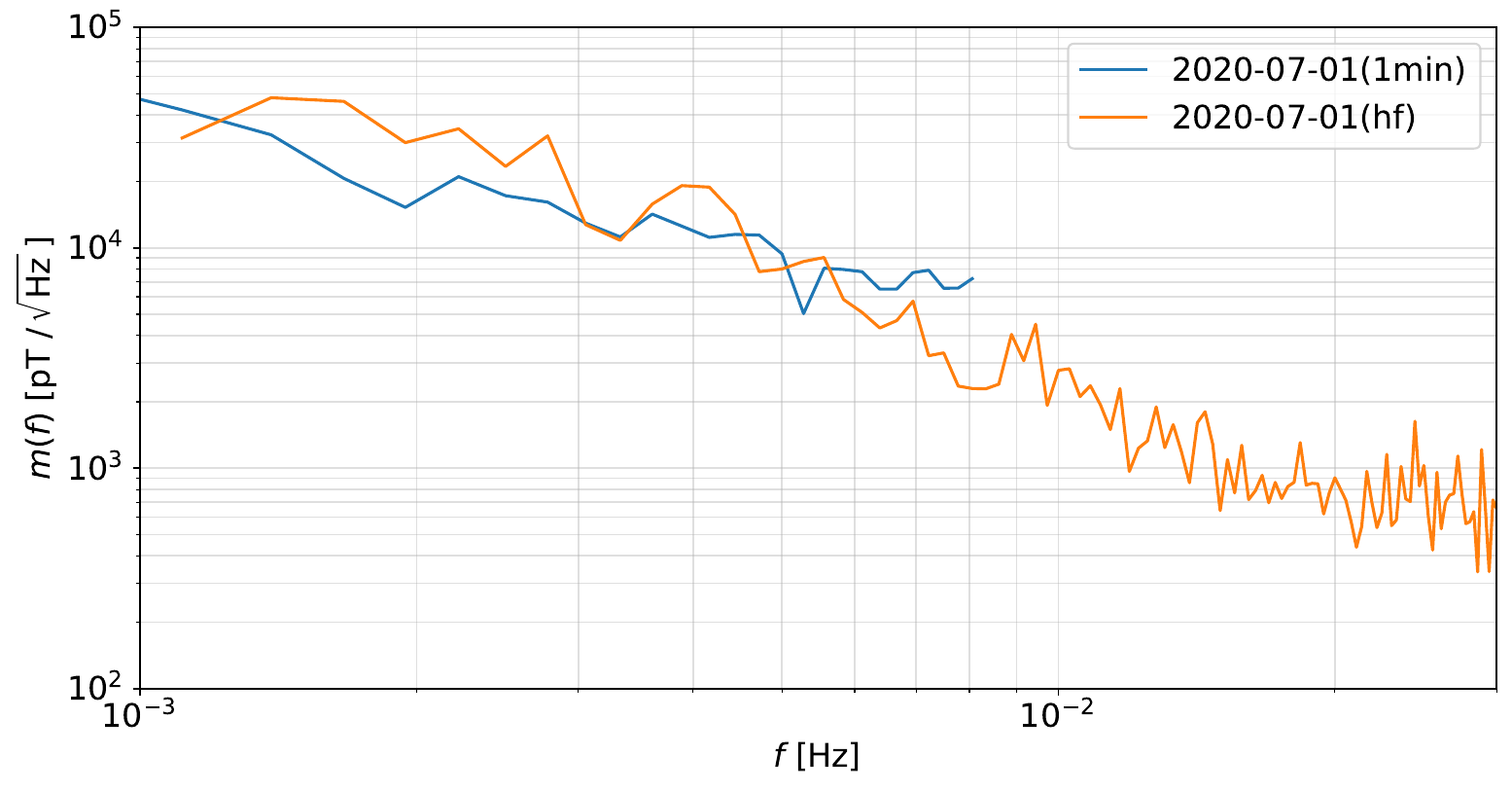}
\includegraphics[width=8cm]{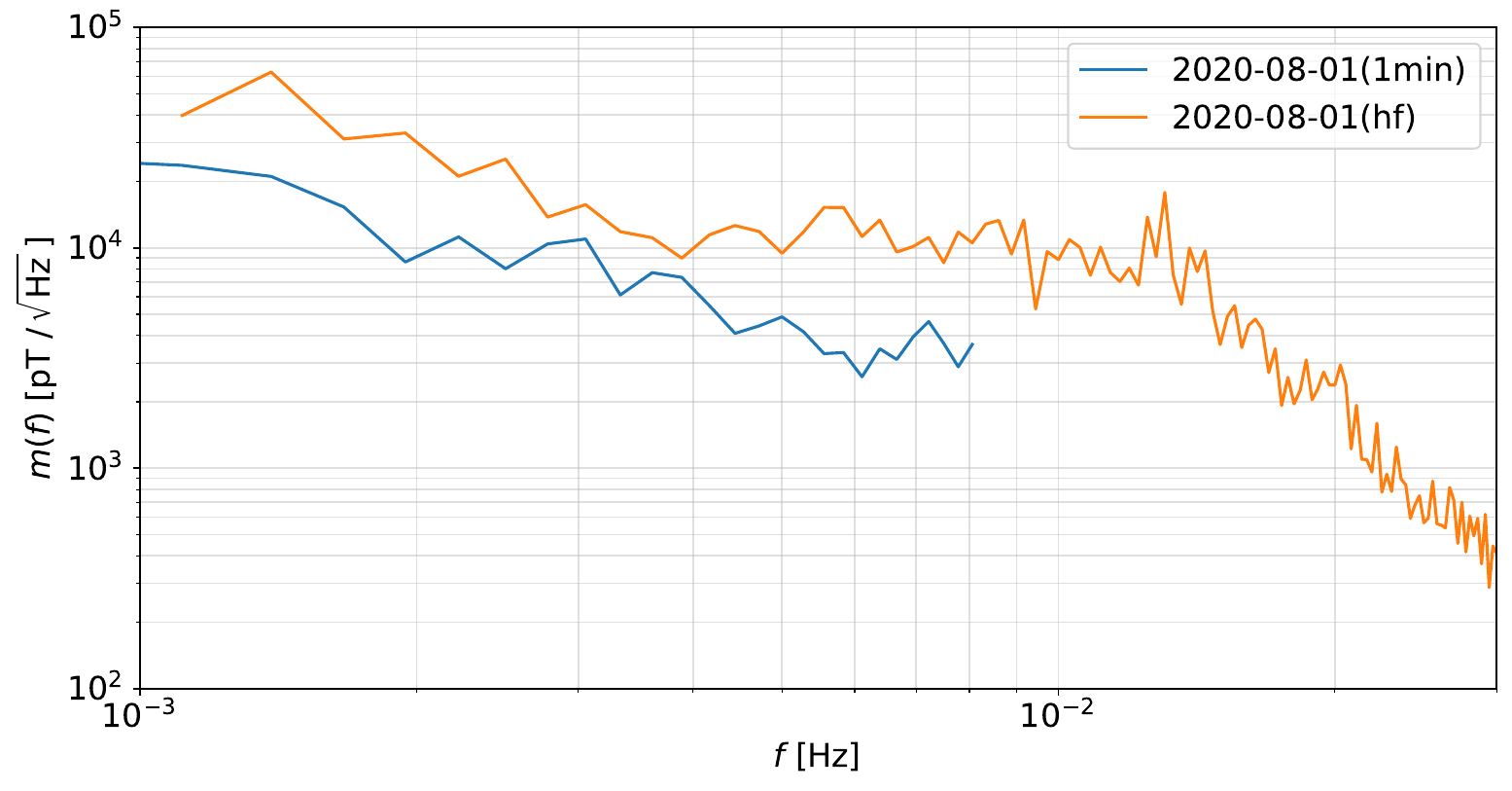}
\includegraphics[width=8cm]{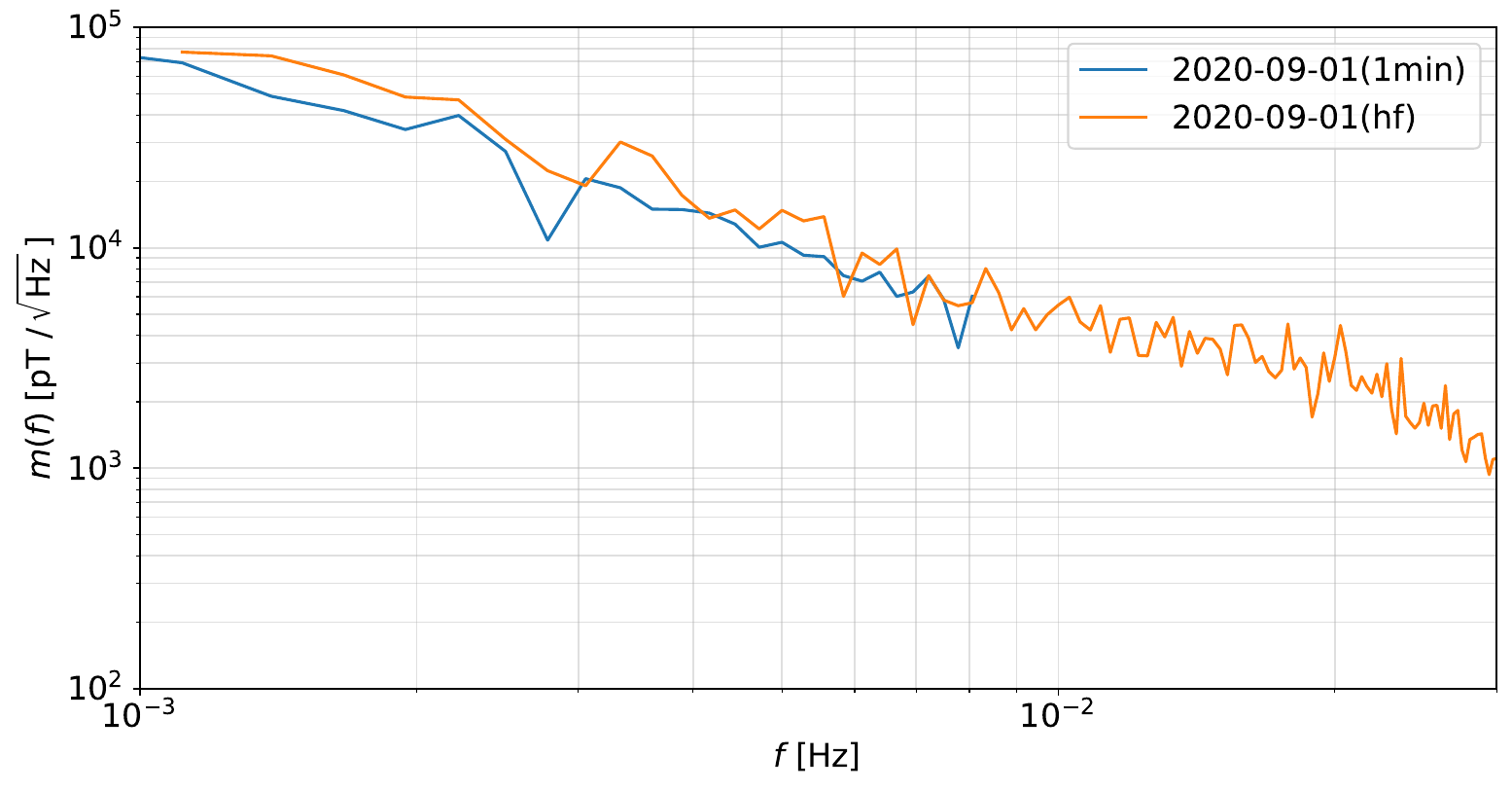}
\includegraphics[width=8cm]{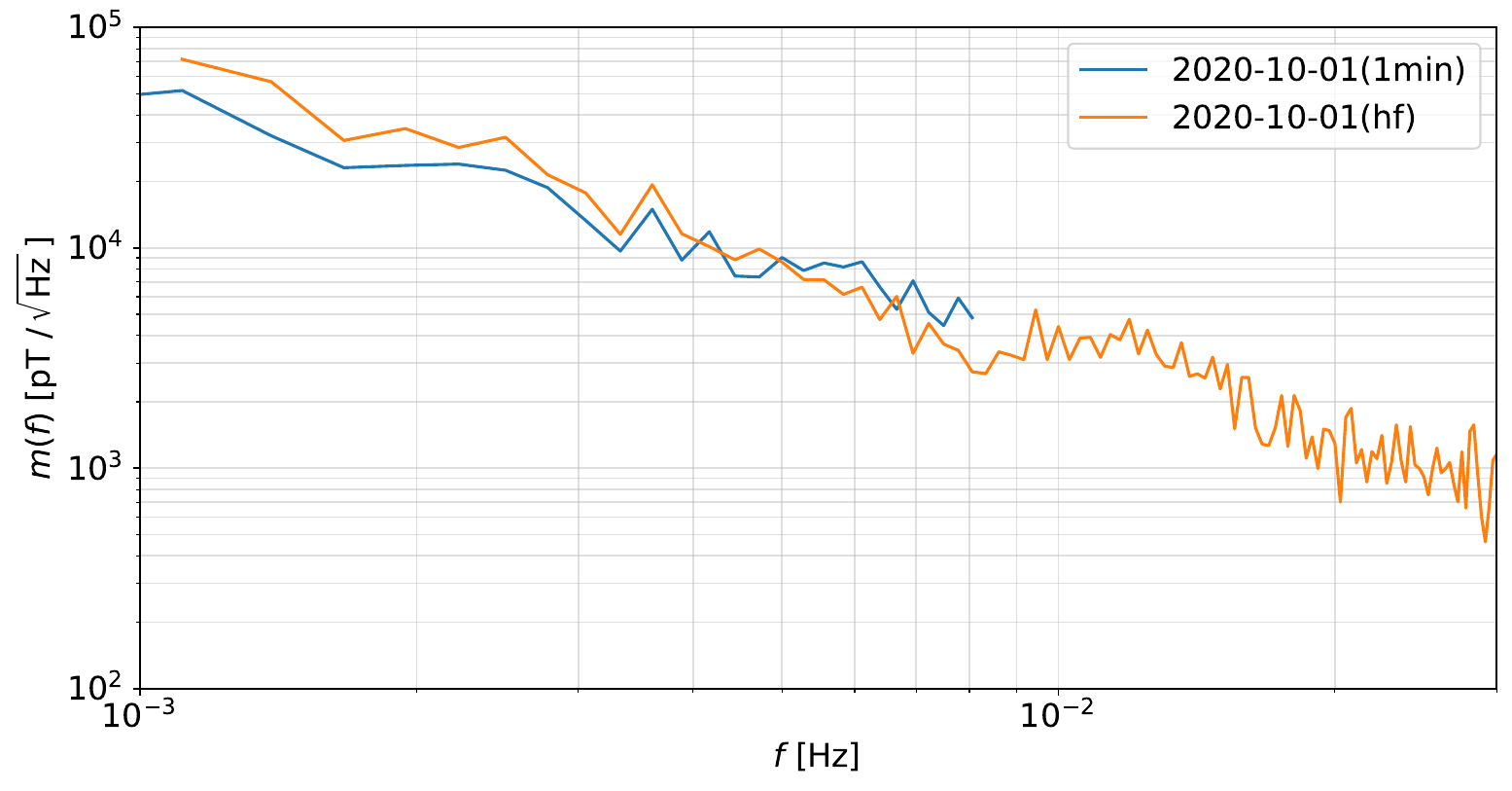}
\includegraphics[width=8cm]{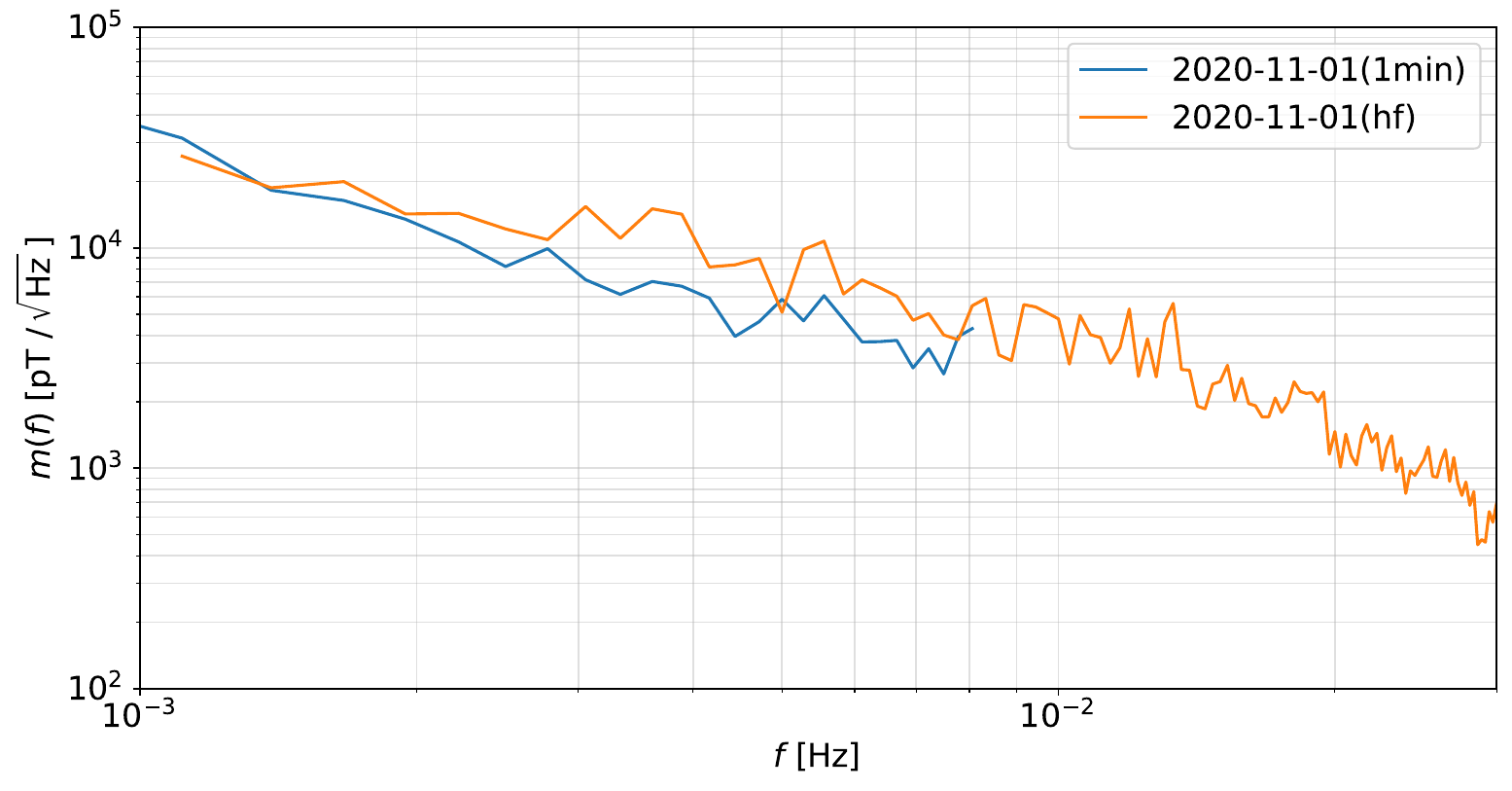}
\includegraphics[width=8cm]{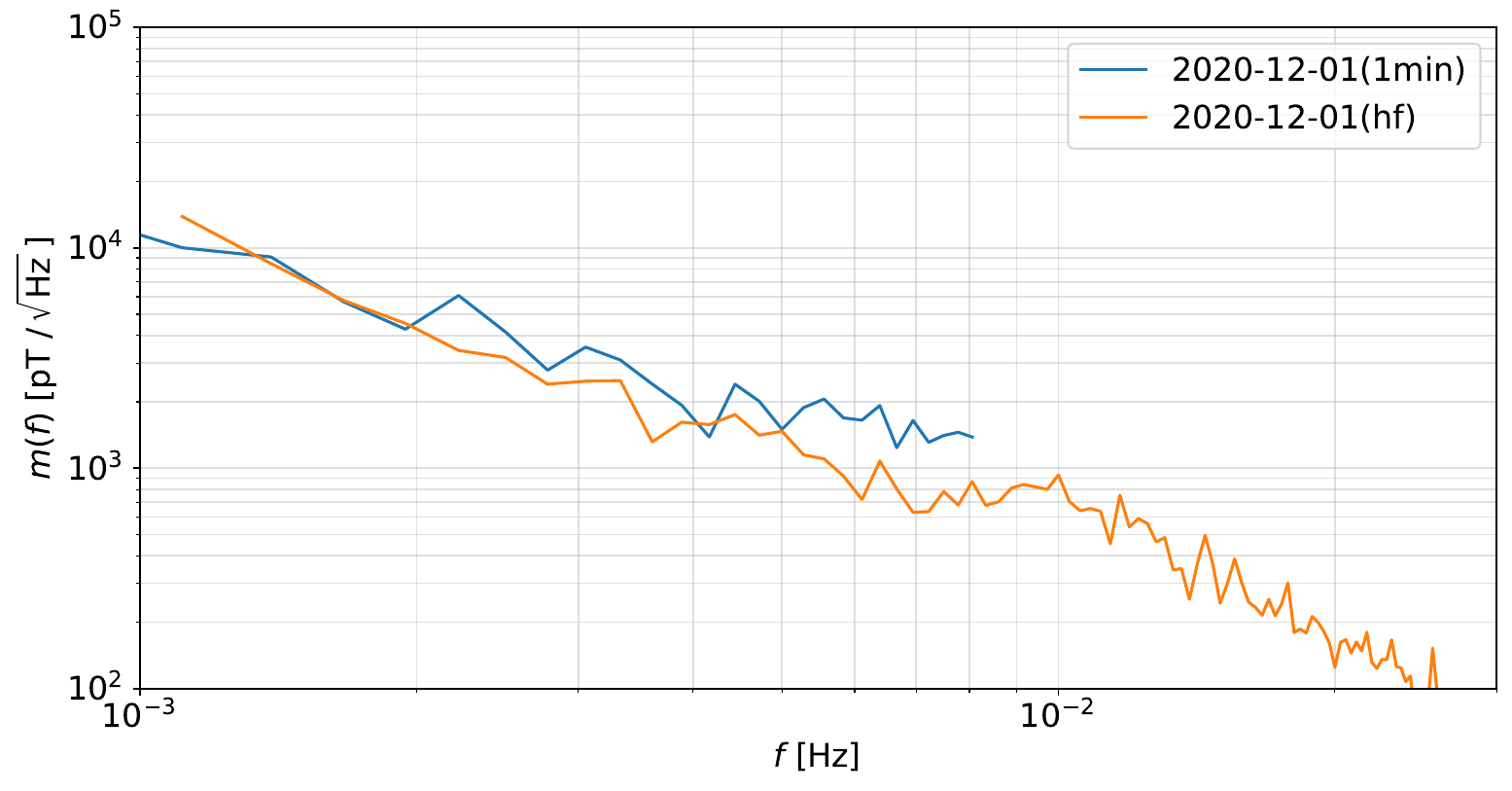}
\caption{Same as Fig.~\ref{fig:calibration-comparison} but the data are from one day in each month, 2020.}
\label{fig:calibration-comparison_2020}
\end{center}
\end{figure*}

\begin{figure*}[h]
\begin{center}
\includegraphics[width=8cm]{calibration-20210101.pdf}
\includegraphics[width=8cm]{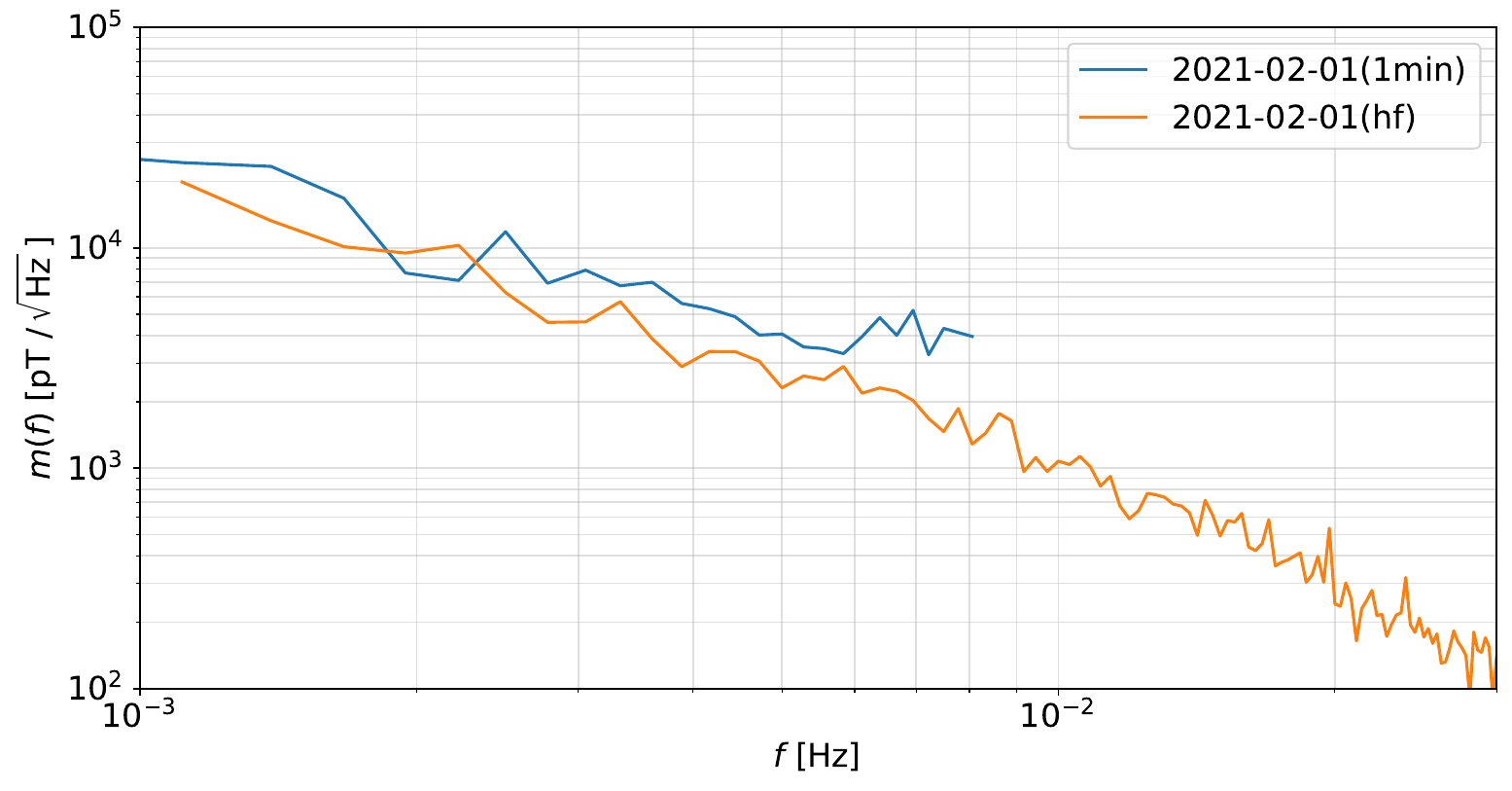}
\includegraphics[width=8cm]{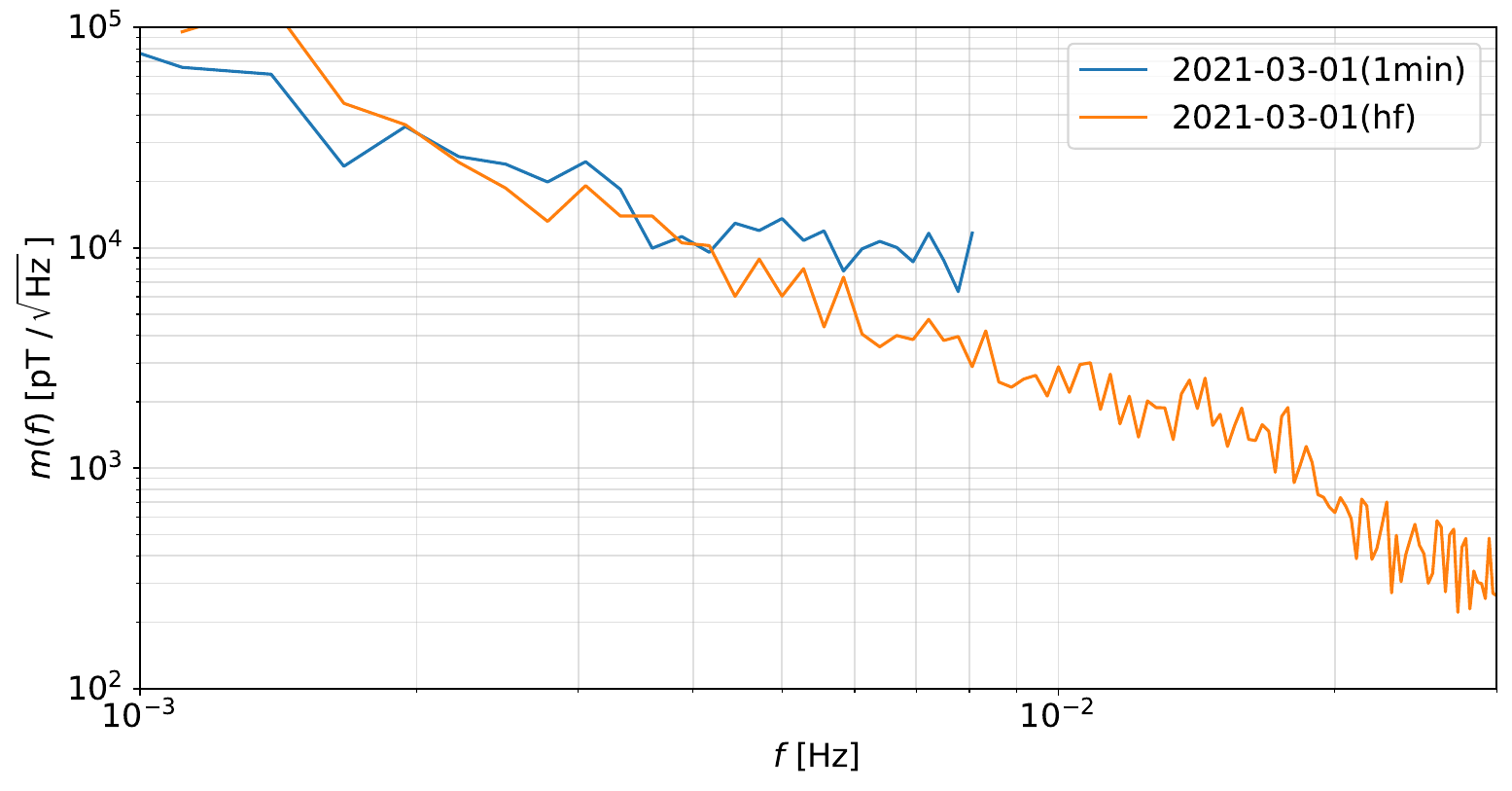}
\includegraphics[width=8cm]{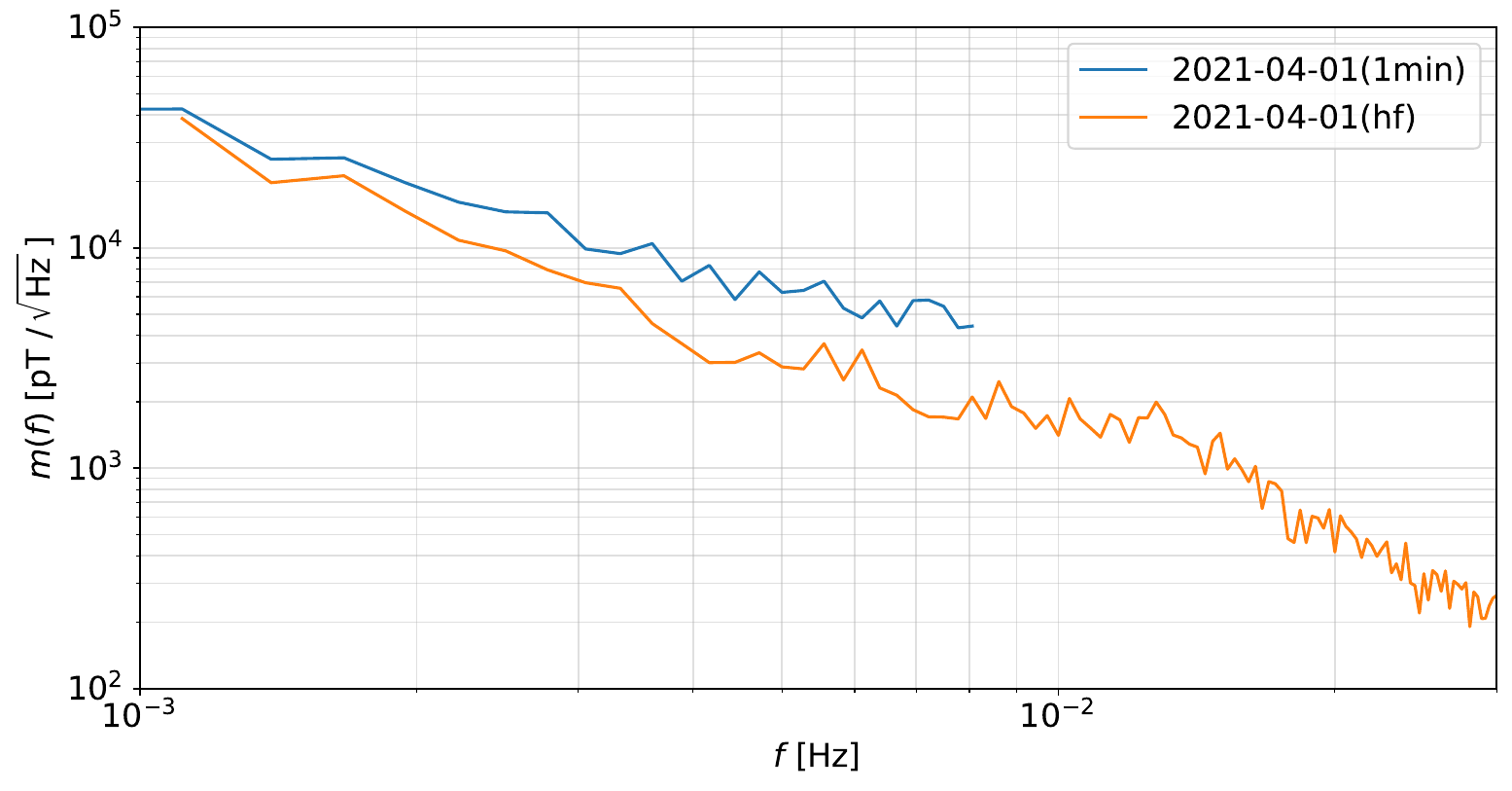}
\includegraphics[width=8cm]{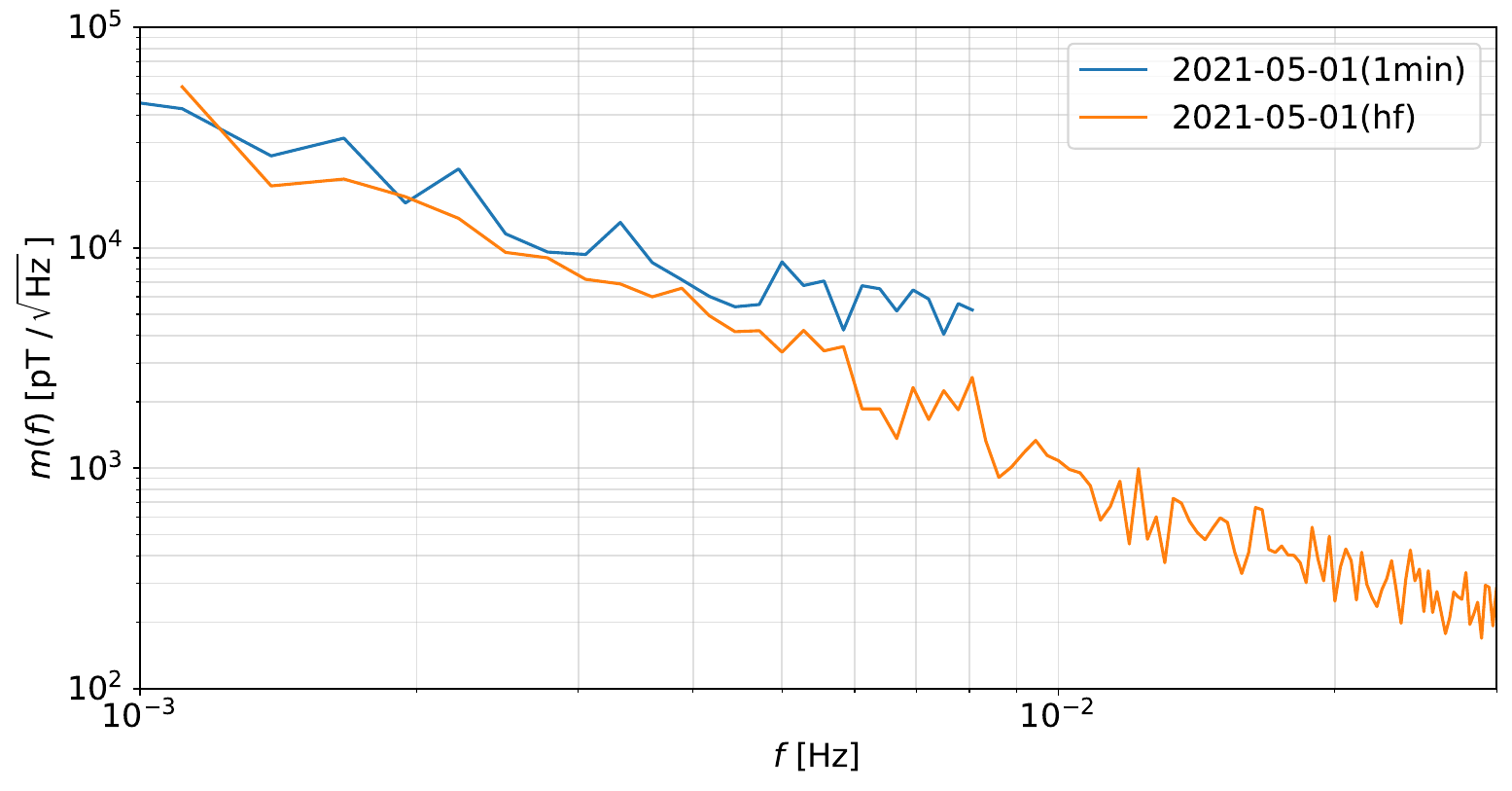}
\includegraphics[width=8cm]{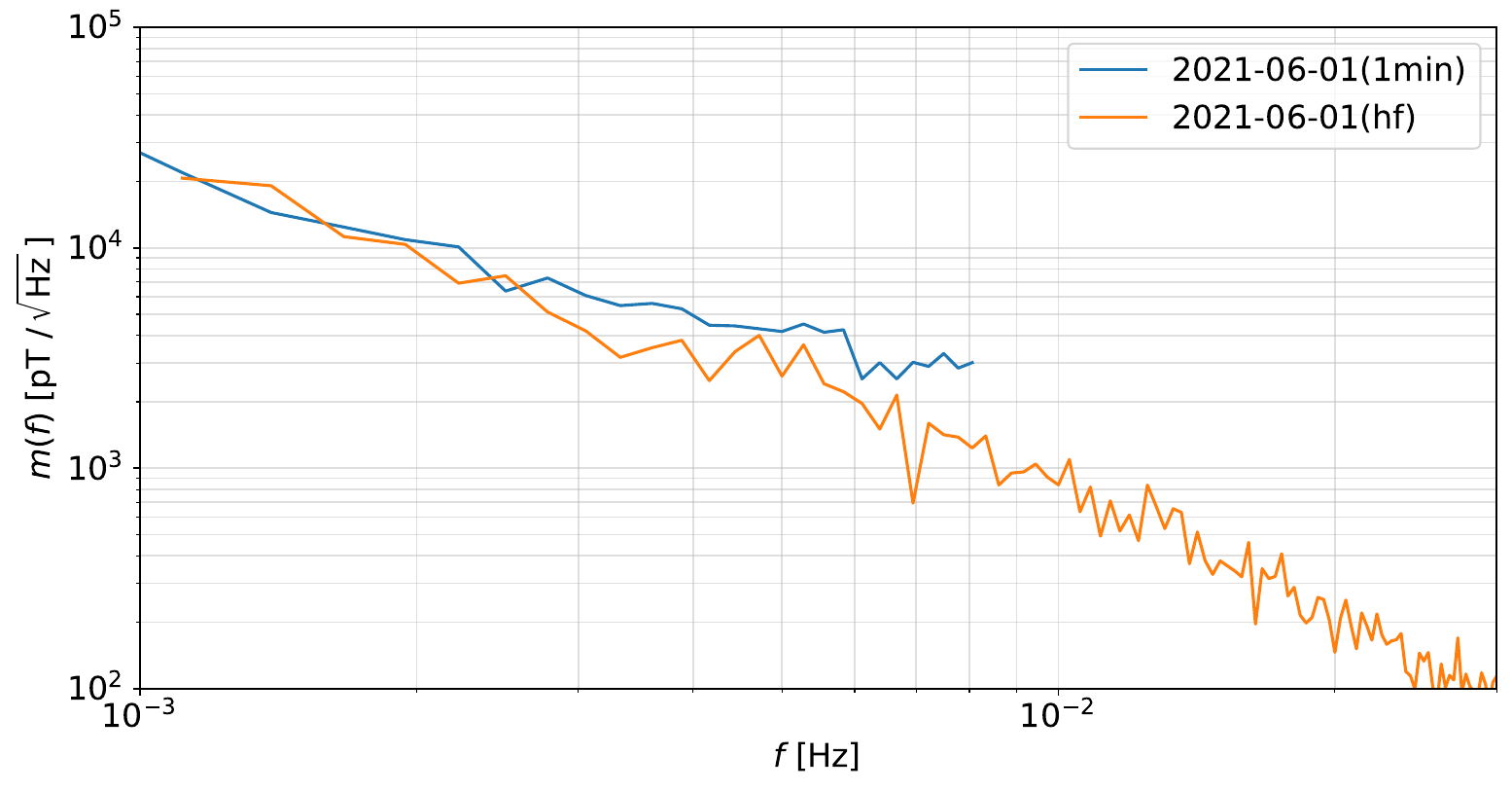}
\includegraphics[width=8cm]{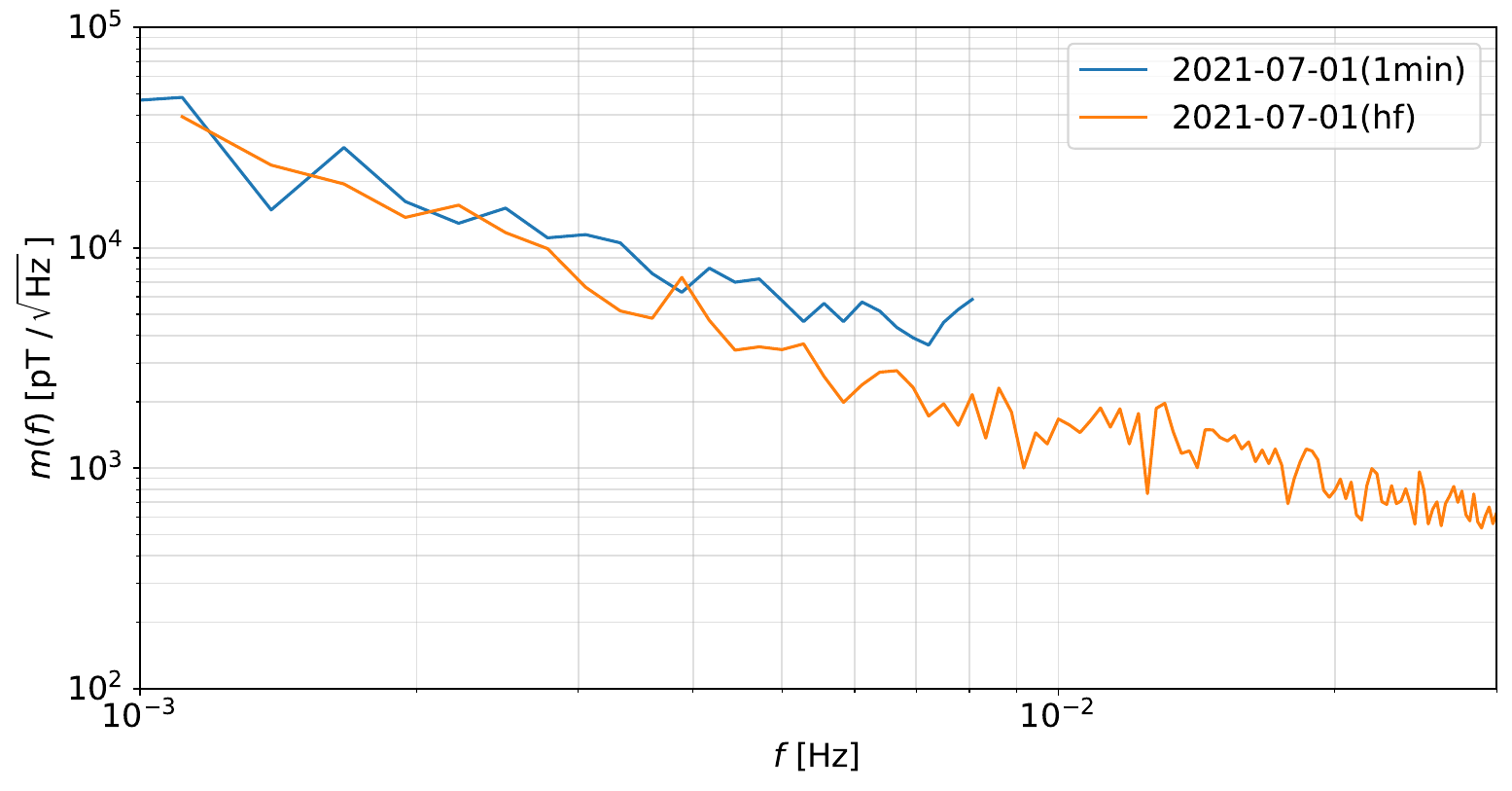}
\includegraphics[width=8cm]{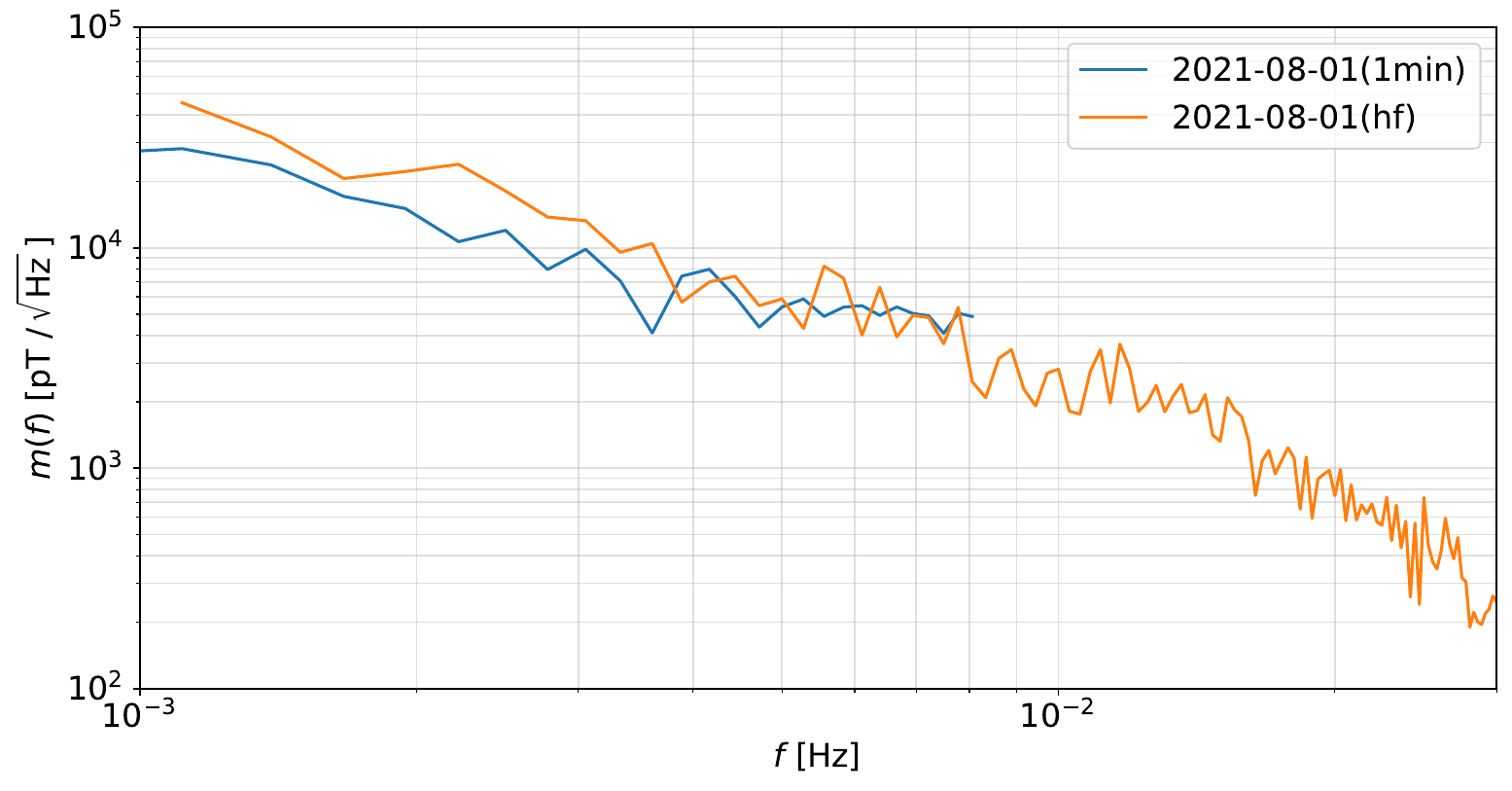}
\includegraphics[width=8cm]{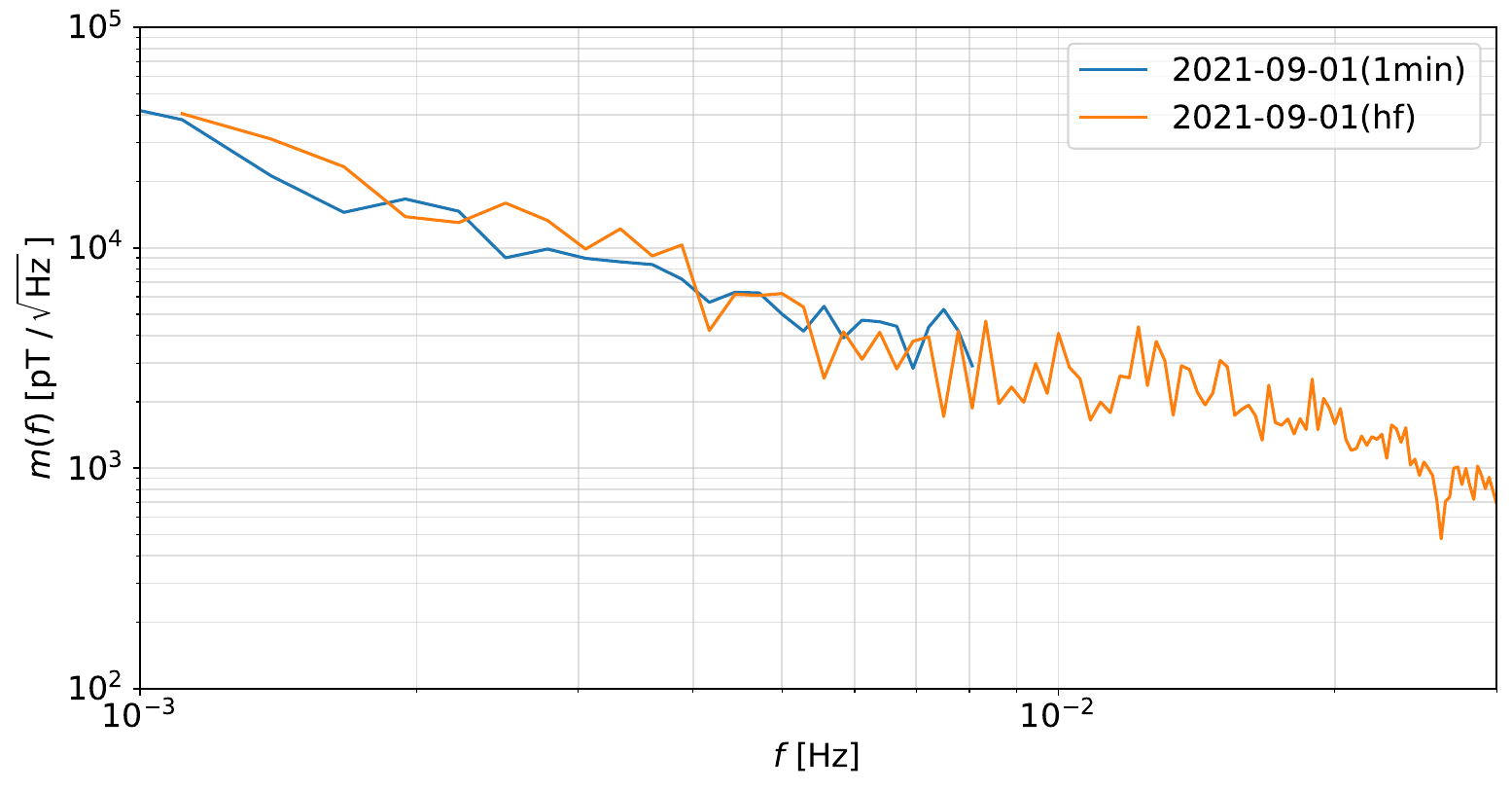}
\includegraphics[width=8cm]{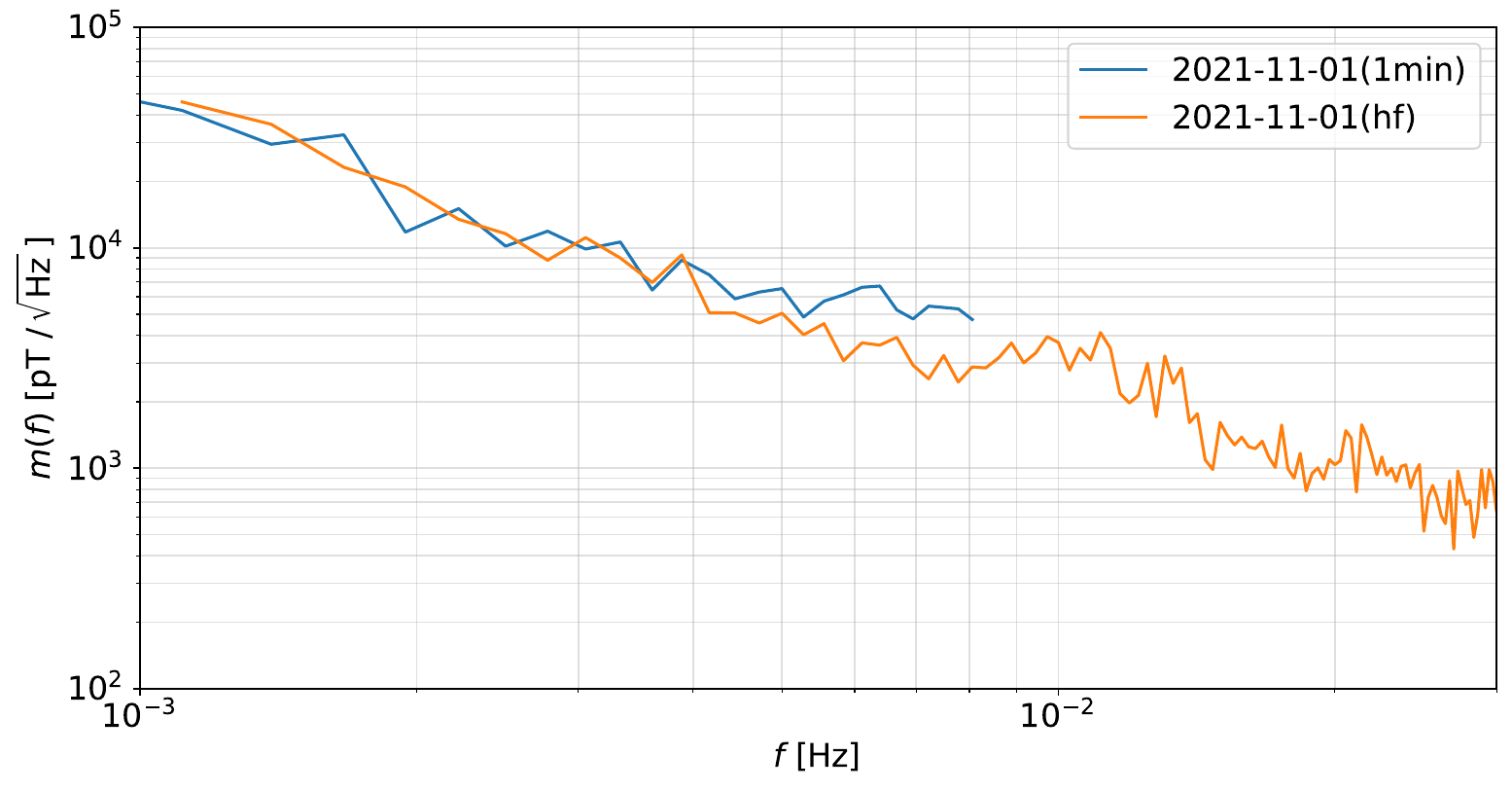}
\includegraphics[width=8cm]{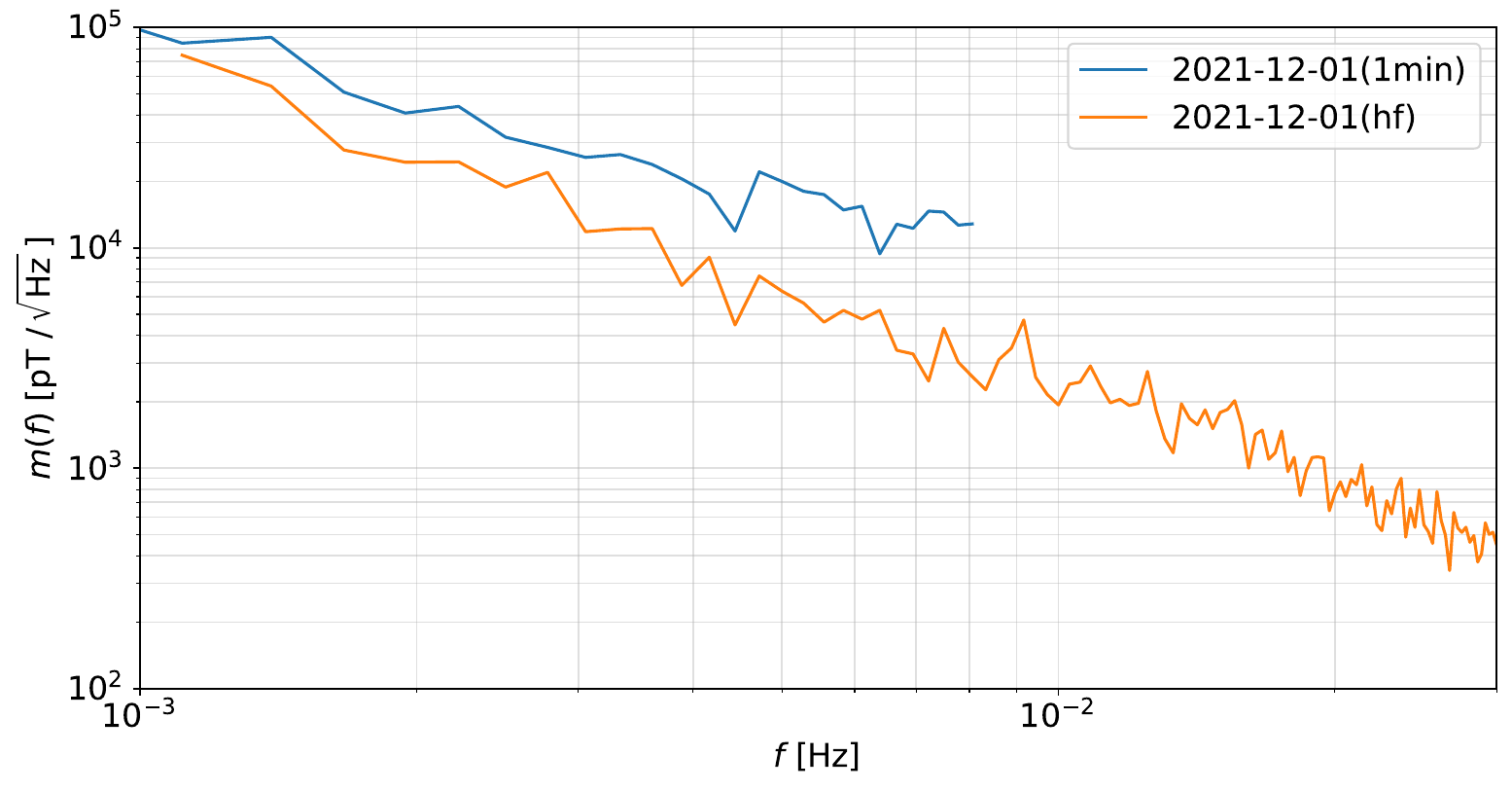}
\caption{Same as Fig.~\ref{fig:calibration-comparison} but the data are from one day in each month, 2021.}
\label{fig:calibration-comparison_2021}
\end{center}
\end{figure*}

\begin{figure*}[h]
\begin{center}
\includegraphics[width=8cm]{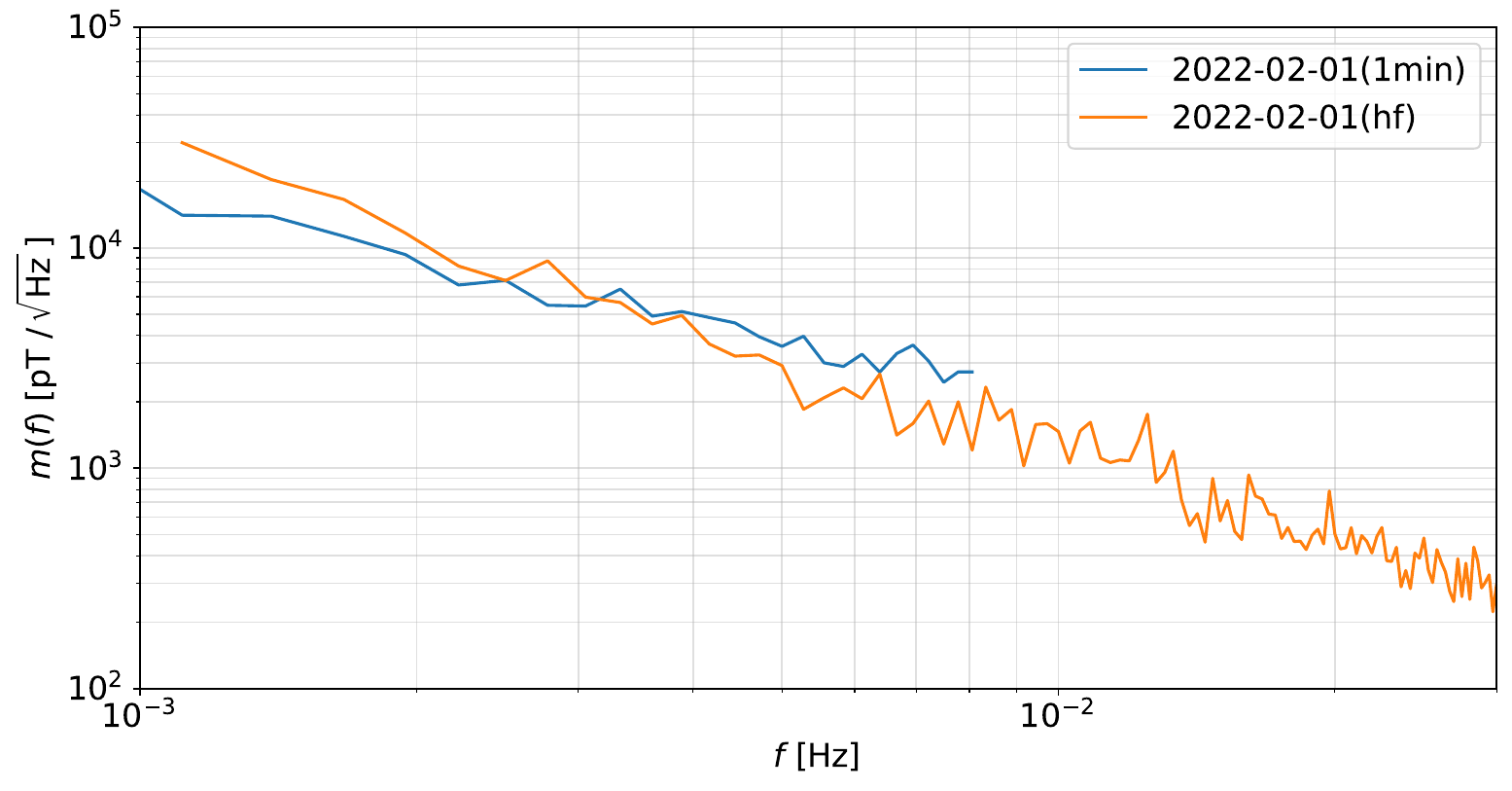}
\includegraphics[width=8cm]{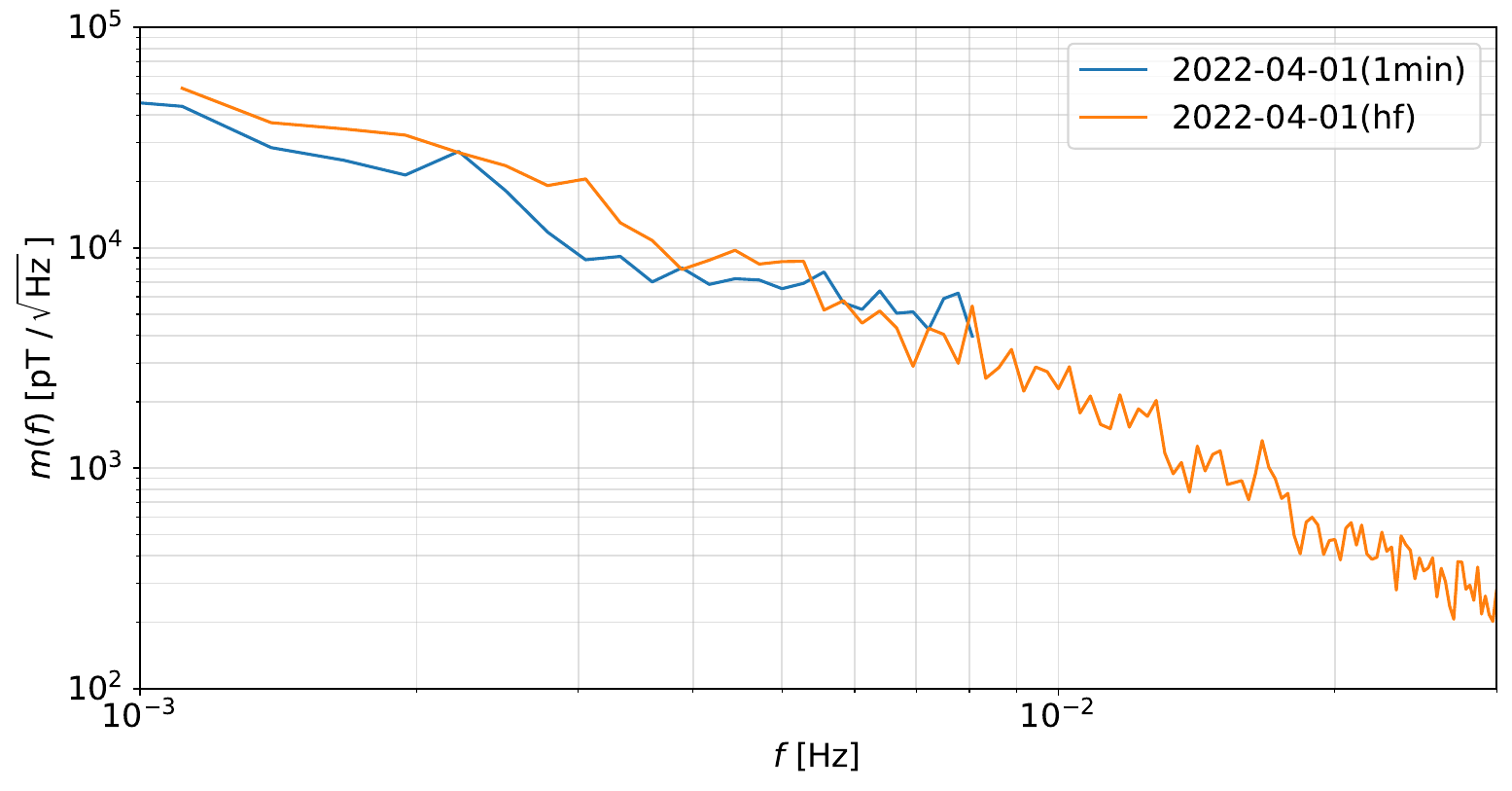}
\includegraphics[width=8cm]{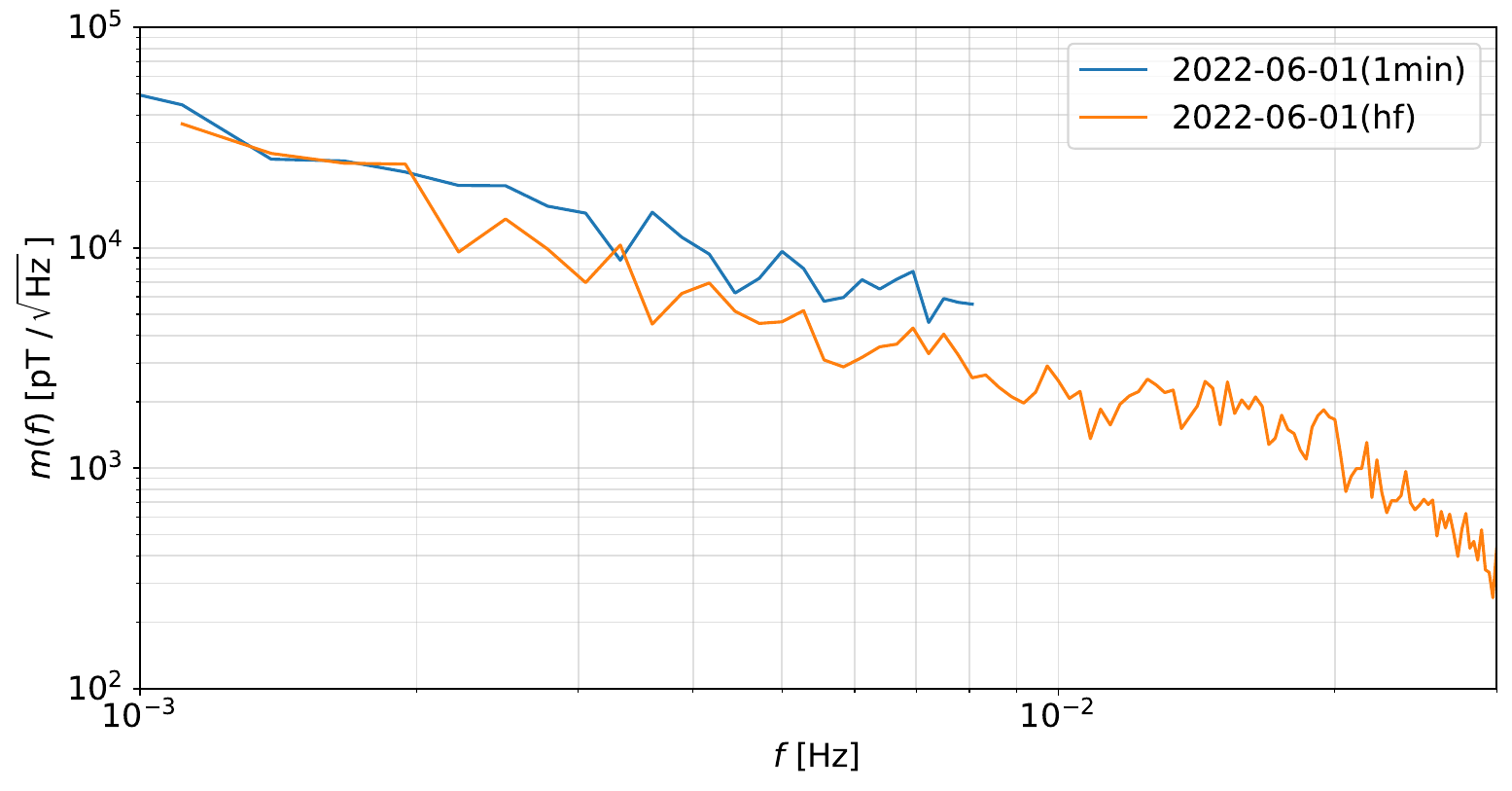}
\includegraphics[width=8cm]{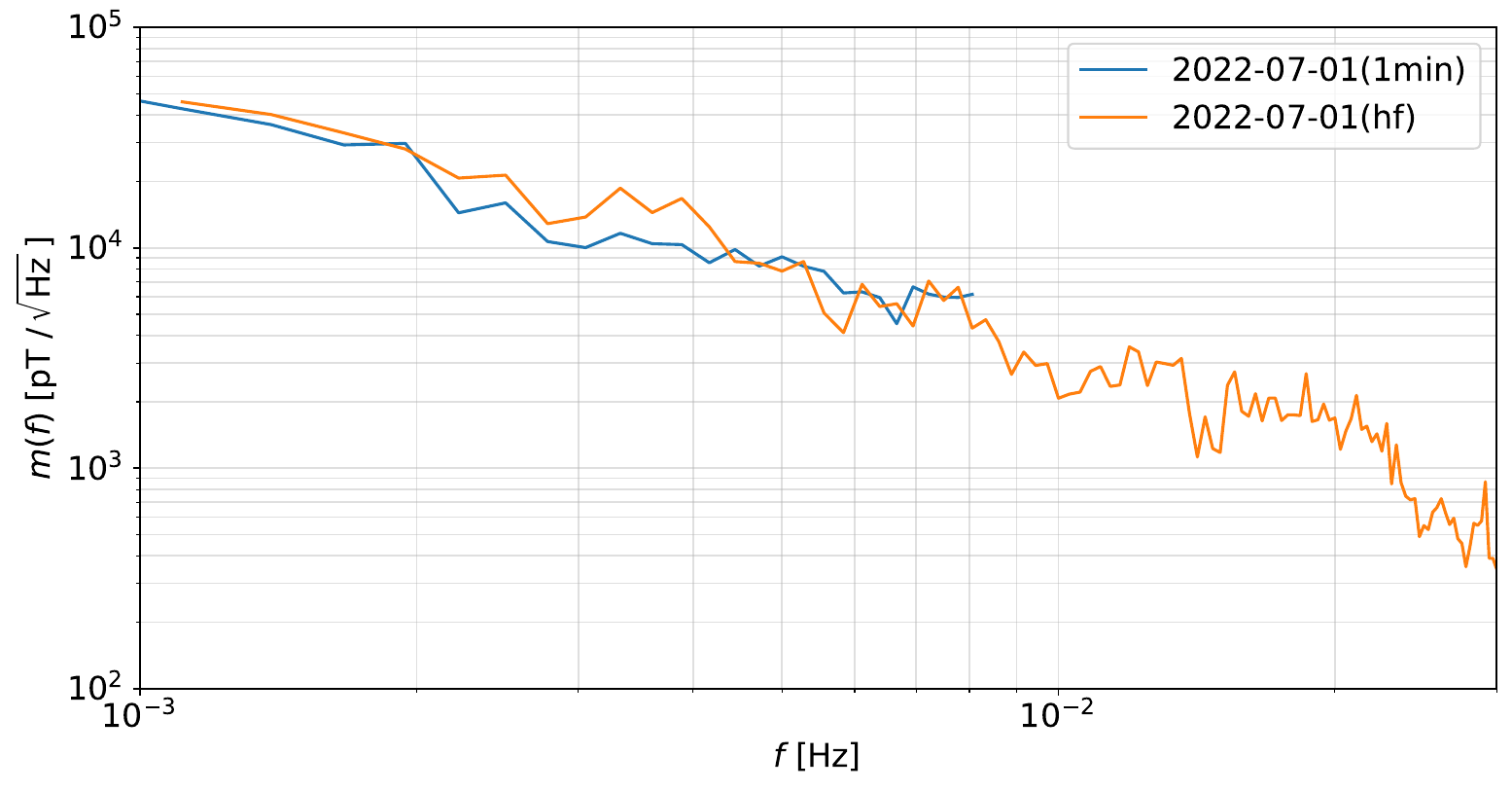}
\includegraphics[width=8cm]{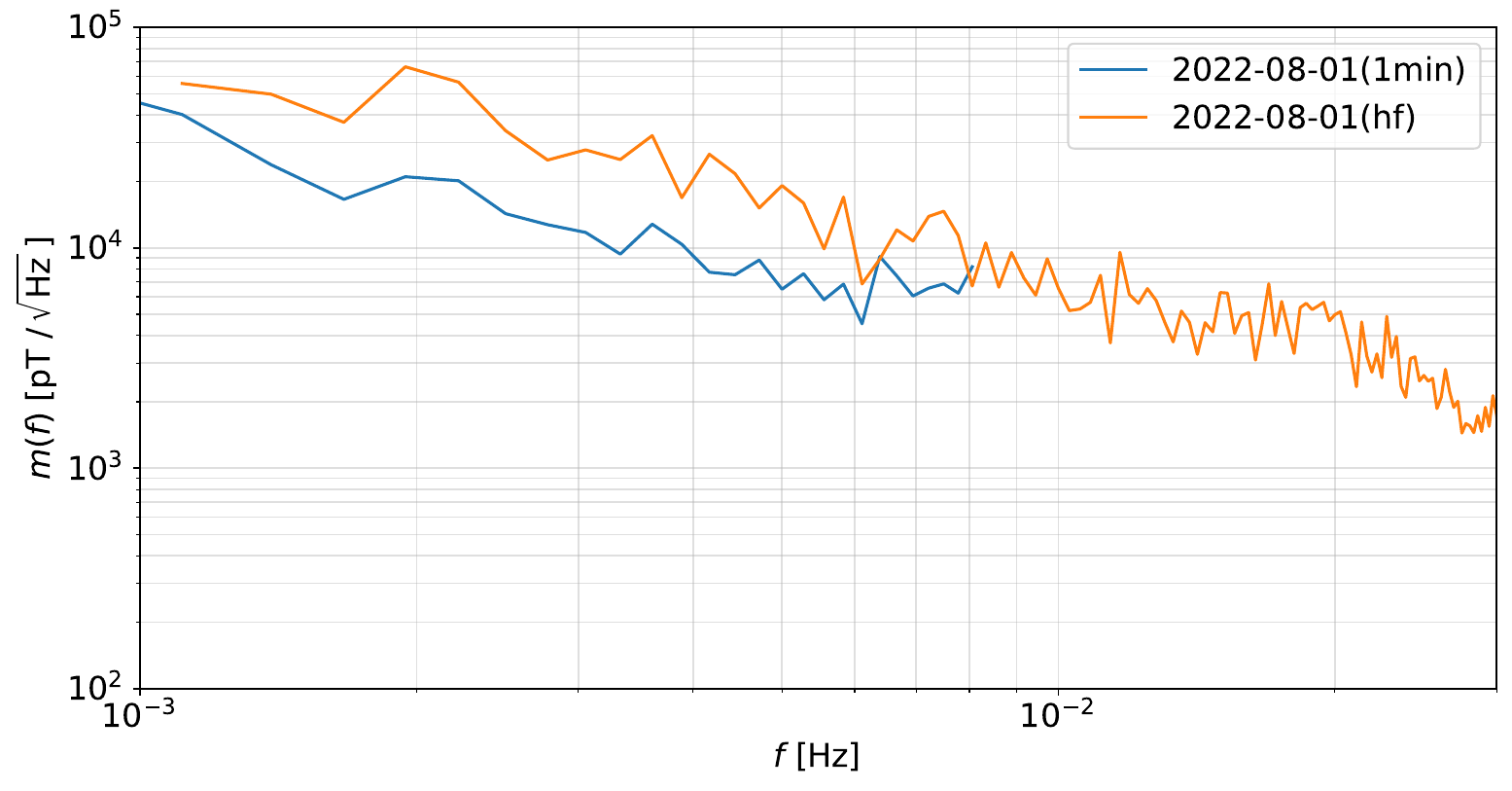}
\includegraphics[width=8cm]{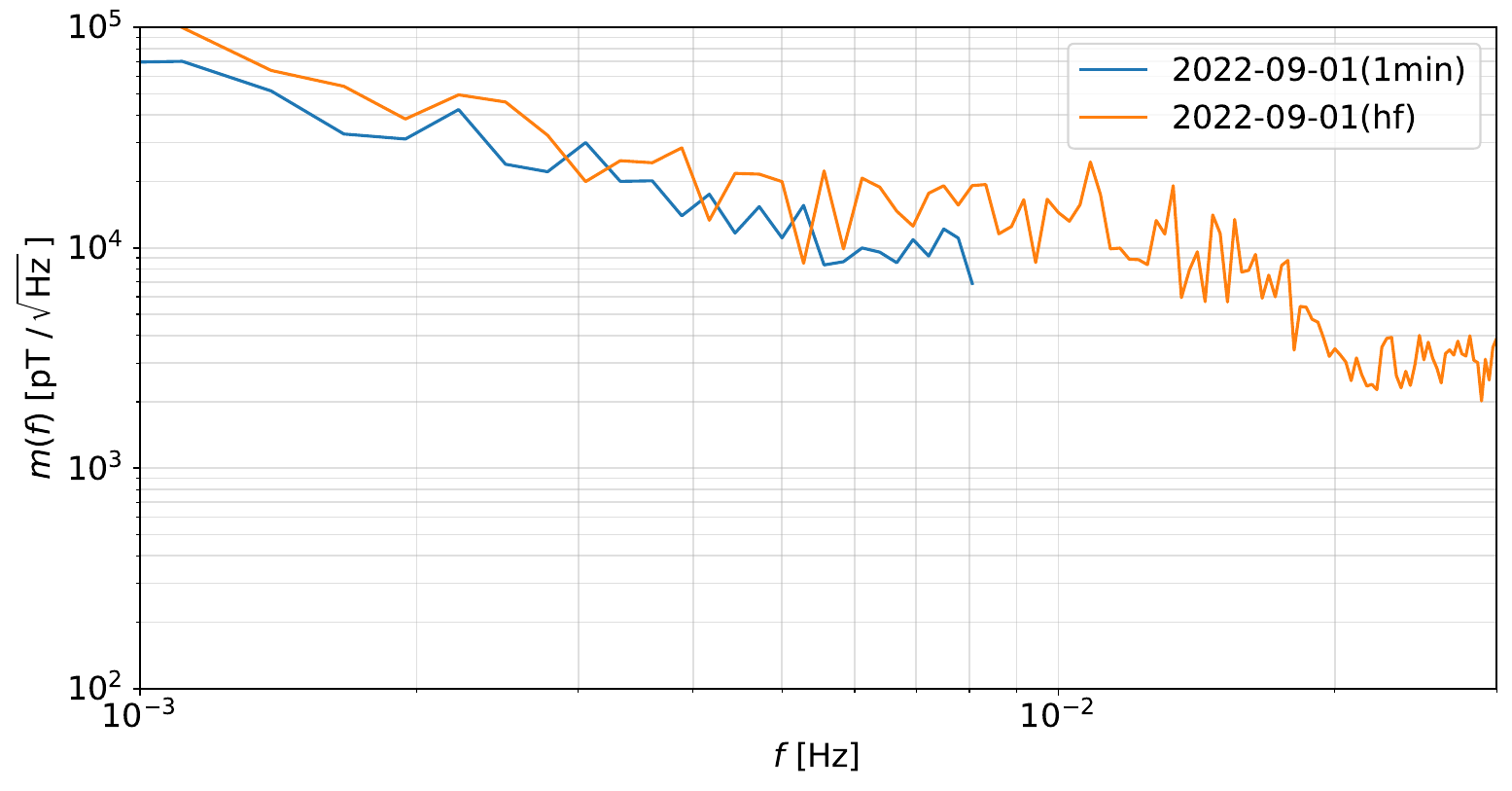}
\includegraphics[width=8cm]{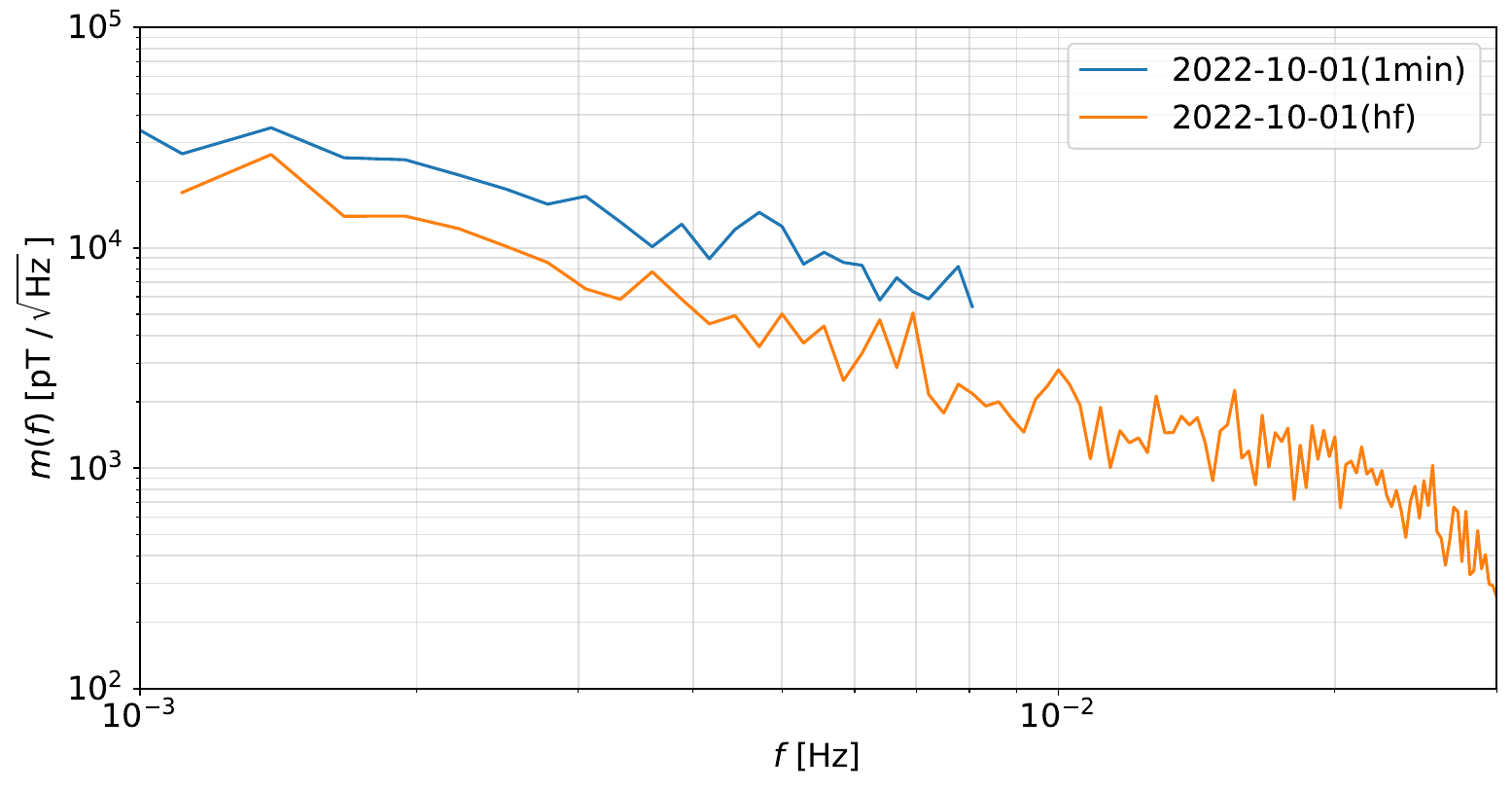}
\caption{Same as Fig.~\ref{fig:calibration-comparison} but the data are from one day in each month, 2022.}
\label{fig:calibration-comparison_2022}
\end{center}
\end{figure*}

The induction coil system at the Eskdalemuir observatory, UK~\cite{Beggan-Musur-2018} is very old. The instruments had been working since the 1970s and the digitizer since the 1990s. The calibration factors provided at the website of British Geological survey~\cite{BGS-data} are listed in Table~\ref{tab:calibration-factor}. However, they were measured long ago and have not been checked during the period of the data we use and until the instruments finally broke in August 2023. For the reliability of the data, we have investigated if the calibration factors are still correct during the period by comparing the magnetic field amplitude, $m(f) \equiv \sqrt{S(f)}$, where $S(f)$ is defined in Eq.~\eqref{eq:sf}, with that measured by another instrument in the same period. However, there is no independent high-frequency ($100\unit{Hz}$) sampling data at the Eskdalemuir observatory. We instead used the 1-minute ($0.017\unit{Hz}$) sampling data taken from INTERMAGNET~\cite{INTERMAGNET} and compare the spectral amplitude. First we pick up a day each year and compare the amplitude from 2013 to 2021 (the data on January 1 are missing in 2014 and 2022). The spectral amplitude is computed by dividing the 1-day data into 24 1-hour data and taking the 3/24-quantile amplitude from the minimum. This is determined by the trade-off between two factors. The lower quantile is more affected  by noise fluctuations and the spectrum is not smooth. The higher quantile is likely to be affected by transient noises and the spectrum happens to be large. The comparison of the 3/24-quantile amplitudes is shown in Fig.~\ref{fig:calibration-comparison}. They are consistent with each other, verifying that the calibration factors are correct. Second, for completeness, we compare the amplitudes in each month (a few months are missing due to non-existence of the data) during the later period (2020-2022) of the data that we use for the axion search. As shown in Figs.~\ref{fig:calibration-comparison_2020}--\ref{fig:calibration-comparison_2022}, they are consistent with each other, again verifying that the calibration factors are correct. 

\begin{table}[t]
\begin{center}
\begin{tabular}{|c| c|}
\hline
frequency (Hz) & calibration factor [mV/nT] \\
\hline
100  &  50.151 \\
50  &  50.282 \\
32  &  50.258 \\
18  &  50.223 \\
10  &  50.209 \\
5.6  &  50.204 \\
3.2  &  50.199 \\
1.8  &  50.200 \\
1.0  &  50.169 \\
0.56  & 50.167 \\
0.32  & 50.157 \\
0.18  & 50.139 \\
0.10  & 50.077 \\
0.056 & 49.939 \\
0.032 & 49.540 \\
0.018 & 48.237 \\
0.010 & 44.639 \\
0.007 & 40.402 \\
0.005 & 34.041 \\
0.003 & 22.464 \\
0.002 & 13.736 \\
0.001 & 4.530 \\
\hline
\end{tabular}
\end{center}
\caption{Calibration factors converting nT to mV.}
\label{tab:calibration-factor}
\end{table}

\section{Noise amplitude distribution}
\label{app:noise-dist}

\begin{figure*}[h]
\begin{center}
\includegraphics[width=10.5cm]{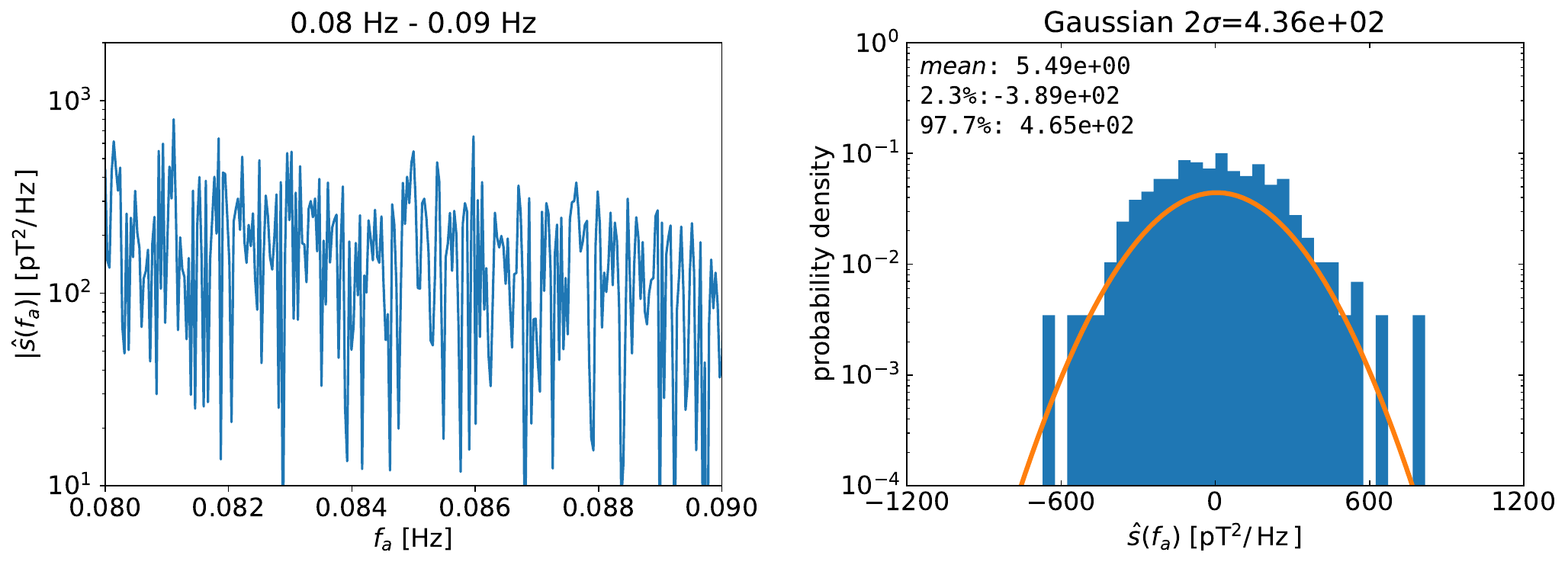}
\includegraphics[width=10.5cm]{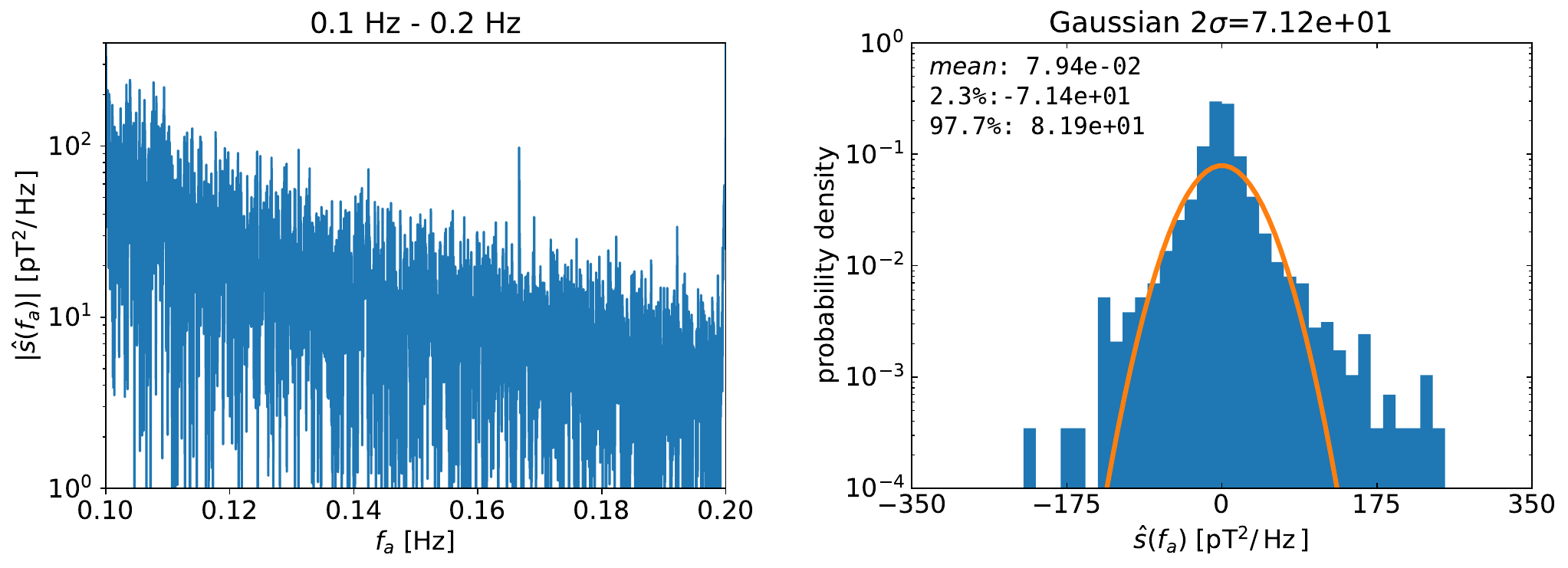}
\includegraphics[width=10.5cm]{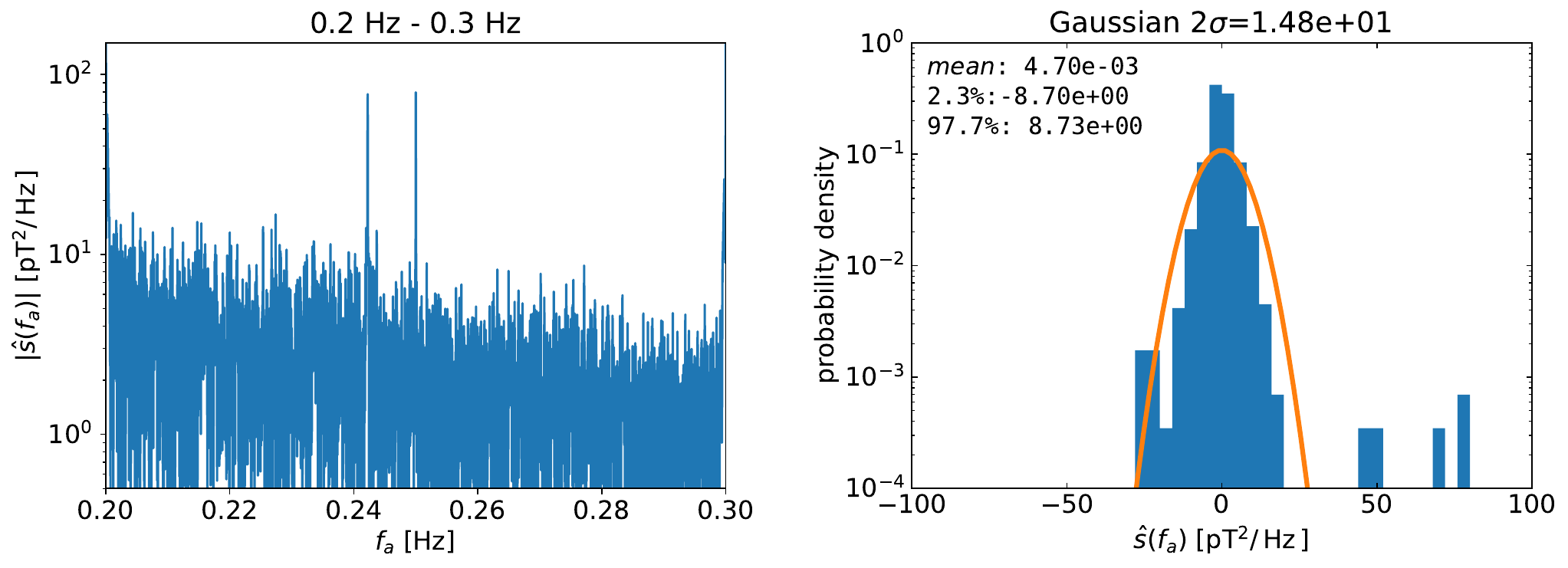}
\includegraphics[width=10.5cm]{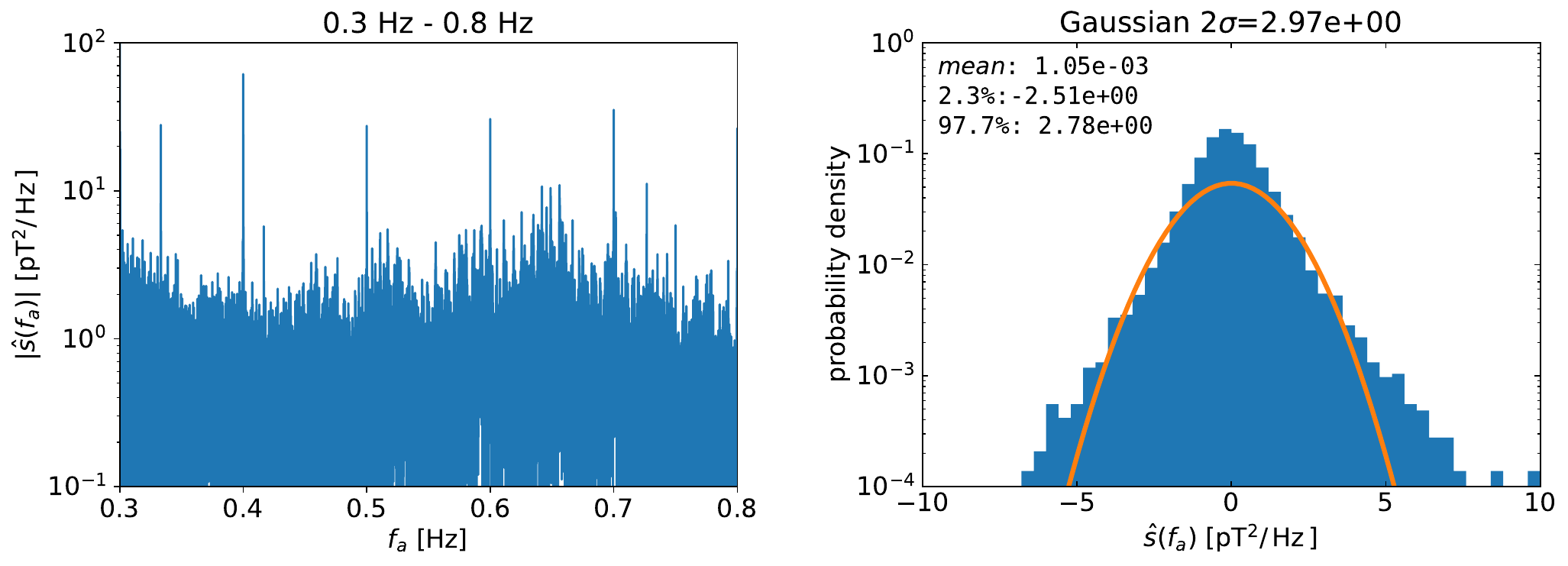}
\includegraphics[width=10.5cm]{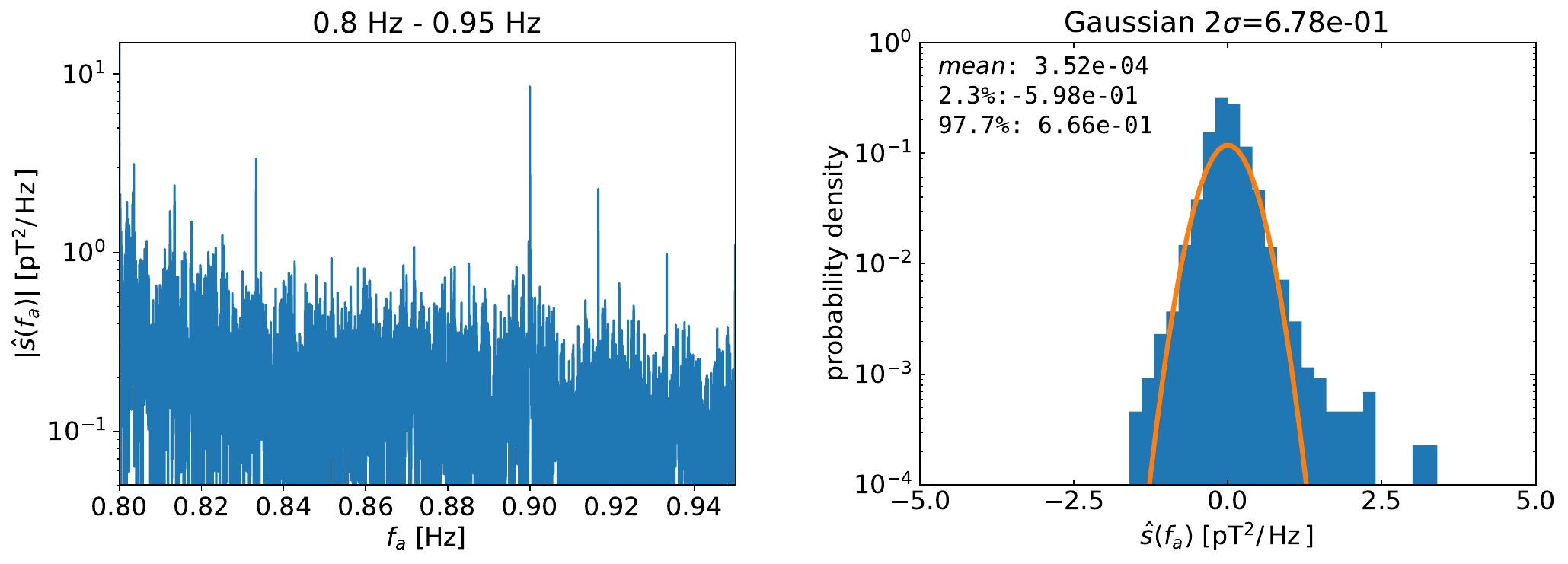}
\caption{Noise amplitude distributions at low frequencies. The three numbers on the right panels are the mean and the 2.3\% and 97.7\% of the cumulative distribution of $\hat{s}(f_{\rm a})$. The numbers above the right panels are twice of the standard deviation when the distributions are fitted by the Gaussian distribution.}
\label{fig:noise-dist-log-lowf}
\end{center}
\end{figure*}

\begin{figure*}[h]
\begin{center}
\includegraphics[width=10.5cm]{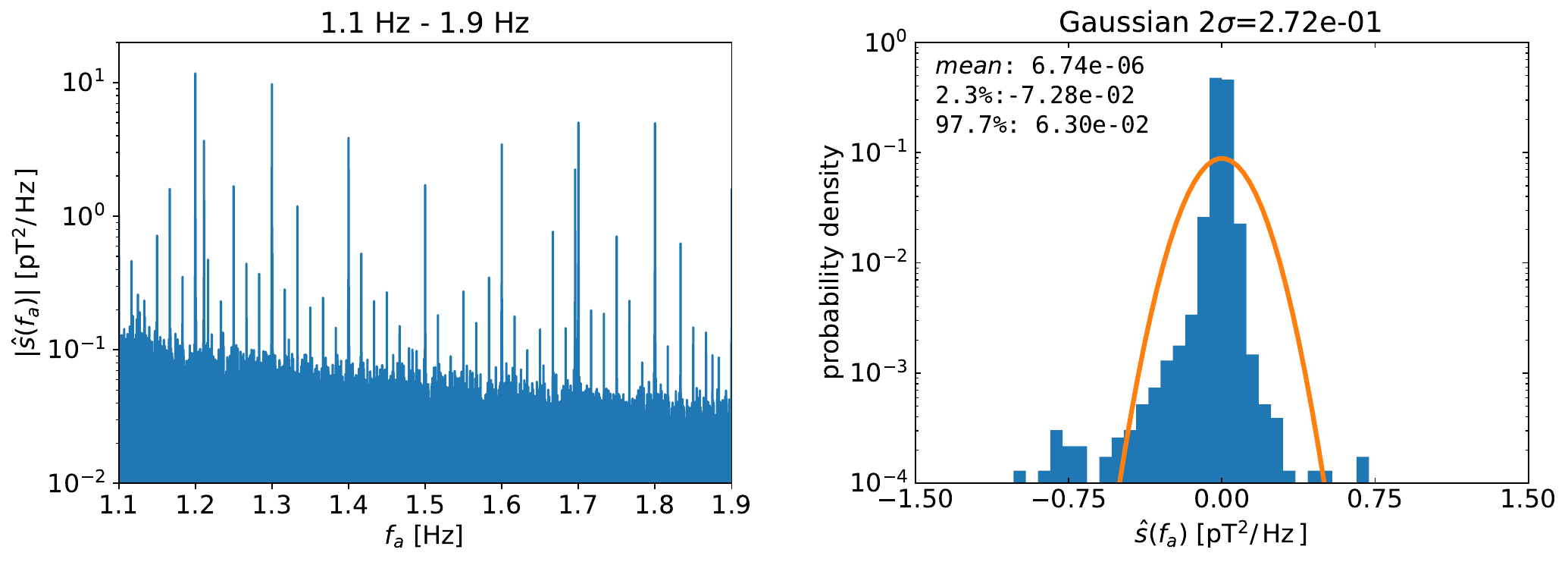}
\includegraphics[width=10.5cm]{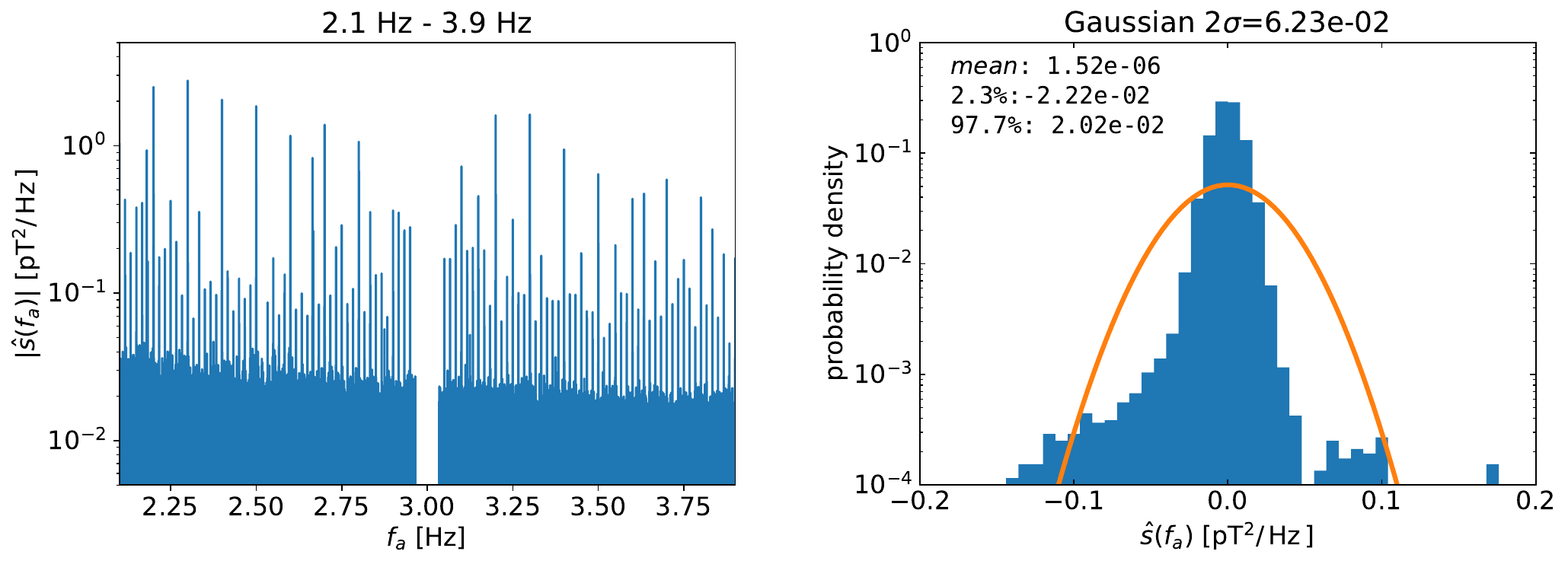}
\includegraphics[width=10.5cm]{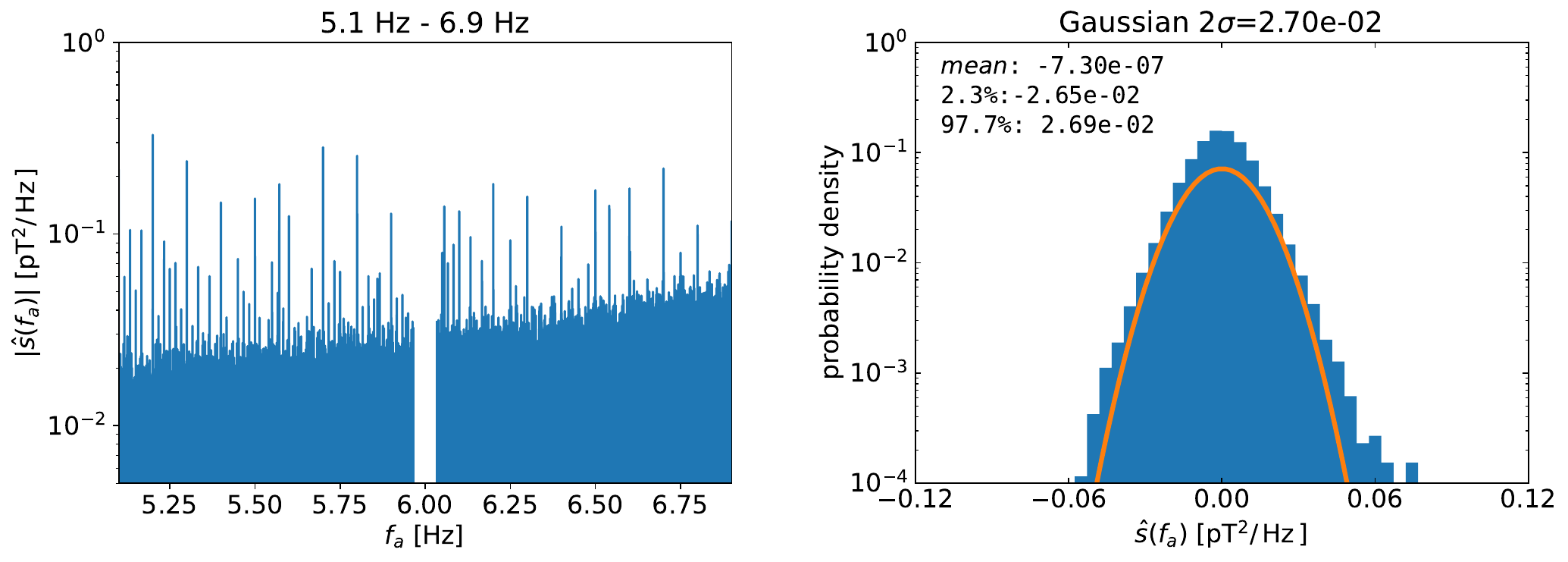}
\includegraphics[width=10.5cm]{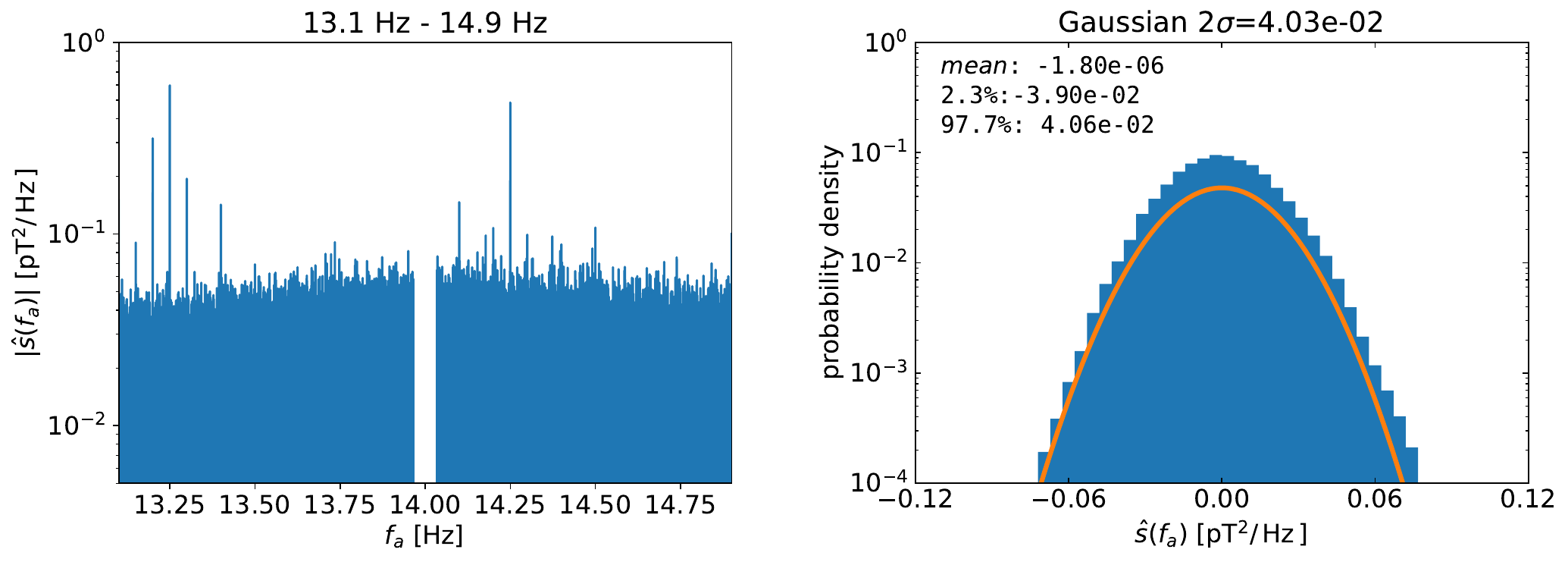}
\includegraphics[width=10.5cm]{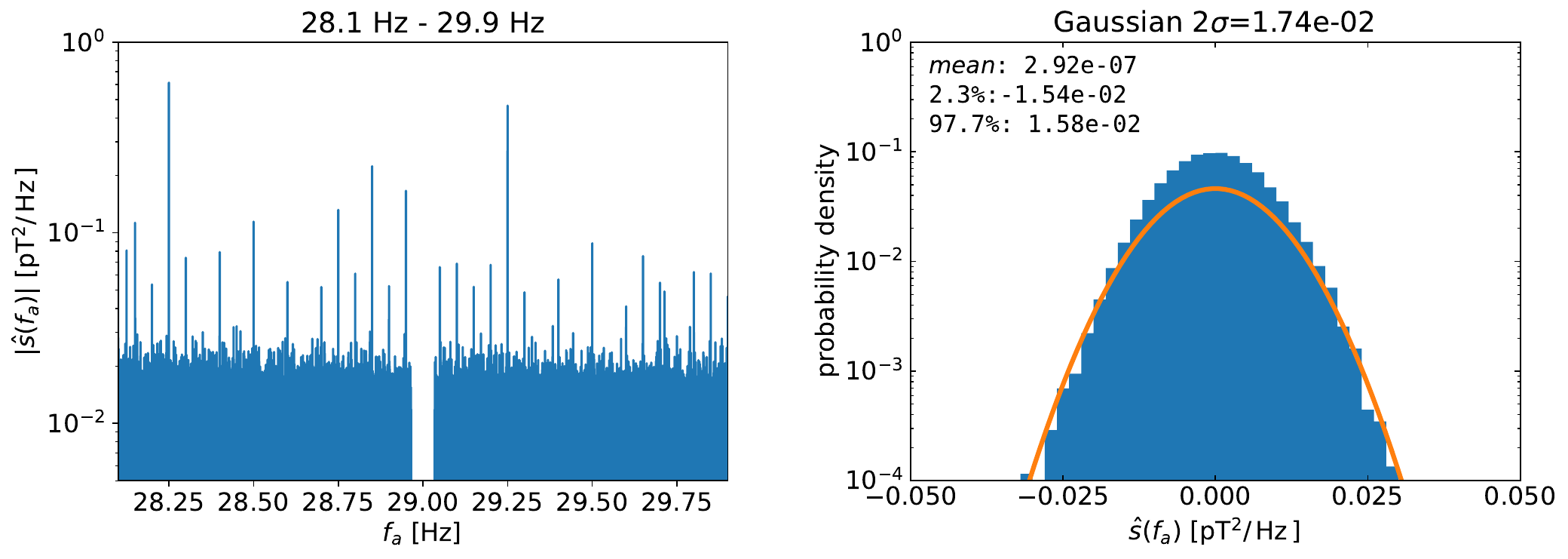}
\caption{Same as Fig.~\ref{fig:noise-dist-log-lowf} but at high frequencies.}
\label{fig:noise-dist-log-highf}
\end{center}
\end{figure*}

Noise properties are crucial to further investigate axion signal candidates and also to obtain an upper limit on the axion coupling strength. Here we show noise amplitude distributions in the low-frequency and high-frequency ranges. In Figs.~\ref{fig:noise-dist-log-lowf} and \ref{fig:noise-dist-log-highf}, the distributions of the noise amplitude are plotted in several frequency intervals at low and high frequencies, respectively. The distributions are well approximated by the Gaussian distribution with zero mean except for the frequency range of $1.1$--$3.9\unit{Hz}$. Particularly, at the frequency above $5.1\unit{Hz}$, the distributions are almost indistinguishable from the Gaussian distribution, though more axion signal candidates were found in the frequency range. In $1.1$--$3.9\unit{Hz}$, there are many strong lines on the spectrum, which broaden the width of the Gaussian distribution when fitted. 

\bibliography{ref}

\end{document}